\documentclass[12pt]{article}

\newlength{\abstractwidth}
\usepackage{epsf}
\usepackage{color}
\usepackage{graphicx}
\usepackage{tikz}
\usetikzlibrary{arrows,shapes,positioning}
\usetikzlibrary{decorations.markings}
\usepackage[rightcaption]{sidecap}
\tikzstyle arrowstyle=[scale=1]
\tikzstyle directed=[postaction={decorate,decoration={markings,
    mark=at position .65 with {\arrow[arrowstyle]{stealth}}}}]
\tikzstyle reverse directed=[postaction={decorate,decoration={markings,
    mark=at position .65 with {\arrowreversed[arrowstyle]{stealth};}}}]
\usetikzlibrary{positioning}

\usepackage{amssymb}
\usepackage{latexsym}

\renewcommand{\thefootnote}{\fnsymbol{footnote}}
\renewcommand{\thanks}[1]{\footnote{#1}}
\newcommand{\starttext}{
\setcounter{footnote}{0}
\renewcommand{\thefootnote}{\arabic{footnote}}}

\newcommand{\bea}{\begin{eqnarray}}
\newcommand{\eea}{\end{eqnarray}}
\newcommand{\be}{\begin{eqnarray}}
\newcommand{\ee}{\end{eqnarray}}

\def\cB{{\cal B}}
\def\cC{{\cal C}}

\def\cF{{\cal F}}
\def\cG{{\cal G}}
\def\cH{{\cal H}}

\def\cL{{\cal L}}
\def\cM{{\cal M}}

\def\cO{{\cal O}}

\def\cR{{\cal R}}

\def\cV{{\cal V}}

\def\mH{\mathfrak{H}}

\def\mS{\mathfrak{S}}

\def\mpp{\mathfrak{p}}

\def\ZZ{{\mathbb Z}}
\def\RR{{\mathbb R}}

\def\CC{{\mathbb C}}

\def\Im{{\rm Im \,}}

\def\half{{1\over 2}}
\def\p{\partial}
\def\DD{\nabla  }
\def\DDb{\bar \nabla }

\def\f{\varphi}

\def\ep{\varepsilon}

\def\no{\nonumber}
\def\sm{\smallskip}

\def\L{R}
\def\w{u}
\def\weight{weight \,}

\begin{document}
\starttext
\setcounter{footnote}{0}

\begin{flushright}
 DAMTP-2016-21\\
2016 April 14 \\
revised version: 2017 September 6 
\end{flushright}

\vskip 0.3in

\begin{center}

{\Large \bf Identities between Modular Graph Forms}\footnote{Research supported in part by the National Science Foundation under grants PHY-13-13986 \\ and PHY-16-19926.}

\vskip 0.3in

{\large \bf Eric D'Hoker$^{(a)}$  and Michael B. Green$^{(b,c)}$ } 

\vskip 0.1in

{ \sl (a) Mani L. Bhaumik Institute for Theoretical Physics} \\
{\sl Department of Physics and Astronomy }\\
{\sl University of California, Los Angeles, CA 90095, USA} 

\vskip 0.05in

{ \sl (b) Department of Applied Mathematics and Theoretical Physics }\\
{\sl Wilberforce Road, Cambridge CB3 0WA, UK}

\vskip 0.05in

{\sl (c) Queen Mary University of London, Centre for Research in String Theory,}\\
{\sl School of Physics, Mile End Road, London, E1 4NS, England}

\vskip 0.07in

{\tt \small dhoker@physics.ucla.edu; M.B.Green@damtp.cam.ac.uk}

\vskip 0.3in

\end{center}

\begin{abstract}

This paper investigates the  relations between modular graph forms, which are generalizations of the modular graph functions that were introduced in earlier papers motivated by the structure of the low energy expansion of genus-one Type II superstring amplitudes.  These modular graph forms are multiple sums associated with decorated Feynman graphs on the world-sheet torus. The action of standard differential operators on these modular graph forms admits an algebraic representation on the decorations. First order differential operators are used to map general non-holomorphic modular graph functions to holomorphic modular forms. This map is used to provide proofs of the identities between modular graph functions for weight less than six conjectured in earlier work, by mapping these identities to relations between holomorphic modular forms which are proven by holomorphic methods. The map is further used to exhibit the structure of  identities at arbitrary weight.
\end{abstract}

\newpage

\setcounter{tocdepth}{2}

\newpage

\baselineskip=15pt
\setcounter{equation}{0}
\setcounter{footnote}{0}

\newpage

\section{Introduction}
\setcounter{equation}{0}
\label{sec:1}

The structure of the low energy expansion of genus-one Type II superstring amplitudes leads one to associate  modular functions with Feynman graphs for a conformal scalar field on a torus. The resulting {\sl modular graph functions} exhibit a rich mathematical structure, which may be related to single-valued elliptic multiple polylogarithms, and single-valued multiple zeta values \cite{D'Hoker:2015qmf}.  Although some properties of the  space of modular graph functions have been investigated,  a deeper understanding of their general mathematical  structure is lacking. 

\sm

The {\sl weight} $w$ of any graph is the number of scalar Green functions in the graph while the number of closed loops is denoted by $L$.    We need only consider connected graphs since a disconnected graph is associated with a modular graph function that factors into the product of two lower weight modular graph functions.  Furthermore connected graphs that are one edge reducible (that become disconnected when one edge is removed) vanish so we need to consider only connected one edge irreducible graphs.   
\sm

There is a unique one-loop modular graph function with a given value of  $w$.  This is the non-holomorphic Eisenstein series~$E_w$ which satisfies the Laplace-eigenvalue equation $\Delta E_w = w(w-1) E_w$, where $\Delta$ is the Laplace-Beltrami operator on the Poincar\'e upper half plane.  Non-holomorphic Eisenstein series  are very familiar objects in number theory (for a historical account see  \cite{Weil:1977}; for general overviews see for example \cite{Siegel,Terras,Zagier:2008}).

\sm

Two-loop modular functions were studied extensively in \cite{D'Hoker:2015foa}, building on  earlier work in \cite{Green:2008uj,Green:1999pv}. They were found to obey systems of inhomogeneous Laplace-eigenvalue equations whose inhomogeneous parts  contain terms linear and quadratic in non-holomorphic Eisenstein series. With the help of group theoretic methods these systems were decoupled onto eigen-spaces with fixed weight and fixed eigenvalue of the Laplacian, and an infinite sequence of linear relations between modular graph functions of one- and two-loop order was shown to emerge \cite{D'Hoker:2015foa}.

\sm

Modular graph functions for three-loops and higher no longer systematically satisfy the type of Laplace-eigenvalue equations exhibited at one- and two-loops and, as a result, our understanding of the corresponding spaces of modular functions is much more limited. Nonetheless, a few isolated relations between the simplest modular graph functions with three and four loops were conjectured, based on the evidence of their matching Laurent expansion near the cusp \cite{D'Hoker:2015foa}. The simplest of these conjectured relations, namely relating a three-loop modular graph function to one- and two-loop modular graph functions, and referred to as the $D_4$ conjecture, was recently proven in \cite{D'Hoker:2015zfa} by direct summation of the lattice sums, thereby  generalizing a procedure used by Zagier at two-loops \cite{Zagier:2014}. The origin of these conjectured relations and their underlying nature remained, however, unexplained.

\sm

The goal of the present paper is to make progress on various fronts by enlarging the space of non-holomorphic modular graph functions to a space of modular forms of general holomorphic and anti-holomorphic modular weights.  This enlarged space will be associated with {\sl decorated graphs} that are described as follows. Each decorated graph consists of a set of  vertices, whose number will be denoted by $V$, joined to each other by {\sl decorated edges}. The decoration of each edge consists of a pair of integers that label respectively the {\sl exponents} of the holomorphic and anti-holomorphic momenta on the torus,   while the vertices in the decorated graph have valence three or higher, as will be explained more explicitly later.  When the integers within each pair are equal to one another on every edge, the graph evaluates to a modular graph function of the type considered in earlier papers \cite{D'Hoker:2015foa,D'Hoker:2015qmf}.  More generally, however, the  integers within each pair  labeling the edges will be unequal.  In that case the sum over all the edges of the holomorphic exponents  may equal the sum of the anti-holomorphic exponents, in which case the graph evaluates to a  $SL(2,\ZZ)$ modular function that lies outside the class of those considered in earlier papers.  If the sum of the holomorphic exponents  is not equal to the sum of anti-holomorphic exponents the graph evaluates to a $SL(2,\ZZ)$  modular form that has unequal holomorphic and anti-holomorphic modular weights.\footnote{Whereas it makes sense to define the weight  of a  modular graph function as the number of modular invariant Green functions in a graph,  this terminology is not useful when considering the more general graphs  with edges that carry unequal values for the holomorphic and anti-holomorphic exponents.  These graphs generally evaluate to modular forms.  As we will see later,  for these more general graphs the concept of the weight of the graph is replaced by the total values of these exponents when summed over all the edges of the graph.}

\sm

We shall refer to this enlarged space as the space of {\sl modular graph forms}.  This space includes the entire space of non-holomorphic modular graph functions as well as the space of holomorphic modular forms, and all variations in between these two. The modular covariant Cauchy-Riemann operators and the Laplace-Beltrami operator on the Poincar\'e upper half plane map the space of modular graph forms into itself, and admit simple algebraic representations on the array of pairs of integers which  decorate the graphs. In addition to the mathematical advantages of this enlargement, it is also motivated by string theory, since the construction of amplitudes with more than four external massless states involves such generalized modular forms (see for example \cite{Green:2013bza}). 

\sm

We will consider the action of modular covariant differential operators on modular graph forms in order to relate the existence of identities between modular graph functions (and more generally identities between modular graph forms) to the existence of  identities between holomorphic modular forms. We shall show that any one-loop subgraph which involves only holomorphic momenta, while being modular by construction, evaluates to a holomorphic dependence on the remaining momenta of the graph, thereby effectively reducing the number of loops in the graph. We shall refer to this process as {\sl holomorphic subgraph reduction}.

\sm

As an application of this formalism and the holomorphic subgraph reduction procedure, we shall provide a simple and short proof of the $D_4$ conjecture  that was proven laboriously in \cite{D'Hoker:2015zfa}, as well as of all the remaining outstanding conjectures of \cite{D'Hoker:2015foa}. These conjectures claim the vanishing of a modular invariant function $F$ which is a polynomial in modular graph functions, with rational coefficients,  homogeneous weight $w$, but different numbers of loops.  For high enough weight $w$, several different identities will exist.  Using a {\sl sieve algorithm} which will be defined and developed in this paper, and which is based on the  holomorphic subgraph reduction procedure, we will demonstrate that  each of the conjectured identities satisfies the very striking covariant derivative condition, 
\bea
(\nabla)^{w-1} F=0
\label{1a2}
\eea
where the modular covariant Cauchy-Riemann operator $\nabla$ is  defined by,
\bea
\nabla = 2 i \tau_2 ^2 \frac{\partial}{\partial \tau}
\label{1a3}
\eea
and the upper half plane has been parametrized by the complex parameter $\tau$ with $\tau_2 = \Im (\tau) >0$. The identity $F=0$ will follow very simply by solving this equation with the boundary condition that $\lim_{\tau_2\to \infty} F=0$.
 
 \sm

Finally, we shall show that the sieve algorithm,  in conjunction with the holomorphic subgraph reduction procedure,  not only provides a method for proving known conjectured relations, but more generally may be used to systematically construct further identities between modular graph  forms.

\sm

The remainder of this paper is organized as follows. In section \ref{sec:2}, we   introduce modular graph forms, discuss their symmetry properties, and exhibit the action of the covariant Cauchy-Riemann operators and the Laplace-Beltrami operator. In section \ref{sec:3}, we illustrate these general concepts for dihedral graphs (graphs with $V=2$). In section \ref{sec:4}, we develop the strategy for proving modular graph function identities, and for relating these identities to the existence of identities between holomorphic modular forms and prove the $D_4$ conjecture by way of example. In section \ref{sec:5}, we discuss and concretely evaluate holomorphic subgraph reduction.  In section \ref{sec:6}, the formalism is used to prove the conjectures proposed in \cite{D'Hoker:2015foa}  for the dihedral modular graph functions $D_{3,1,1}$ and $D_5$. In section \ref{sec:10}, the formalism is applied to  trihedral graphs (graphs with $V=3$), and is used to prove the $D_{2,2,1}$ conjecture of \cite{D'Hoker:2015foa} in section \ref{sec:11}. A discussion of future avenues of investigation is given in section \ref{sec:12}, while various technical aspects are relegated to two appendices.

\section{Modular Graph Forms}
\setcounter{equation}{0}
\label{sec:2}

We begin this section with a brief review of modular graph functions, 
and then motivate and define their generalization to modular graph forms.

\subsection{Review of modular graph functions}
\label{sec:21}

An oriented torus $\Sigma $ with complex structure modulus $\tau$ may be represented in the complex plane by the quotient $\CC / \Lambda$, where $\Lambda $ is the lattice given by $\Lambda = \ZZ  \oplus \tau \ZZ$. The torus may be parametrized by a complex coordinate $z$ or equivalently by two real coordinates $\alpha, \beta \in \RR/\ZZ$ related  by $z=\alpha + \beta \tau$, and the volume form on $\Sigma$ is normalized to $d^2z = {i \over 2} dz \wedge d\bar z = d \alpha \wedge d \beta $. The moduli space $\cM_1$ of orientable tori, or equivalently of genus-one Riemann surfaces, may be represented by a fundamental domain for the action of $PSL(2, \ZZ)$ on the Poincar\'e upper half plane, and may be parametrized by $\tau = \tau_1, + i \tau_2$ with $\tau_1, \tau_2 \in \RR$. When a choice is required, we shall choose  $\cM_1= \{ \tau, ~ 0 < \tau_2, ~ 1 \leq |\tau|, ~ |\tau_1 | \leq \half \}$.

\sm

The fundamental ingredient in the construction of modular graph functions is the scalar Green function on $\Sigma$, defined to obey,
\bea
\label{3a1}
 \p _{\bar z} \p_z \, G(z|\tau) = - \pi \delta ^{(2)} (z) + { \pi \over \tau_2}
 \hskip 1in 
 \int _\Sigma d^2 z \, G(z|\tau)=0\, .
\eea
The Dirac delta function $\delta ^{(2)}(z)$ is normalized so that $\int _\Sigma d^2 z \, \delta ^{(2)} (z)=1$.
For our purpose, it is convenient to express the Green function $G(z|\tau)$ as a Fourier sum,
\bea
\label{3a2}
G(z|\tau) = \sum _{p \in \Lambda } ' { \tau _2 \over \pi |p|^2} \, e^{2 \pi i (n \alpha  - m \beta )}\, .
\eea
The integers $m,n$ parametrize the discrete momenta $p=m +n \tau  \in \Lambda$ of the Fourier modes on the torus, while the prime on the summation symbol indicates that the contribution from $p=0$ is not to be included in the sum. 
\sm 

 The generating function of modular graph functions with  $N$ vertices is given by,
\bea
\label{3a3}
\cB_N (s_{ij}|\tau) = \prod _{k=1}^N \int _\Sigma { d^2 z_k \over \tau_2} \,
\exp \left ( \sum _{1 \leq i < j \leq N} s_{ij} G(z_i-z_j|\tau) \right )\, .
\eea
This expression is inspired by string theory, where it governs the amplitude for the scattering of four gravitons when $N=4$. For higher $N$, the formula generates only a subset of the graphs that contribute to the low energy expansion of the $N$-graviton superstring amplitude. 

\sm

In string theory, the parameters $s_{ij}$ are related to the momenta of the incoming and outgoing string states and obey certain interrelations. Momentum conservation in the string amplitude requires the relations, 
\bea
\sum_{1\le i < j \le N} s_{ij}=0\, .
\eea
As a result of these relations,  the zero mode of the Green function in (\ref{3a3}) cancels out of the expression (\ref{3a3}) and the string amplitude is conformal invariant, and independent of the choice of metric for given modulus $\tau$. The expression for an individual graph function, obtained by expanding $\cB_N (s_{ij} |\tau)$ in powers of $s_{ij}$, does depend on the zero mode, but choosing to set it to zero  by imposing the normalization condition (\ref{3a1}) will considerably simplify the structure of  the modular graph functions and the relations between them.  
  
\sm

To generate all modular graph functions we shall drop all such interrelations between the $s_{ij}$ and view them as independent complex parameters, but retain the normalization of (\ref{3a1}). With this set-up, modular graph functions $\cB_N(s_{ij}|\tau)$ form a subspace of the modular functions $\cB_{N+1}(s_{ij}|\tau)$, obtained by setting $s_{i\, N+1}=0$ for all $i =1, \dots, N$.
For any fixed value of $\tau$ with $\tau_2 >0$, the integrals over $z_k$ are absolutely convergent for $|s_{ij}| < 1$ and may be analytically continued throughout $\CC$. The resulting $\cB_N(s_{ij}|\tau)$ is invariant under the action of the modular group $PSL(2, \ZZ)$ on $\tau$ provided the variables $s_{ij}$ are  invariant under $PSL(2,\ZZ)$.

\subsubsection{Graphical expansion}

The series expansion of $\cB_N (s_{ij}|\tau)$ in a multiple power series in the variables $s_{ij}$ may be organized graphically in terms of Feynman graphs, for a scalar quantum field on the torus~$\Sigma$, with at most $N$ interaction vertices. Each term in this expansion corresponds to a specific graph, and will evaluate to a specific non-holomorphic modular function, whence the terminology {\sl modular graph function}. The building block of the modular graph functions is the scalar Green function $G(z_i-z_j|\tau)$, which we associate  with an edge in the graph between vertices $z_i$ and $z_j$.  We shall use the following graphical representation,
\bea
\tikzpicture[scale=1.7]
\scope[xshift=-5cm,yshift=-0.4cm]
\draw (1,0) -- (2.5,0) ;
\draw (1,0) [fill=white] circle(0.05cm) ;
\draw (2.5,0) [fill=white] circle(0.05cm) ;
\draw(3.7,0) node{$= ~ ~ G(z_i-z_j|\tau)\, . $};
\draw (1,-0.25) node{$z_i$};
\draw (2.5,-0.25) node{$z_j$};
\endscope
\endtikzpicture
\label{fig1}
\eea
The integration over the position of the vertex $z$ on which $r$  Green functions end will be denoted by an unmarked  filled black dot, in contrast with an unintegrated vertex $z_i$ which will be represented by a marked unfilled white dot. The basic ingredient in the graphical notation is depicted in the graph below,
\bea
\tikzpicture[scale=1.7]
\scope[xshift=-5cm,yshift=-0.4cm]
\draw (2,0.7) -- (1,0) ;
\draw (2,0.7) -- (1.5,0) ;
\draw (2,0.7) -- (2.5,0) ;
\draw (2,0.7) -- (3,0) ;
\draw (2,0.2) node{$\cdots$};
\draw [fill=black]  (2,0.68)  circle [radius=.05] ;
\draw (1,0)    [fill=white] circle(0.05cm) ;
\draw (1.5,0) [fill=white] circle(0.05cm) ;
\draw (2.5,0)    [fill=white] circle(0.05cm) ;
\draw (3,0) [fill=white] circle(0.05cm) ;
\draw(5.2,0) node{$\displaystyle= \ \int _\Sigma {d^2 z \over  \tau_2} \, \prod _{i=1}^r G(z-z_i|\tau) \, .$};
\draw (1,-0.25) node{$z_1$};
\draw (1.5,-0.25) node{$z_2$};
\draw (2.57,-0.25) node{$z_{r-1}$};
\draw (3.1,-0.25) node{$z_r$};
\endscope
\endtikzpicture
\label{fig2}
\eea
The coefficient of the monomial obtained as the product over all $ i < j $ of $s_{ij} ^{\nu_{ij}}$ in the power series expansion of $\cB_N (s_{ij}|\tau)$ is given as follows, 
\bea
\cC_\Gamma (\tau) = \prod _{k=1}^N \int _\Sigma { d^2 z_k \over \tau_2} \, \prod _{1 \leq i < j \leq N} 
G(z_i-z_j |\tau)^{\nu_{ij}}\, .
\eea
The  graph $\Gamma$ has $N$ vertices labelled $ i =1,\cdots,  N$ and $\nu_{ij}$ edges between vertices $i$ and $j$. The total number of edges in the graph, also referred to as the weight $w$ of $\Gamma$, is given by, 
\bea
w = \sum _{1 \leq i < j \leq N} \nu_{ij}\, .
\eea
By construction, the integral $\cC_\Gamma (\tau)$ is modular invariant, as it arises from the expansion in powers of $s_{ij}$ of the modular invariant generating function $\cB_N(s_{ij}|\tau)$, and therefore associates with a graph $\Gamma$ a non-holomorphic modular function $\cC_\Gamma (\tau) $.

\subsubsection{Fourier representation}

The integration over the position $z_i$ of the vertex $i$  enforces momentum conservation on all the Green functions that emanate from this vertex. Carrying out the integrations over the vertex positions $z _i$ for $i=1,\dots, N$ gives a constrained multiple sum representation for the graphs, studied for the case $N=4$ in \cite{Green:2008uj,Green:1999pv,D'Hoker:2015foa}, and given explicitly as follows,  
\bea
\label{gengraph}
\cC_\Gamma (\tau) =  \sum_{p_1,\dots,p_w \in \Lambda} ' ~ \prod_{\alpha =1}^w
  {\tau_2\over  \pi|p_\alpha |^2}\, \prod_{i =1}^N
  \delta \left ( \sum_{\alpha =1}^w \Gamma _{i \alpha} \, p_\alpha \right )\, .
\eea 
The Kronecker $\delta$ symbol takes the value 1 when its argument vanishes and zero otherwise, and the coefficients   $\Gamma _{i \alpha}$ are given as follows,
\bea
\Gamma _{i \alpha} = \left \{ 
\matrix{\pm 1 & \hbox{if edge $\alpha$ ends on vertex $i$}\, , \cr & \cr
0  & \hbox{otherwise}\, , \cr}
\right .
\eea 
the sign being determined by the orientation of the momenta. Throughout, the Kronecker $\delta$-function on the momenta $p_\alpha$  will be shorthand for the following product, 
\bea
\label{3a9a}
\delta \left ( \sum_{\alpha =1}^w \Gamma _{i \alpha} \, p_\alpha \right )
=
\delta \left ( \sum_{\alpha =1}^w \Gamma _{i \alpha} \, m_\alpha \right )
\delta \left ( \sum_{\beta =1}^w \Gamma _{i \beta} \, n_\beta \right ) \, ,
\eea
where $p_\alpha = m_\alpha  + n_\alpha \tau $ and $m_\alpha, n _\alpha \in \ZZ$. In particular, the Kronecker $\delta$-functions do not depend on the modulus $\tau$. By translation invariance on the torus, one of the momentum conservation Kronecker $\delta$-functions is redundant, but we shall leave it in place in order to make the symmetry of the graph more manifest. 

\sm

The following additional points should be noted.   For given weight $w$, all the information on the graph $\Gamma$ is encoded in the coefficients $\Gamma_{i \alpha}$, the other parts of the sum in (\ref{gengraph}) being the same for all modular graph functions of weight $w$. The momentum sum representation of the modular graph function $\cC_\Gamma( \tau)$ makes its modular invariance manifest. The modular graph function is  invariant under an arbitrary relabeling of the vertices of its graph.  

\subsubsection{Irreducible graphs}
\label{irred}

Without loss of generality, we may focus on {\sl irreducible graphs}.  Any disconnected graph is reducible, and its modular graph function factorizes into the product of the modular graph functions of its connected components. Any graph which becomes disconnected upon cutting any one of its edges (also referred to as one-edge reducible graphs), including when a single edge ends on any one of its vertices, is also reducible and, with our normalization of~$G$, its associated modular graph function vanishes. Finally, any graph which becomes disconnected upon removing any one of its vertices is also reducible, and its associated modular graph function factorizes into the modular graph functions of the corresponding connected components. Any graph which is not reducible by the above criteria is {\sl irreducible}.

\subsection{Motivating modular graph forms}

As described in the Introduction, we shall enlarge the space of non-holomorphic modular graph functions, considered in earlier papers, to the space of {\sl modular graph forms}. The enlarged class of graphs are {\sl decorated} by assigning a pair of integers $(a_r, b _r)$ to each edge. We shall denote the total number of decorated edges by $R$, where $r=1,\cdots, R$ labels the decorated edges. The integers $a_r$ and $b _r$ are, respectively,  the exponents of the holomorphic momentum $p_r$ and the anti-holomorphic momentum $\bar p_r$ associated with the edge~$r$. For a general modular graph form, the integers $a_r$ and $b_r$ need not be equal to one another.  

\sm

We shall define the integers $a$ and $b$ by the sum of the exponents $a_r$ and $b_r$,
\bea
\label{ab}
a = \sum _{r=1}^R a_r 
\hskip 1in
b = \sum _{r=1}^R b_r 
\eea
and refer to the pair $(a,b)$ as the {\sl weight} of the modular graph form, which generalizes the definition of the weight $w$ associated with modular graph functions.  For the special case of modular graph functions we have $a_r=b_r$ for all $r$, so that $a=b$, and this number coincides with the {\sl weight} $w$ which has been introduced earlier. 

\sm

The weight $(a,b)$ encompasses also the assignment of the modular weight. Since modular graph forms are generally non-holomorphic, there is no natural assignment of a modular weight, as one may always multiply by a power of $\tau_2$, which is a modular form of modular weight $(-1,-1)$. But the difference of the holomorphic and the anti-holomorphic modular weights has a canonical meaning and, for a modular graph form of weight $(a,b)$, is given by the difference $a-b$, as will be explained in detail in the sections below.

\sm

Within the enlarged class of modular graph forms three distinct types are notable:
\begin{description}
\itemsep=-0.05in
\item[$(i)$] If $a=b$  the modular graph form has modular weight $(0,0)$.\footnote{Here the notation $(u,u')$ denotes the holomorphic and anti-holomorphic modular weights of a modular form, as reviewed in Appendix~\ref{app:AA}.}  While a modular function, it belongs to the class of modular graph functions considered in earlier papers only when  $a_r=b_r$ for all $r$.
\item[$(ii)$]   If  $a \ne 0$ and $ b_r=0$ for all $r$ the graph evaluates to a  holomorphic modular form multiplied by a power of $\tau_2$  (and if $ a_r = 0$ for all $r$ and $b \ne 0$ the graph evaluates to an anti-holomorphic modular form multiplied by a power of $\tau_2$).
\item[$(iii)$]  If  $a \ne b$ the graph evaluates to a non-holomorphic  modular form rather than a non-holomorphic modular function.
\end{description}
The definitions of reducible and irreducible  graphs, given in section \ref{irred} for modular graph functions,  readily carries over  to modular graph forms. 
Next, we shall elaborate on the physical and mathematical motivations for this enlargement.

\sm

One reason for including graphs of type $(i)$  is that they necessarily arise in closed Type II superstring theory when considering one loop scattering amplitudes with $N>4$ gravitons.   
In such amplitudes there are $N-4$ factors of the holomorphic derivative of the Green function  $\p_z G$, as well as  $N-4$ factors of the  anti-holomorphic derivative $\p_{\bar z} G$ in the graphs that enter the low energy expansion.  The Fourier representation of $\p_z G$ is given by,
\bea
\label{pG}
\p_z G(z|\tau) = - \sum _{ p \in \Lambda} ' { 1 \over p} \, e^{2 \pi i (n \alpha - m \beta)}
\eea
using the earlier notation where $z=\alpha + \beta \tau$ and $\alpha, \beta \in \RR/\ZZ$.
Allowing for arbitrary numbers of such derivatives and their complex conjugates alters the exponents of the momenta in (\ref{gengraph}), and renders the exponent of a holomorphic factor $p_r$ generally different from the exponent of its complex conjugate $\bar p_r$. Including derivatives on $G$  only adds positive powers of $p_r$ and $\bar p_r$ to the summand. However,  any decorated edge includes the effect of a chain of bivalent vertices.  The net  exponents of $p_r$ and $\bar p_r$  in such a chain may be arbitrary, mutually independent, positive or negative integers. Indeed, the effect of including a bivalent vertex connecting two derivative Green functions may be illustrated using the Fourier sum representation for two factors of $\p_zG$, and we find, 
\bea
\int _\Sigma { d^2 z \over \tau_2} \, \p_z G(z-z_i|\tau) \, \p_z G(z-z_j|\tau) 
= 
 \sum _{ p \in \Lambda} ' { 1 \over p^2} \, e^{2 \pi i n (\alpha_i - \alpha _j) - 2 \pi i m ( \beta_i - \beta_j)}
 \eea
for $z_i = m_i  + n_i \tau $, $z_j = m_j  + n_j\tau $, $z_i = \alpha _i + \beta _i \tau$, and $z_j = \alpha _j + \beta _j \tau$. We see that the inclusion of bivalent vertices can decrease the exponents of $p_r$ and $\bar p_r$. Therefore, an edge labeled $r$ in a generalized graph is specified by its momenta $p_r, \bar p_r $, and by a pair of integers $(a_r, b _r)$ which specify the {\sl exponents}  of the holomorphic and anti-holomorphic momenta. 
We shall graphically represent these exponents by a  {\sl decorated edge} as follows, 
\bea
\tikzpicture[scale=1.7]
\scope[xshift=-5cm,yshift=-0.4cm]
\draw[thick] (0,0) -- (2,0) ;
\draw (0.65,-0.15) [fill=white] rectangle (1.35,0.15) ;
\draw (1,0) node{$a_r, b _r $};
\draw[fill=white] (0,0)  circle [radius=.05] ;
\draw[fill=white] (2,0)  circle [radius=.05] ;
\draw (3.5,0) node{$\approx \hskip 0.15in (p_r) ^{- a_r} \, (\bar p _r) ^{- b _r}$};
\endscope
\endtikzpicture
\label{fig3}
\eea
The precise normalization will be given shortly. The use of the decoration labels allows us to include all the effects of bivalent vertices in these labels.  Our  graphs, which have decorated edges, therefore do not have any further bivalent vertices. 

\sm

String theory has a further source of contributions in which holomorphic and anti-holomorphic momenta arise with independent exponents. Worldsheet Weyl fermion fields have inherently holomorphic (or anti-holomorphic) dependence on $\tau$, tied directly to their worldsheet chirality.  An edge associated with a Weyl fermion carries a Szeg\H o kernel instead of a scalar Green function. For odd spin structure, the Szeg\H o kernel precisely coincides with the derivative $\p_zG$ of the scalar Green function, so that its inclusion will be taken into account by the same arguments given above for $\p_z G$. For even spin structure the Szeg\H o kernel  is obtained by shifting the lattice sum in (\ref{pG})  by one of the three even half characteristics, and the inclusion of spin fields and fermion vertex operators will introduce further modifications to the lattice sums which may be reduced to the modular graph forms introduced here upon using Fay's trisecant formula \cite{Fay} and bosonization (see, for example,  section VII of \cite{D'Hoker:1988ta}).

\sm

The mathematical motivation for introducing modular graph forms is readily seen from their structure. The space of modular graph forms includes all non-holomorphic modular functions as they were defined 
in \cite{D'Hoker:2015qmf} by setting $a_r = b _r$ for all edges $r$, as well as  all  holomorphic modular forms (modulo factors of $\tau_2$) by setting $b_r=0$ for all $r$, and all variations in between. In addition, we shall show below that the first order Cauchy-Riemann operators map the space of modular graph forms into itself, as does the Laplacian.

\subsection{Labeling decorated graphs and modular graph forms}
\label{sec:23}

The easiest way to represent a graph is often ... by the graph itself. Still, we shall need to exhibit the labels on the decorated graph in a manner that can be conveniently used to evaluate the derivatives and utilize the symmetries of the associated modular graph form. 

\sm

In any  decorated  graph every decorated edge must begin and end on distinct vertices.
Furthermore, in a decorated graph, only vertices with valence three or higher occur, as all bivalent vertices will be represented by the labels of the decorated graph. We denote the total number of vertices by $V$, and the total number of decorated edges by $\L$.

\subsubsection{The case of arbitrary $V \geq 2$}

For general $V\geq 2$, each vertex is labeled by an integer $i = 1, \cdots, V$. The edges are ordered lexicographically starting with the edges leaving at vertex $i =1$ to vertex $j =2$, then to vertex $j =3$ and so on, followed by edges leaving at vertex $i =2$ to vertex $j =3$ and then to vertex $j =4$ and so on. 
In the explicit calculations later in this paper we will only consider families of  graphs for which the number of vertices $V$ is 2 or 3, but general features of arbitrary graphs and their associated modular graph forms will be considered.

\sm

In order to  illustrate this labeling for a general graph, we may first consider a sub-graph with edges joining two vertices, labelled $i<j$ as illustrated by the following figure.
\bea
\tikzpicture[scale=1.5]
\scope[xshift=-5cm,yshift=-0.4cm]
\draw[thick] (-3,0.035) -- (-1.5,0.035);
\draw[thick] (-3,-0.035) -- (-1.5,-0.035);
\draw[fill=white] (-3,0)  circle [radius=.05] ;
\draw[fill=white] (-1.5,0)  circle [radius=.05] ;
\draw (-3, 0.3) node{$i$};
\draw (-1.5, 0.3) node{$j$};
\draw (-2.25,-0.3) node{$ A_{ij}, B_{ij} $};
\draw (-0.8,0) node{$=$};
\draw (0, 0.3) node{$i$};
\draw (4, 0.3) node{$j$};
\draw[thick] (0,0)  ..controls (2, 0.95) ..  (4,0) ;
\draw[thick] (0,0)  ..controls (2, -0.8)  ..  (4,0) ;
\draw[thick] (0,0)  ..controls (2, 0.2)   ..  (4,0);
\draw (1.3,0.46) [fill=white] rectangle (2.65,0.80) ;
\draw (2,0.64) node{$a_{ij\,1}, b _{ij\, 1}\, $};
\draw (1.3,-0.03) [fill=white] rectangle (2.65,0.31) ;
\draw (2,0.14) node{$a_{ij\, 2}, b _{ij\, 2} $};
\draw (1.3,-0.69) [fill=white] rectangle (2.65,-0.35) ;
\draw (2,-0.5) node{$a_{ij\, \mu_{ij}}, b _{ij\, \mu_{ij}} $};
\draw[fill=white] (0,0)  circle [radius=.05] ;
\draw[fill=white] (4,0)  circle [radius=.05] ;
\endscope
\endtikzpicture
\label{fig5}
\eea
This subgraph has $\mu_{ij}$ decorated edges joining vertices labeled $i,j$.  The set of exponents $\{a_{ij\, \alpha}\}$ of holomorphic momenta $p_{ij \, \alpha}$  ($\alpha=1,\dots,\mu_{ij}$) will be collected in a row matrix $A_{ij}$ and the set of exponents $\{b_{ij\, \alpha}\}$ of anti-holomorphic momenta $\bar p_{ij \, \alpha}$  in a row matrix $B_{ij}$,
\bea
A_{ij} & = & 
\left [ \matrix{
a_{ij \, 1} & a_{ij \, 2 } & \cdots & a_{ij \, \mu_{ij}} \cr } \right ]
\no \\
B_{ij} & = & 
\left [ \matrix{
b_{ij \, 1} & \, b_{ij \, 2 } & \cdots & b_{ij \, \mu_{ij}} \cr} \, \right ]\, .
\eea
This labeling generalizes so that for a general graph consisting of $V$ vertices and $\L$ decorated edges there are $V(V-1)/2$ sets of row matrices of the form $A_{ij}$ and the same number of the form $B_{ij}$.
The labels are assembled into a $ 2 \times \L$  matrix, in lexicographical order, for which we shall use the following notation, 
\bea
\label{3d1}
\left [ \matrix{A \cr B \cr} \right ]  = \left [ 
\matrix{A_{12} \cr B_{12} \cr } \Bigg |  \cdots \Bigg | 
\matrix{A_{1V} \cr B_{1V} \cr } \Bigg | 
\matrix{A_{23} \cr B_{23} \cr } \Bigg |\cdots \Bigg | 
\matrix{A_{2V} \cr B_{2V} \cr } \Bigg | \cdots \Bigg | 
\matrix{A_{V-1 \, V} \cr B_{V-1 \, V} \cr }
\right ] \, .
\eea
The vertical bars indicate the separation between the subsets of edges stretched between a given pair of vertices. We shall often replace the composite labels $(ij \, \mu_{ij})$ on the exponents $a$ and $b$ by a single enumeration label $r = 1, \cdots, \L$.  

\sm

Once the full graph $\Gamma$ has been assembled in terms of subgraphs between pairs of vertices, all vertices are then integrated over $\Sigma$.  To a decorated graph $\Gamma$ with connectivity matrix $\Gamma _{i \, r}$ and with exponents given in (\ref{3d1}), we associate a modular graph form, given by the following expression,
\bea
\label{3b5}
\cC \left [ \matrix{A \cr B \cr} \right ]  (\tau) =  \sum_{p_1,\dots,p_\L \in \Lambda} ' ~ \prod_{r =1}^\L
  {(\tau_2/\pi)^{\half a_r + \half b_r} \over  (p_r) ^{a_r} ~ (\bar p _r) ^{b_r} }\, \prod_{i =1}^V
  \delta \left ( \sum_{s =1}^\L \Gamma _{i \, s} \, p_s \right )\, .
\eea 
Under complex conjugation,  the modular graph forms behave as follows, 
\bea
\label{3d2}
\cC \left [ \matrix{A \cr B \cr} \right ]  (\tau) ^* 
= \cC \left [ \matrix{B \cr A \cr} \right ]  (\tau)
\eea
thereby amounting to swapping the upper and lowers rows of the matrix of exponents. 

\subsubsection{The cases $V=0,1$}

Every irreducible graph with $V=0$ is necessarily a one-loop graph, and is completely specified by a single pair of exponents $(a,b)$. It will be convenient to use the general notation already introduced above to represent the modular graph forms for $V=0$ as follows,
\bea
\cC \left [ \matrix{ a & 0 \cr  b & 0  \cr} \right ]  (\tau) 
= \sum_{p_1, p_2 \in \Lambda} ' 
  {(\tau_2/\pi)^{\half a + \half b} \over  (p_1) ^a ~ (\bar p _1) ^b }\, 
  \delta \left ( p_1+p_2 \right )
=  \sum_{p \in \Lambda} ' 
  {(\tau_2/\pi )^{\half a + \half b} \over  p ^a ~ \bar p  ^b } \, .
\eea 
For $a=b$ we recover the Eisenstein series, while for $a\not=b$ the graphs correspond to derivatives of Eisenstein series, as we shall see shortly below in ({\ref{Es1}).

\sm

Every graph with $V=1$ is reducible. If the valence of the vertex is odd the corresponding modular graph function vanishes. If the valence is even the modular graph function factorizes into modular graph functions with $V=0$.

\subsection{Modular properties}

Under a modular transformation in $PSL(2,\ZZ)$, 
\bea
 \tau \, \to \, \tau' = { \alpha \tau + \beta \over \gamma \tau + \delta} 
\eea
where $\alpha, \beta , \gamma, \delta \in \ZZ$ and $\alpha \delta - \beta \gamma =1$, 
the exponents $A,B$ are unchanged and a general modular graph form of (\ref{3b5}) 
transforms as follows, 
\bea
\cC \left [ \matrix{A \cr B \cr} \right ]  (\tau') 
= \left ( { \gamma \tau+\delta \over \gamma  \bar \tau +\delta } \right ) ^{\half a - \half b}
\cC \left [ \matrix{A \cr B \cr} \right ] (\tau) \, .
\eea
The weight $(a,b)$ of the modular graph form was defined earlier  in (\ref{ab}).
When $a=b$ the modular graph form is a modular function with vanishing modular weight;
for $A=B$ this function is real and corresponds to a modular graph function reviewed in section \ref{sec:21}, while for $A\not=B$ it will generally be complex. For $B=0$, the modular graph form is proportional to a holomorphic modular form of weight $a$.

\sm

The modular graph form corresponding to a general graph with arbitrary exponents $A,B$ for which $a \not= b$ will be neither a modular function, nor will it be holomorphic.  In these cases, there exists no unique natural normalization for the overall $\tau_2$ factor. The convention adopted in (\ref{3b5}) to define $\cC$ is natural for its symmetry under complex conjugation. But there are two other natural normalizations that will be useful, defined as follows,
\bea
\label{3b6}
\cC ^+ \left [ \matrix{A \cr B \cr} \right ]  (\tau) & = &  \sum_{p_1,\dots,p_\L \in \Lambda} ' ~ \prod_{r =1}^\L
  {(\tau_2)^{ a_r} \, \pi ^{- \half a_r - \half b_r}\over  (p_r) ^{a_r} ~ (\bar p _r) ^{b_r} }\, \prod_{i =1}^V
  \delta \left ( \sum_{s =1}^\L \Gamma _{i \, s} \, p_s \right )
  \no \\
  \cC ^- \left [ \matrix{A \cr B \cr} \right ]  (\tau) & = &  \sum_{p_1,\dots,p_\L \in \Lambda} ' ~ \prod_{r =1}^\L
  {(\tau_2)^{ b_r} \, \pi ^{-\half a_r - \half b_r} \over  (p_r) ^{a_r} ~ (\bar p _r) ^{b_r} }\, \prod_{i =1}^V
  \delta \left ( \sum_{s =1}^\L \Gamma _{i \, s} \, p_s \right ) \, .
\eea 
They are related to $\cC$ by a suitable power of $\tau_2$, given as follows, 
\bea
\cC ^\pm \left [ \matrix{A \cr B \cr} \right ]  (\tau) = (\tau_2)^{\pm \half (a-b)} \, 
\cC  \left [ \matrix{A \cr B \cr} \right ]  (\tau) \, .
\eea
By construction, $\cC^+$ and $\cC^-$ are modular forms with respective modular weights $(0,b-a)$ and $(a-b,0)$ and the following associated transformation properties,
\bea
\cC^+ \left [ \matrix{A \cr B \cr} \right ]  (\tau') 
& = &  \left (  \gamma \bar \tau+\delta  \right ) ^{b-a}\,
\cC^+ \left [ \matrix{A \cr B \cr} \right ] (\tau) 
\no \\
\cC^- \left [ \matrix{A \cr B \cr} \right ]  (\tau') 
& = &  \left ( \gamma \tau +\delta  \right ) ^{a-b}\,
\cC^- \left [ \matrix{A \cr B \cr} \right ] (\tau) \, .
\eea
When $B=0$ all dependence on the momenta $p_r$  is holomorphic in $\tau$ and the normalization $\cC^-$  is manifestly holomorphic, while when $A=0$ it is $\cC^+$ which  is manifestly anti-holomorphic. Note that the complex conjugation properties of $\cC^\pm$ coincide with those of $\cC$ in (\ref{3d2}).

\subsection{Cauchy-Riemann operators}

We shall now introduce holomorphic and anti-holomorphic derivatives on the functions and forms  $\cC$, which are covariant under modular transformations.  There are several closely related and commonly used definitions of covariant derivatives acting on modular forms that are useful in different contexts.   A brief review is given in appendix \ref{app:AA} of three  such derivatives that map  modular forms of weight $(\w, \w')$ into forms of weight $(\w-2,\w')$, $(\w+1, \w'-1)$  and $(\w,\w'-2)$, respectively.   The complex conjugate derivatives interchange the action on $\w$ and $\w'$.  The third case is particularly useful because when $\w=0$ the covariant derivative reduces to the Cauchy-Riemann operator,  $\DD=2i \tau_2^2 \p_\tau$, which maps a weight $(0, \w')$ form into  a weight $(0,\w'-2)$ form without the need for a connection. In particular, on the forms $\cC^+$ of weight $(0,b-a)$ and $\cC^-$ of weight $(a-b,0)$ the respective derivatives  $\DD$ and $\bar \DD$ act as follows,
\bea
\label{CR}
\DD = + 2 i \tau _2^2 \p_\tau ~ : ~ (0,b-a) & \to & (0,b-a-2)
\no \\
\DDb = - 2 i \tau _2^2 \p_{\bar \tau} ~ : ~ (a-b,0) & \to & (a-b-2,0) \, .
\eea
Explicit formulas for the action of the first order derivatives on modular graph forms may be obtained easily by noting that the Kronecker $\delta$-function factors in (\ref{3b5}) and (\ref{3b6}) are independent of $\tau$ and $\bar \tau$ in view of their definition in (\ref{3a9a}). The derivative $\DD$ then only acts on the momenta $p_r$ with holomorphic dependence in the denominators, and on the overall multiplicative factors of $\tau_2$. To evaluate these, we use the derivative of each factor,  
\bea
\DD \left ( { \tau _2 \over p_r} \right ) =  \tau_2^2 \, { \bar p_r \over (p_r )^2} 
\hskip 1in 
\DDb \left ( { \tau _2 \over \bar p_r} \right ) =  \tau_2^2 \, {  p_r \over (\bar p_r )^2} \, .
\eea
As a result, we have, 
\bea
\label{3b7}
\DD \, \cC ^+ \left [ \matrix{A \cr B \cr} \right ]  (\tau) \! & = &  
\sum_{p_1,\dots,p_\L \in \Lambda} ' 
\left ( \sum _{t =1} ^\L a_t \, \tau_2 \, { \bar p_t \over p_t} \right ) 
\prod_{r =1}^\L
  {(\tau_2)^{ a_r} \pi ^{-\half a_r - \half b_r} \over  (p_r) ^{a_r}  (\bar p _r) ^{b_r} }\, \prod_{i =1}^V
  \delta \left ( \sum_{s =1}^\L \Gamma _{i \, s} \, p_s \right )
\no \\
\DDb \,   \cC ^- \left [ \matrix{A \cr B \cr} \right ]  (\tau) \! & = &  
\sum_{p_1,\dots,p_\L \in \Lambda} '  
\left ( \sum _{t =1} ^\L b_t \, \tau_2 \, { p_t \over \bar p_t} \right ) 
\prod_{r =1}^\L
  {(\tau_2)^{ b_r} \pi ^{- \half a_r - \half b_r}\over  (p_r) ^{a_r}  (\bar p _r) ^{b_r} }\, \prod_{i =1}^V
  \delta \left ( \sum_{s =1}^\L \Gamma _{i \, s} \, p_s \right ) \, .
  \qquad \quad
\eea 
The presence of the extra $\tau_2$-factor  inside the first parenthesis on the right side of (\ref{3b7}) precisely compensates  for the change in modular weight due to the presence of the extra denominator momenta. 
As a result, the action of the  derivatives may be expressed in terms of an algebraic action on the 
exponents of the $\cC$-functions, and we find,
\bea
\label{3b8}
\DD \, \cC^+  \left [ \matrix{A \cr B \cr} \right ]  & = & 
   \sum _{r =1}^\L a_r  \, \cC^+ \left [ \matrix{A + S_r \cr B - S_r \cr} \right ]
   \no \\
\DDb  \, \cC^- \left [ \matrix{A \cr B \cr} \right ]  & = & 
 \sum _{r =1}^\L b_r  \, \cC^- \left [ \matrix{A - S_r  \cr B+ S_r \cr} \right ] \, .
\eea
In the above formulas, the $\L$-dimensional row-vector $S_r$ is defined  
to have zero entry in each slot, except for slot $r$ where the value of the entry is 1,
\bea
\label{3b9}
S_r = \Big [ \, \underbrace{ 0, \cdots, 0}_{r -1} , \, 1, \, \underbrace{0, \cdots, 0}_{\L-r } \, \Big ] \, .
\eea
The action of $S_r$ on the exponents is by a shift which implements the differentiation rule of (\ref{3b6}), and  $A\pm S_r$ is  understood in the sense of addition of the row vectors $A$ and $S_r$.

\subsection{Laplacian on modular graph functions}

A modular covariant Laplace-Beltrami operator may be defined on modular forms of any weight. In the sequel it will suffice, however, to consider the Laplacian $\Delta$ acting on modular functions, namely of vanishing modular weight, on which the Laplacian is given by,  
\bea
\Delta = 4 \tau_2^2 \p_\tau \p _{\bar \tau} = \tau_2^2 ( \p_{\tau_1}^2 + \p_{\tau _2}^2)
\eea
for $\tau = \tau _1 + i \tau _2$ and $\tau_1, \tau_2 \in \RR$, and $\tau_2 >0$.  The Laplacian acting on modular functions
$\cC$, namely with $a=b$,  may be expressed in terms of the Cauchy-Riemann  operators $\DD$ and $\DDb$ acting on modular forms as in (\ref{CR}) by the relation,
\bea
\label{3g1}
\Delta = \DDb (\tau_2)^{-2} \DD \, .
\eea
Acting on modular functions $\cC$, the  modular weights work out as follows. When $\cC$ has  modular weight $(0,0)$, then $(\tau_2)^{-2} \DD \cC$ has modular weight $(2,0)$, which is mapped back to modular weight $(0,0)$ by $\DDb$. This relation allows us to express the  action of the Laplacian algebraically, using the relations (\ref{3b8}), and we find, 
\bea
\label{1c4}
 ( \Delta +a  ) \, \cC \left [ \matrix{A \cr B \cr} \right ]  =
\sum _{r,r' =1}^\L a_r  b_{r'}  \, \cC \left [ \matrix{A +S_r - S_{r'}  \cr B -S_r +S_{r'}  \cr} \right ] \, .
\eea
Since $a=b$ here, we have $\cC^\pm = \cC$, for which the same relations above hold.

\subsection{Momentum conservation identities}

The presence of a product of momentum conservation Kronecker $\delta$-functions in the Eisenstein summation representing the modular graph forms $\cC$ in (\ref{3b5}) and (\ref{3b6}) implies a set of linear relations between modular graph forms for different exponents. To exhibit these relations, it suffices to state the following obvious relations, 
\bea
\left ( \sum_{r =1}^\L \Gamma _{k \, r} \, p_{r} \right ) 
\prod_{i =1}^V  \delta \left ( \sum_{s =1}^\L \Gamma _{i \, s} \, p_s \right )=0
\eea
valid for each $k=1,\cdots, V$, along with their complex conjugate. Inserting the first factor into any Eisenstein summation may be represented conveniently in terms of the shift operators $S_r$ of (\ref{3b9}). The resulting $V$ relations, and their complex conjugates, may be conveniently formulated as the following identities on the exponents,
\bea
\label{2g1}
\sum _{r =1} ^\L \Gamma _{k \, r} \, \cC \left [ \matrix{ A - S_r \cr B \cr} \right ] & = &
\sum _{r =1} ^\L \Gamma _{k \, r} \, \cC \left [ \matrix{ A  \cr B - S_r \cr} \right ] = 0
\eea
valid for each $k=1,\cdots, V$. Multiplication by suitable factors of powers of $\tau_2$ shows that the  identities of (\ref{2g1}) hold equally well when replacing $\cC$ by either $\cC^+$ or $\cC^-$. Finally, recalling that translation invariance on the torus renders one of the momentum conservation  Kronecker $\delta$-functions redundant, it is clear that only $V-1$ of the above $V$ pairs of relations will be linearly independent.

\section{Dihedral modular graph forms}
\setcounter{equation}{0}
\label{sec:3}

The labeling of general graphs and their associated modular graph forms, as described in section \ref{sec:2},  is rather cumbersome. Fortunately,  we shall need here only families of  graphs for which the number of vertices $V$ is 2 or 3. For low values of $V$, more concrete parametrizations are available, and the symmetry properties under permutations of the vertices and edges, which  are manifest from looking at the graphs, may be spelled out in more detail. 

\subsection{Labeling dihedral graphs}

In this section, we shall discuss the simplest general class of decorated graphs, namely dihedral graphs for which $V=2$. The exponents for a general graph with $\L$ decorated edges may be labeled as follows,
\bea
\label{3e1}
 \left [ \matrix{ A \cr B \cr} \right ] = 
\left [ \matrix{
a_1 & a_2 & \cdots & a_\L \cr 
b_1 & b_2 & \cdots & b_\L \cr } \right ]
\eea
and we have the following graphical representation, 
\bea
\tikzpicture[scale=1.5]
\scope[xshift=-5cm,yshift=-0.4cm]
\draw[thick] (-3,0.035) -- (-1,0.035);
\draw[thick] (-3,-0.035) -- (-1,-0.035);
\draw[fill=black] (-3,0)  circle [radius=.05] ;
\draw[fill=black] (-1,0)  circle [radius=.05] ;
\draw (-2,-0.3) node{$ A, B$};
\draw (-0.25,0) node{$=$};
\draw[thick] (0.5,0) node{$\bullet$} ..controls (2, 0.95) ..  (3.5,0) node{$\bullet$};
\draw[thick] (0.5,0) node{$\bullet$} ..controls (2, -0.8) ..  (3.5,0) node{$\bullet$};
\draw[thick] (0.5,0) node{$\bullet$} ..controls (2, 0.2) ..  (3.5,0) node{$\bullet$};
\draw (1.6,0.5) [fill=white] rectangle (2.4,0.79) ;
\draw (2,0.64) node{$a_1, b _1 $};
\draw (1.6,0) [fill=white] rectangle (2.4,0.28) ;
\draw (2,0.14) node{$a_2, b _2 $};
\draw (1.6,-0.65) [fill=white] rectangle (2.4,-0.35) ;
\draw (2,-0.5) node{$a_\L, b _\L $};
\draw (2,-0.2) node{$\cdots$};
\endscope
\endtikzpicture
\label{fig4}
\eea
The associated modular graph form may be read off from (\ref{3b5}), and is given by,
\bea
\label{3e2}
\cC  \left [ \matrix{ A \cr B \cr} \right ]  (\tau) 
=  \sum_{p_1,\dots,p_\L \in \Lambda} ' ~ \prod_{r =1}^\L
  {(\tau_2/\pi )^{\half a_r + \half b_r} \over  (p_r) ^{a_r} ~ (\bar p _r) ^{b_r} }\, 
  \delta \left ( \sum_{s =1}^\L \, p_s \right ) \, .
\eea 
By translation invariance on the torus only a single momentum conservation Kronecker $\delta$-function is required on the sum. The modular graph form $\cC$  is invariant under permutations $\sigma \in \mS_\L$ of the edges, where $\mS_\L$ is the group of permutations of $\L$ objects. A permutation $\sigma$ acts upon momenta by $\sigma (p_r) = p_{\sigma (r)}$ and $\sigma (\bar p_r)= \bar p _{\sigma (r)}$ and upon exponents by, 
\bea
\label{1b4}
\sigma  \left [ \matrix{ A \cr B \cr} \right ] = 
\left [ \matrix{
a_{\sigma (1)} & a_{\sigma (2)} & \cdots & a_{\sigma (\L)} \cr 
b_{\sigma (1)}  & b_{\sigma (2)} & \cdots & b_{\sigma (\L)} \cr } \right ] \, .
\eea
For dihedral graphs there is only a single momentum conservation relation, along with its complex conjugate, and as a result, we have the following linear relations between their associated modular graph forms,
\bea
\label{rels}
\sum _{r =1} ^\L  \cC \left [ \matrix{ A - S_r \cr B \cr} \right ] (\tau) & = &
\sum _{r =1} ^\L  \cC \left [ \matrix{ A  \cr B - S_r \cr} \right ] (\tau) = 0 \, .
\eea

\subsection{Graphs for $\L=1, 2$ and Eisenstein series}

The class of dihedral graphs contains  many of the modular graph functions that were defined and studied earlier, in particular in \cite{D'Hoker:2015foa}. It will be useful to discuss these special cases by increasing value of the number of decorated edges $\L$. Graphs for which $\L=1$  vanish by momentum conservation,
\bea
\label{3a9}
\cC \left [ \matrix{ a_1 \cr  b_1 \cr } \right ] (\tau)=0
\eea
for all $a_1, b_1$. Moving on to higher values of $\L$, we proceed as follows.

\sm

Graphs for which  $\L=2$ actually contain only bivalent vertices. Therefore they might have been discussed as graphs with  $V=0$, and are not strictly speaking dihedral. Nonetheless, it will be convenient to discuss them together with dihedral graphs, as we shall do here. By inspection of (\ref{3e2}), we readily identify the general modular function for $\L=2$ with the non-holomorphic Eisenstein series,
\bea
\label{Es0}
\cC \left [ \matrix{ a ~ 0 \cr  a ~ 0 \cr } \right ] (\tau) 
=   \cC \left [ \matrix{ a_1 ~ a_2 \cr  a_1 ~ a_2  \cr } \right ] (\tau) =  E_a (\tau)
\eea
where $a_1, a_2 \in \ZZ$, \,  $a=a_1+a_2$,  with $a >1$, and the non-holomorphic Eisenstein series $E_a$ is defined by,
\bea
\label{Es}
E_a(\tau) = \sum _{ p \in \Lambda}' \left ( { \tau_2 \over \pi \, p \, \bar p} \right )^a
= \sum _{(m,n) \not= (0,0)}  \left ( { \tau _2 \over \pi |m + n \tau|^2} \right )^a \, .
\eea
Cauchy-Riemann derivatives of the Eisenstein series may be calculated with the help of (\ref{3b8}), and yield the following modular forms,
\bea
\label{Es1}
\DD ^k \, \cC \left [ \matrix{ a ~ 0 \cr  a ~ 0 \cr } \right ] (\tau) 
= {  (a+k-1)! \over (a-1)! } \, \cC^+ \left [ \matrix{ a+k & 0 \cr  a-k & 0 \cr } \right ] (\tau) \, .
\eea
The action of weight-changing operators on non-holomorphic Eisenstein series (as well as the connection with holomorphic Eisenstein series) is well known  dates back to Maass \cite{Maass} (see also \cite{Pribitkin} for a short article).
For $a=k$ the multiple derivative evaluates as follows,  
\bea
\label{Gk}
\cC^+ \left [ \matrix{ 2k & 0 \cr  0 & 0 \cr } \right ] (\tau) 
= (\tau_2)^{2k} \, \cC^- \left [ \matrix{ 2k & 0 \cr  0 & 0 \cr } \right ] (\tau)
= (\tau_2)^{2k} G_{2k}(\tau)
\eea
where $G_{2k}$ is the {\sl holomorphic} Eisenstein series defined by,
\bea
\label{Gs}
G_{2k} (\tau) = \sum _{p \in \Lambda } ' { 1 \over \pi ^k \, p^{2k}} \, .
\eea
More generally, we have, 
\bea
\cC \left [ \matrix{ a_1 & a_2 \cr  b_1 & b_2  \cr } \right ] (\tau) 
= (-)^{a_2+b_2} \cC \left [ \matrix{ a_1+a_2  & 0 \cr  b_1+b_2 & 0 \cr } \right ] (\tau) \, .
\eea
These forms vanish whenever the sum of all exponents is an odd integer, and are given by derivatives of Eisenstein series $E_k$ when the sum of all exponents is even.

\subsection{Graphs for general $\L \geq 3$}

A general class of modular graph functions was defined in \cite{D'Hoker:2015foa} by,
\bea
C_{a_1, \cdots , a_\L} (\tau) = \sum _{ p_1, \cdots, p_\L \in \Lambda } \delta _{p,0} \prod _{r=1}^\L \left ( {\tau_2 \over \pi |p_r |^2} \right )^{a_r}
\eea
where $p=p_1 + \cdots + p_\L$. The function is invariant under permutations of the exponents $a_r$ for $r=1, \cdots, \L$. This class of modular graph functions corresponds to dihedral graphs, and may be identified as follows,
\bea
\label{3c2}
C_{a_1 , \cdots, a_\L} =
\cC \left [ \matrix{ a_1 & a_2 & \cdots & a_\L \cr  a_1 & a_2 & \cdots & a_\L \cr } \right ]
\eea
where $w=a_1 + \cdots + a_\L$ equals the weight of the graph.

\sm

The two-loop modular graph functions $C_{a,b,c}$ with $\L=3$ were investigated in great detail in \cite{D'Hoker:2015foa}. In particular, it was shown that the action of the Laplace operator $\Delta$ admits an algebraic representation of the exponents $a,b,c$, given by,
\bea
&&
\Big ( \Delta - a(a-1) -b(b-1)-c(c-1) \Big ) C_{a,b,c} 
\no \\ && \hskip 0.3in =
ab \left ( C_{a+1,b-1,c} + \half C_{a+1,b+1,c-2} - 2 C_{a+1,b,c-1} \right ) 
\no \\ && \hskip 0.5in 
+ \hbox{ 5 permutations of } a,b,c
\eea
along with the boundary  relations 
\bea
C_{a,b,0} & = &  E_aE_b-E_{a+b}
\no \\ 
C_{a,b,-1} & = &  E_{a-1}E_b+ E_a E_{b-1} \, .
\eea
These relations, along with $\mS_3$ permutation symmetry and methods of generating functions,   were used in \cite{D'Hoker:2015foa} to gain a thorough understanding of the space of all such two-loop modular graph functions. In particular, it was found that for every odd weight $w\geq 3$, there exists a linear relation among the functions $C_{a,b,c}$ with $w=a+b+c$, the Eisenstein series $E_w$, and the Riemann zeta function $\zeta(w)$, with integer coefficients. 

\sm

For $\L \geq 4$,  the Laplace operator no longer maps the space of modular graph functions $C_{a_1, \cdots, a_R}$ into itself, and the powerful methods available for $\L=3$ no longer apply directly. It is for this reason that the results for $\L\geq 4$ obtained in \cite{D'Hoker:2015foa} were all at the conjectural level, and applied only to a few isolated cases, specifically for the following dihedral graphs, 
\bea
\label{3c4}
D_4  =  \cC \left [ \matrix{ 1 \, 1 \, 1 \, 1 \cr  1 \, 1 \, 1 \, 1 \cr } \right ]
\hskip 0.5in 
D_5 = \cC \left [ \matrix{ 1 \, 1 \, 1 \, 1 \, 1 \cr  1 \, 1 \, 1 \, 1 \, 1 \cr } \right ]
\hskip 0.5in
D_{3,1,1} = \cC \left [ \matrix{ 2 \, 1 \, 1 \, 1  \cr  2 \, 1 \, 1 \, 1  \cr } \right ] \, .
\eea
These functions were conjectured in \cite{D'Hoker:2015foa} to satisfy polynomial relations with modular functions with lower numbers of loops. It will be the object of the next section to outline a strategy for proving these conjectures, a task we shall accomplish in subsequent sections.

\section{Strategy}
\setcounter{equation}{0}
\label{sec:4}

The primary goals that we will achieve in this paper are as follows: 
\begin{enumerate}
\itemsep=-0.05in
\item to prove the four identities between modular graph functions conjectured in 
\cite{D'Hoker:2015foa};
\item  to understand the origin of these identities better by relating them to identities between holomorphic modular graph forms;
\item to develop an efficient method for finding and establishing identities between modular graph forms, in general, beyond those proposed in~\cite{D'Hoker:2015foa}. 
\end{enumerate}
 In this section, we begin by giving the precise statement of point 1. above, and then proceed to developing a strategy for proving not only the identities conjectured in \cite{D'Hoker:2015foa}, but more generally for  finding and proving identities at any weight and any modular weight, and understanding their relation with certain identities between holomorphic modular forms.

\subsection*{Theorem 1}

{\sl The non-holomorphic modular functions $F_4, F_5, F_{3,1,1}$ and $F_{2,2,1}$ defined by,
\bea
\label{4a1}
F_4 & = & D_4 - 24 C_{2,1,1} - 3 E_2^2 + 18 E_4
\no \\
F_5 & = & D_5 - 60 C_{3,1,1} - 10 E_2 C_{1,1,1} + 48 E_5 - 16 \zeta (5)
\no \\
 40 F_{3,1,1} & = & 40 D_{3,1,1} -  300  C_{3,1,1} - 120 E_2 E_3 + 276 E_5 -  7  \zeta (5)
\no \\
 10 F_{2,2,1} & = & 10 D_{2,2,1} - 20 C_{3,1,1} + 4 E_5 - 3 \zeta (5)
\eea
vanish identically, proving the conjectures of \cite{D'Hoker:2015foa}.  }

\sm

The dihedral modular graph functions $C_{2,1,1}$, $C_{3,1,1}$, $D_4$, $D_5$, $D_{3,1,1}$ were defined in (\ref{3c2}) and (\ref{3c4}), while the trihedral modular graph function $D_{2,2,1}$ will be defined in section~\ref{sec:11}. 
Assigning weight $w$ to the non-holomorphic Eisenstein series $E_w$ as well as to the zeta-value $\zeta (w)$, we see that the first identity is homogeneous of weight 4, while the last three are homogeneous of weight~5. 

\sm

The identities between modular graph functions of  Theorem 1 were conjectured in \cite{D'Hoker:2015foa} on the basis of their matching Laurent series near the cusp. Further evidence was provided there by showing that their lowest order exponential corrections also vanish. A  proof of the weight four conjecture was given recently in \cite{D'Hoker:2015zfa} by direct summation of the lattice sums generalizing a procedure used earlier in \cite{Zagier:2014} for the modular graph function $C_{1,1,1}$.

\subsection{Strategy for proving modular graph function identities}
\label{sec:41}

It was shown in \cite{D'Hoker:2015qmf}  that a modular graph function $F$  for an irreducible graph of weight $w \geq 2$  has polynomial growth for the cusp at $\infty$, governed by a Laurent polynomial,
\bea
\label{4b1}
F (\tau) = \sum _{k=1-w} ^w f_k \,  \tau _2 ^k + \cO(e^{- 2 \pi \tau_2})
\eea
as $\tau _2 \to \infty$,
where the coefficients $f_k$ are constants.\footnote{More precisely, $\pi^{-k}f_k$ are rational multiples of single-valued-multiple-zeta-values for $1-w \leq k \leq w-1$, while $\pi^{-w}f_w$ is a rational number, but these features will not be of direct relevance here.} 
The space of modular graph functions for weight $w$ is closed under addition. The product of two modular graph functions $F_1$ and $F_2$ of respective weights $w_1$ and $w_2$ is a modular function of weight $w=w_1+w_2$ for which the Laurent polynomial is of the general form (\ref{4b1}). The identities we seek are given by the vanishing of a modular function $F$ which is a homogeneous polynomial of weight $w$ in modular graph functions, and admits a Laurent polynomial of the form (\ref{4b1}). The conjectures of (\ref{4a1}) were made in \cite{D'Hoker:2015foa} on the basis that the Laurent polynomial for their respective $F$-function vanishes.  

\sm

The strategy in this paper for proving  identities of the form $F=0$ will not rely on matching Laurent series. Instead, it will be closer in spirit to the methods used in \cite{D'Hoker:2015foa} to study the functions $C_{a,b,c}$ using the Laplace operator. Since, for $R \geq 4$, the Laplacian no longer maps to functions in the same space we are led to working with the enlarged space of modular graph forms. 

\sm

In the sequel, we shall develop and illustrate a sieve algorithm, based on holomorphic subgraph reduction,  for modular graph functions $F$ of any given weight $w$ whose fundamental criterion is the relation,
\bea
\label{FF}
(\nabla )^{w-1} F=0 \, .
\eea
In particular, to prove the four identities of Theorem 1, the sieve algorithm and holomorphic subgraph reduction will be used to produce the following relations, 
\bea
\label{FFF}
(\nabla )^3 F_4=(\nabla )^4 F_5=(\nabla )^4 F_{3,1,1}=(\nabla )^4 F_{2,2,1}=0 \, .
\eea 
The strategy  for obtaining these results will be outlined in subsection \ref{sec:43}, illustrated for the case of $F_4$ in subsection \ref{sec:44}, and formulated generally in terms of the sieve algorithm and holomorphic subgraph reduction  in subsection \ref{sec:46}. 

\sm

The key result we shall then use to prove the Theorem is the following lemma.

\subsection*{Lemma 1}

{\sl Let $F$ be a non-holomorphic modular function with polynomial growth  near the cusp at $\infty$. If $F$ satisfies the differential equation,
\bea
(\DD)^n F=0
\label{4b2}
\eea
for some arbitrary integer $ n \geq 1$, then $F$ is constant as a function of $\tau$.}

\sm

Having established the relations (\ref{FFF}), the assumptions of the Lemma will be satisfied with $n=w-1$ and, as a result of the Lemma, the functions $F_4, F_5, F_{3,1,1}$ and $F_{2,2,1}$ must be constant. Knowing that their value at the cusp $\tau_2 \to \infty$  vanishes from the results of \cite{D'Hoker:2015foa} then completes the proof of the Theorem.

\subsection{Proof of Lemma 1} 

We prove Lemma 1, in two different ways: the first by relating the problem to the Laplacian and its spectral properties, the second by more explicit calculation and solution of the differential equation (\ref{4b2}) subject to modular invariance as well as slightly stronger assumptions on $F$, which hold true for modular graph functions.

\subsubsection{Proof of the Lemma by Laplacian methods}

The key ingredient  is a factorization identity between the Laplacian $\Delta= 4 \tau_2^2 \p_{\bar \tau} \p_\tau$ acting on modular functions and the Cauchy-Riemann operators $\DD=2 i \tau_2^2 \p_\tau $ and $\DDb=-2i\tau_2^2 \p_{\bar \tau}$,\footnote{Consideration of the operator on the left of (\ref{factor}) was suggested by Deligne \cite{Deligne} for use  in \cite{D'Hoker:2015zfa}. As a simple check on the identity (\ref{factor}), one may verify that, upon acting on $\tau_2^\alpha$, both sides evaluate to $\tau_2^\alpha$ multiplied by  $\prod _{s=1}^n (\alpha^2 -\alpha - s^2 +s)$. A full proof of the  identity (\ref{factor})  may be provided by induction on $n$, commencing with (\ref{3g1}). }
\bea
\label{factor}
\prod _{s=1}^n \Big ( \Delta - s(s-1) \Big ) 
= \DDb ^n \, (\tau_2)^{-2n}  \, \DD ^n \, .
\eea
Combining the assumption of  Lemma 1, $\nabla ^n F=0$, with the factorization formula (\ref{factor}), we conclude that $F$ must satisfy the following relation,
\bea
\label{B2}
\prod _{s=1}^n \Big ( \Delta - s(s-1) \Big ) F=0 \, .
\eea
Equation (\ref{B2}) provides a weaker condition on $F$ than the original condition $\nabla ^n F=0$ did, so that the space of solutions of (\ref{B2}) will contain the space of solutions of (\ref{4b2}), and a solution to (\ref{B2}) need not be, in general, a solution to $\nabla ^n F=0$. 

\sm

Next,  we use the assumption that $F$ has polynomial growth at the cusp to solve (\ref{B2}). Indeed, it is well-known \cite{Terras, Zagier:2008} that the general $SL(2,\ZZ)$-invariant solution with polynomial growth  to the  eigenvalue equation $\Delta f =  s(s-1)  f$ for real $s>1$  is given by the non-holomorphic Eisenstein series $f=E_s$, uniquely up to a multiplicative constant. For $s=1$, the corresponding solution $f$ is an arbitrary constant. Iterating the solution process for the $n$ factors in (\ref{B2}), we deduce that $F$ is given by, 
\bea
\label{B3}
F (\tau) = c_0 + \sum _{s=2} ^n c_s E_s (\tau)
\eea
for a set of as yet undetermined constants $c_s$ with $s=0,2,\cdots, n$. 

\sm

To enforce the original condition $\nabla ^n F=0$ on the expression for $F$  in (\ref{B3}), we enforce the vanishing of the Laurent polynomial of $\nabla ^n F$ on  (\ref{B3}). The Laurent polynomial of $E_s$ has only two non-vanishing terms, one proportional to $\tau_2^s $ and the other proportional to $\tau_2^{1-s}$.
Thus, the contribution to the Laurent polynomial of $\nabla ^n F$ with the highest power of $\tau_2$ arises from the $s=n$ term in (\ref{B3}). Its vanishing requires $c_n=0$. Repeating the argument for successive values of $s$ similarly shows that $c_s=0$ for all $s=2, \cdots, n$, so that $F=c_0$ is constant, which completes the proof of the lemma.

\subsubsection{Proof by direct calculation}

A  special case of Lemma 1, based on assumptions that are sufficient for the considerations of this paper, may be proven by direct resolution of (\ref{4b2}), subject to the conditions of modular invariance. We shall consider modular functions $F$ of weight $w$ which satisfy the equation $(\nabla )^{w-1} F=0$ of (\ref{FF}) and admit the following  expansion   near the cusp $\tau \to \infty$, (derived in \cite{D'Hoker:2015qmf} for any modular graph function of weight $w$),
\bea
\label{B5}
F (\tau, \bar \tau) = \sum _{k=1-w} ^ w \tau_2^k \, \f_k (q, \bar q)
\eea
where $\f_k(q,\bar q)$ is an entire function of $q=e^{2 \pi i \tau}$ and $\bar q$.  By inspection, its derivative of arbitrary order $n$ has the following expansion,
\bea
\label{B6}
\nabla ^n F (\tau, \bar \tau)= \sum _{k=1-w+n}^{w+2n} \tau _2 ^k \, \f_k ^{(n)} (q,\bar q)
\eea
where $\f_k ^{(0)} (q, \bar q) = \f_k (q, \bar q)$. The coefficients $\f_k ^{(n)}$ then obey a recursion relations in $n$,  
\bea
\label{B7}
\f_{w+2n} ^{(n)} & = & 2 i \p_\tau \f _{w+n-2} ^{(n-1)}
\no \\
\f_k ^{(n)} ~ & = & 2 i \p_\tau \f_{k-2} ^{(n-1)} + (k-1) \f_{k-1} ^{(n-1)} 
\no \\
\f_{1-w+n} ^{(n)} & = & (n-w) \f ^{(n-1)} _{n-w}
\eea
where the middle line holds for $2-w+n \leq k \leq w+2n-1$.

\sm

Enforcing the condition $\nabla ^{w-1}F=0$ amounts to requiring $\f_k ^{(w-1)}=0$ for $0 \leq k \leq 3w-2$. The third recursion relation in (\ref{B7}) may then be solved algebraically, and we have,
\bea
\label{B8}
\f_{-k} ^{(w-1-k)} = 0 \hskip 1in 0 \leq k \leq w-1
\eea
so that $\f_{1-w}=0$. The first two recursion relations of (\ref{B7}) may be analyzed by induction on $n$,
and impose the conditions $\p_\tau ^{3w-3-k} \f_k ^{(w-2)} =0$  for $0 \leq k \leq 3w-4$.
Proceeding by induction to all remaining values of $n=w-1, w-2, \cdots, 1$, and taking into account the truncation at $k=1$  in the second recursion relation of (\ref{B7}), we find the following conditions, 
\bea
\p_\tau ^{\ell +m} \f ^{(w-1-m)} _{3w-2-2m-\ell} =0
& \hskip 1in & 1 \leq m \leq w-1
\no \\
&& 
0 \leq \ell \leq 3w-2-m
\eea
Setting $m=w-1$ produces the following relations,
\bea
\p_\tau ^{2w-1-k} \f  _k =0
\hskip 1in
1-w \leq k \leq w \, .
\eea
We conclude that $\f_k $ must be a polynomial in $\tau$ of degree $2w-k-2$ with coefficients which are functions of $\bar \tau$. Since $\f_k $ is real and polynomial of degree $2w-k-2$ in $\tau$ it must also be polynomial in $\bar \tau$ of degree $2w-k-2$.  Modular invariance requires $F$ to be constant.

\subsection{Strategy for relating to holomorphic modular forms}
\label{sec:43}

While the preceding subsection provides a strategy for proving  identities between modular graph functions, it does not give much insight as to the deeper origin of these identities, and  the key relation $(\nabla )^{w-1}F=0$. In the present subsection, we shall bridge this gap by relating the existence of identities between non-holomorphic modular graph functions to the existence of identities involving only holomorphic modular graph forms. Since the ring of holomorphic modular forms for the torus is polynomial and generated by the holomorphic Eisenstein series $G_4$ and $G_6$, its structure is very tight, and results in a wealth of identities. 

\sm

The basic strategy for relating identities between non-holomorphic modular functions and holomorphic modular forms is very simple, but the detailed execution of the idea is more involved. As may be seen from the action of the differential operator $\DD$ on an arbitrary modular graph form in (\ref{3b8}), 
\bea
\label{4c1}
\DD \, \cC^+  \left [ \matrix{A \cr B \cr} \right ]  = 
\sum _{r =1}^\L a_r  \, \cC^+ \left [ \matrix{A + S_r \cr B - S_r \cr} \right ] \, .
\eea
$\DD$ raises the total exponent $a$ by one, and lowers the total exponent $b$ by one. Here, the total exponents $a$ and $b$ were defined in (\ref{ab}) and respectively correspond to the sum of the exponents of momenta holomorphic in $\tau$  and anti-holomorphic in $\tau$.  Therefore, the action of $\DD^b$ on $\cC^+$ will reduce the $b$-exponent to zero.  Naively, one may be tempted to conclude that such a form will be holomorphic but this is, of course, generally false since the vanishing $b$-sum may be achieved when some $b_r$ are positive while other $b_r$ are negative. If, however, all modular graph functions with negative exponents could be eliminated, then the result of the {\sl descent} by $b$ operators $\DD$ will indeed be a holomorphic form. 

\sm

Therefore, the basic criterion for reduction to holomorphic modular graph forms is the requirement that,
upon the iterative action by $b$ copies of $\DD$,  all modular graph forms with negative exponents are cancelled. As will be explained in the illustration below, and in the next section, the basic tool for the elimination of negative exponents will be by {\sl holomorphic subgraph reduction}, namely the fact that the modular subgraphs with only holomorphic exponents may be evaluated and simplified by the methods of holomorphic modular forms.

\subsection{Illustration by proving the  $D_4$ conjecture}
\label{sec:44}

Proving the $D_4$ conjecture with the help of the above methods will help illustrating these tools and make the principle of holomorphic subgraph reduction more concrete.  These tools will be key to all other proofs in this paper.\footnote{We could have used an even simpler case of $C_{1,1,1}$ for the purpose of illustration, but since we already know that $C_{1,1,1}= E_3+\zeta (3)$ from \cite{Zagier:2014,D'Hoker:2015foa}, this example is really too simple.} We begin by recalling the form of the three-loop graph $D_4$, and evaluating its first derivative using (\ref{4c1}),
\bea
\label{4d1}
D_4  =  \cC \left [ \matrix{ 1 \, 1 \, 1 \, 1 \cr  1 \, 1 \, 1 \, 1 \cr } \right ]
\hskip 1in
\DD \, D_4  =  4 \, \cC^+ \left [ \matrix{ 2 \, 1 \, 1 \, 1 \cr  0 \, 1 \, 1 \, 1 \cr } \right ] \, .
\eea
Applying the operator $\DD$ once again, we find,
\bea
\label{4d2}
\DD^2   D_4  =  
12 \, \cC^+ \left [ \matrix{ 2 \, 2 \, 1 \, 1 \cr  0 \, 0 \, 1 \, 1 \cr } \right ]
+ \, 8 \, \cC^+ \left [ \matrix{ 3 ~~~~ 1 \, 1 \, 1 \cr  -1 ~~ 1 \, 1 \, 1 \cr } \right ] \, .
\eea
The momentum conservation identities of (\ref{rels}), applied to the lower exponents of the second  term on the right side of the above equation, allow us to eliminate its negative exponent entry, which results in the following expression for   the second derivative,
\bea
\label{4d4}
\DD^2   D_4  
=   12 \, \cC^+ \left [ \matrix{ 2 \, 2 \, 1 \, 1 \cr  0 \, 0 \, 1 \, 1 \cr } \right ]
- 24 \, \cC^+ \left [ \matrix{ 3 \, 1 \, 1 \, 1 \cr  0 \, 0 \, 1 \, 1 \cr } \right ] \, .
\eea
It is easy to see that the action of one further $\DD$ derivative on either term on the right side will produce negative lower exponents which cannot all be eliminated with the help of the momentum conservation relations of (\ref{rels}). It is at this point that we appeal to the reduction of holomorphic subgraphs before proceeding to acting with further derivatives.

\subsubsection{A simple holomorphic subgraph reduction}

Since the exponents in the first two  lower  slots in both terms on the right side of (\ref{4d4})   vanish, there exists a one-loop subgraph for each modular graph form above, which involves only momenta holomorphic in $\tau$. To see this explicitly, we evaluate both modular graph forms using (\ref{3b6}), and combine the results under a single summation, 
\bea
\label{4d5}
\cC^+ \left [ \matrix{ 2 \, 2 \, 1 \, 1 \cr  0 \, 0 \, 1 \, 1 \cr } \right ]
- 2 \, \cC^+ \left [ \matrix{ 3 \, 1 \, 1 \, 1 \cr  0 \, 0 \, 1 \, 1 \cr } \right ]
= 
\sum _{p_1, \cdots, p_4 \in \Lambda}' 
 \left ( { 1 \over p_1 ^2 \, p_2^2 } -  { 1 \over p_1^3 \, p_2} -  { 1 \over p_1 \, p_2^3} \right ) 
{ \tau_2^6 \, \delta _{p,0} \over \pi ^4 |p_3|^2 \, |p_4|^2}
\eea
where $p=p_1+p_2+p_3+p_4$ and we have used the symmetry under the interchange of labels $1$ and $2$ to symmetrize the summand. We partition the sum into the contributions from $p_1+p_2=0$ and from $p_1+p_2 \not= 0$. The contribution from $p_1+p_2=0$ readily evaluates to $3 \tau_2^4 G_4 E_2$, where $E_2$ is the non-holomorphic Eisenstein series, defined in (\ref{Es}), while $G_4$ is the holomorphic modular form defined in (\ref{Gs}). The contribution from  $p_1+p_2 \not= 0$ is evaluated using the following identity between fractions, 
\bea
\label{4d6}
{ 1 \over p_1 ^2 \, p_2^2 } -  { 1 \over p_1^3 \, p_2} -  { 1 \over p_1 \, p_2^3}
=  - { 1 \over p_1^3 (p_1+p_2) } - { 1 \over p_2^3 (p_1+p_2)} \, .
\eea
Collecting the contributions from both parts of the summation, we have,
\bea
\label{4d7}
\cC^+ \left [ \matrix{ 2 \, 2 \, 1 \, 1 \cr  0 \, 0 \, 1 \, 1 \cr } \right ]
- 2 \, \cC^+ \left [ \matrix{ 3 \, 1 \, 1 \, 1 \cr  0 \, 0 \, 1 \, 1 \cr } \right ]
= 3 \tau_2^4 G_4 E_2 
-  \sum _{p_1, \cdots, p_4}' 
 { 2 \,  (1- \delta _{p_1+p_2,0}) \over p_1^3 (p_1+p_2) } \, 
{ \tau_2^6 \, \delta _{p,0} \over \pi ^4 |p_3|^2 \, |p_4|^2} \, .
\eea
Next, we change variables from $p_2$ to $p_2' = p_1 +p_2$, and use the summation variables $p_1, p_2', p_3$ and $p_4$. Excluding $p_2=0$ from the summation, as we are instructed to do, now requires $p_1 \not = p_2'$, and $p_2'\not= 0$ by the insertion of the factor $(1-\delta _{p_1+p_2,0})$. The second term on the right side of (\ref{4d7}) thus reduces to the sum,
\bea
-  \sum _{p_2', p_3, p_4 \in \Lambda }' ~ \sum _{p_1 \not = p_2' \in \Lambda } '
 { 2 \over p_1^3 \, p_2' } \, 
{ \tau_2^6 \, \delta _{p_2'+p_3+p_4,0} \over \pi ^4 |p_3|^2 \, |p_4|^2} \, .
\eea
The summation over $p_1$ may be performed with the help of the fact that the unconstrained sum (namely ignoring the restriction $p_1\not= p_2'$) vanishes by parity, so that we have,
\bea
\sum _{p_1 \not = p_2' \in \Lambda } ' { 1 \over p_1^3} = - { 1 \over (p_2')^3} \, .
\eea
Our ability to perform this basic holomorphic sum provides the simplest illustration of the general tool of {\sl holomorphic subgraph reduction} which will be discussed more generally in section \ref{sec:5} for dihedral graphs, and in section \ref{sec:10} for trihedral graphs.

\sm

The resulting sum is reduced to a modular graph function of the same holomorphic weight, namely six, but with only two loops. It may be identified as follows,
\bea
  \sum _{p_2', p_3, p_4 \in \Lambda }' 
{ 2\, \tau_2^6 \, \delta _{p_2'+p_3+p_4,0} \over \pi^4 (p_2')^4  |p_3|^2 \, |p_4|^2}
= 2 \, \cC^+ \left [ \matrix{ 4 \, 1 \, 1 \cr  0 \, 1 \, 1 \cr } \right ]
\eea
so that we have the following holomorphic subgraph reduction formula,
\bea
\label{4d8}
\cC^+ \left [ \matrix{ 2 \, 2 \, 1 \, 1 \cr  0 \, 0 \, 1 \, 1 \cr } \right ]
- 2 \, \cC^+ \left [ \matrix{ 3 \, 1 \, 1 \, 1 \cr  0 \, 0 \, 1 \, 1 \cr } \right ]
= 
2 \, \cC^+ \left [ \matrix{ 4 \, 1 \, 1 \cr  0 \, 1 \, 1 \cr } \right ]
+
3 \, \cC^+ \left [ \matrix{ 4 \, 0 \cr  0 \, 0 \cr } \right ] \, \cC^+ \left [ \matrix{ 2 \, 0 \cr  2 \, 0 \cr } \right ] \, .
\eea

This identity may be represented graphically by depicting an edge $r$ with a purely holomorphic momentum dependence, namely with $b_r=0$, by a dashed line, and the exponent $a_r$ on the edge by $a_r-1$ square decorations on the edge, as shown in the figure below.
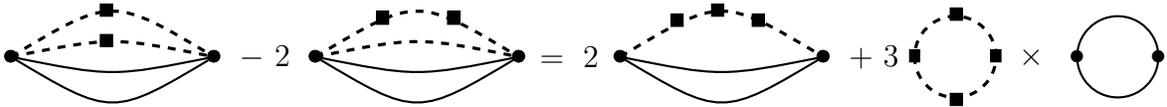
\begin{center}
\tikzpicture[scale=1.35]
\scope[xshift=-5cm,yshift=-0.4cm]
\draw[very thick, dashed] (0,0) node{$\bullet$} ..controls (1, 0.6) ..  (2,0) node{$\bullet$};
\draw[very thick, dashed] (0,0) node{$\bullet$} ..controls (1, 0.2) ..  (2,0) node{$\bullet$};
\draw[thick] (0,0) node{$\bullet$} ..controls (1, -0.2) ..  (2,0) node{$\bullet$};
\draw[thick] (0,0) node{$\bullet$} ..controls (1, -0.6) ..  (2,0) node{$\bullet$};
\draw (1,0.52) [fill=black] rectangle (0.88,0.4) ;
\draw (1,0.22) [fill=black] rectangle (0.88,0.1) ;
\draw (2.5,0) node{$- ~ 2$};
\draw[very thick, dashed] (3,0) node{$\bullet$} ..controls (4, 0.6) ..  (5,0) node{$\bullet$};
\draw[very thick, dashed] (3,0) node{$\bullet$} ..controls (4, 0.2) ..  (5,0) node{$\bullet$};
\draw[thick] (3,0) node{$\bullet$} ..controls (4, -0.2) ..  (5,0) node{$\bullet$};
\draw[thick] (3,0) node{$\bullet$} ..controls (4, -0.6) ..  (5,0) node{$\bullet$};
\draw (3.72,0.32) [fill=black] rectangle (3.6,0.45) ;
\draw (4.42,0.32) [fill=black] rectangle (4.3,0.45) ;
\draw (5.5,0) node{$= ~ 2$};
\draw[very thick, dashed] (6,0) node{$\bullet$} ..controls (7, 0.6) ..  (8,0) node{$\bullet$};
\draw[thick] (6,0) node{$\bullet$} ..controls (7, -0.2) ..  (8,0) node{$\bullet$};
\draw[thick] (6,0) node{$\bullet$} ..controls (7, -0.6) ..  (8,0) node{$\bullet$};
\draw (6.62,0.3) [fill=black] rectangle (6.5,0.42) ;
\draw (7.42,0.3) [fill=black] rectangle (7.3,0.42) ;
\draw (7.02,0.4) [fill=black] rectangle (6.9,0.52) ;
\draw (8.5,0) node{$+ ~ 3$};
\draw[very thick, dashed] (9.3,0)  circle [radius=.4] ;
\draw (9.37,0.36) [fill=black] rectangle (9.25,0.48) ;
\draw (9.37,-0.36) [fill=black] rectangle (9.25,-0.48) ;
\draw (8.85,-0.05) [fill=black] rectangle (8.95,0.07) ;
\draw (9.65,-0.05) [fill=black] rectangle (9.75,0.07) ;
\draw (10.1,0) node{$\times ~ $};
\draw[thick] (10.9,0)  circle [radius=.4] ;
\draw (10.5,0) node{$\bullet$}; 
\draw (11.3,0) node{$\bullet$}; 
\endscope
\endtikzpicture
\end{center}
As a result,  we have the following simplified expression for the second derivative of (\ref{4d4}),
\bea
\label{4d9}
\DD^2   D_4 
= 24 \, \cC^+ \left [ \matrix{ 4 \, 1 \, 1 \cr  0 \, 1 \, 1 \cr } \right ]
+ 36 \, \cC^+ \left [ \matrix{ 4 \, 0 \cr   0 \, 0 \cr } \right ] \, \cC^+ \left [ \matrix{ 2 \, 0 \cr  2 \, 0 \cr } \right ] \, .
\eea
This formula will be the starting point of the remaining application of the holomorphic subgraph  reduction identities.

\subsubsection{A simple sieve algorithm}

The other polynomials in modular graph functions of weight four are $E_2^2$, $C_{2,1,1}$, and $E_4$ and they indeed enter into the $D_4$ conjecture. The second order derivatives  of these functions may be computed in analogous fashion, and we find, 
\bea
\DD ^2 E_2^2 & = & 
8 \, \cC^+ \left [ \matrix{ 3 \, 0 \cr   1 \, 0 \cr } \right ] ^2 
+ 12  \, \cC^+ \left [ \matrix{ 4 \, 0 \cr   0 \, 0 \cr } \right ]  \, \cC^+ \left [ \matrix{ 2 \, 0 \cr   2 \, 0 \cr } \right ] 
\no \\
\DD ^2 C_{2,1,1} & = & 
6 \, \cC^+ \left [ \matrix{ 4 \, 1 \, 1 \cr  0 \, 1 \, 1 \cr } \right ] 
+ 6 \, \cC^+ \left [ \matrix{ 3 \, 2 \, 1 \cr  1 \, 0 \, 1 \cr } \right ] 
+ 4 \, \cC^+ \left [ \matrix{ 6 \, 0 \cr   2 \, 0 \cr } \right ] 
\no \\
\nabla ^2 E_4 & = & 20 \, \cC^+ \left [ \matrix{ 6 \, 0 \cr   2 \, 0 \cr } \right ] \, .
\eea
We can form three linear  homogeneous combinations of weight 4 from these four modular graph functions,  in which the holomorphic subgraph obstruction, proportional to $G_4$,  has been removed, and they are $D_4-3 E_2^2$ as well as $C_{2,1,1}$ and $E_4$. To these three combinations, we now apply one further derivative, to obtain, 
\bea
\DD ^3 \left ( D_4 - 3 E_2^2 \right ) & = & 
432 \, \cC^+ \left [ \matrix{ 7 \, 0 \cr   1 \, 0 \cr } \right ] 
- 288 \, \cC^+ \left [ \matrix{ 4 \, 0 \cr   0 \, 0 \cr } \right ] \, \cC^+ \left [ \matrix{ 3 \, 0 \cr   1 \, 0 \cr } \right ] 
\no \\
\DD^3 C_{2,1,1} & = & 
108  \, \cC^+ \left [ \matrix{ 7 \, 0 \cr   1 \, 0 \cr } \right ] 
- 12 \, \cC^+ \left [ \matrix{ 4 \, 0 \cr   0 \, 0 \cr } \right ] \, \cC^+ \left [ \matrix{ 3 \, 0 \cr   1 \, 0 \cr } \right ] 
\no \\
\DD^3 E_4 & = & 120  \, \cC^+ \left [ \matrix{ 7 \, 0 \cr   1 \, 0 \cr } \right ] \, .
\eea
From these three combinations, we can form two linear combinations in which the holomorphic subgraph obstruction, proportional to $G_4$,  has been removed, and they are $E_4$ as well as $D_4 - 3 E_2^2 - 24 C_{2,1,1}$.  The  $\DD^3 $ derivative of this combination is given by,
\bea
\DD ^3 \left ( D_4  - 3 E_2^2 -24 C_{2,1,1} \right ) = 
-2160 \, \cC^+ \left [ \matrix{ 7 \, 0 \cr   1 \, 0 \cr } \right ] \, .
\eea
Finally, including  the term $18\, \DD ^3 E_4$ to form $\DD^3 F_4$ (see (\ref{4a1})), we find,
\bea
\DD ^3 F_4=0 \, .
\eea
Using Lemma 1 we conclude that $F_4$ must be constant. Invoking the results of \cite{D'Hoker:2015foa}
that $F_4 \to 0$ as $\tau _2 \to \infty$, we conclude that $F_4=0$. 

\sm

To summarize, the method used here is based on the application of the holomorphic subgraph reduction identities, as well as on the principle of a {\sl sieve algorithm} to systematically eliminate the holomorphic subgraph obstructions as one takes higher and higher $\DD$-derivatives. We started out with four monomials of weight four, namely $D_4, E_2^2, C_{2,1,1}$ and $E_4$, eliminated one obstructing combination upon taking two derivatives, and one more upon taking three derivatives. The {\sl sieve algorithm} will be extended in the sequel to identities between modular graph functions of higher weight.

\subsection{Algebraic reduction identities}
\label{sec:45}

In addition to the holomorphic subgraph reduction identities, illustrated in the preceding section, and to be discussed generally in section \ref{sec:5} for dihedral graphs, and in section \ref{sec:10} for trihedral graphs, 
there is also a set of algebraic reduction identities. We shall  here  discuss these identities for dihedral graphs, relegating the case of trihedral graphs to section \ref{sec:10}. 

\sm

The reduction applies when two zero exponents occur in the same pair of indices, in which case the sum can be reduced as follows. When $\L \geq 3$, we have, 
\bea
\label{4e1}
\cC^+ \left [ \matrix{a_1 \, \cdots \, a_{\L-1} ~  0 \cr b_1 \, \cdots \, b_{\L-1} ~ \, 0  \cr} \right ] 
= \prod _{r = 1 }^{\L-1} \cC^+ \left [ \matrix{a_r ~  0 \cr b_r ~  \, 0  \cr} \right ] 
- \cC^+ \left [ \matrix{a_1 \, \cdots \, a_{\L-1}  \cr b_1 \, \cdots \, b_{\L-1}   \cr} \right ] \, .
\eea
This relation is the generalization of the relation $C_{a,b,0} = E_a E_b - E_{a+b}$ for the $C_{a,b,c}$-functions of (\ref{3c2}), which were established and used extensively in \cite{D'Hoker:2015foa}. To prove (\ref{4e1}), we write,
\bea
\cC^+ \left [ \matrix{a_1 \, \cdots \, a_{\L-1} ~ 0 \cr b_1 \, \cdots \, b_{\L-1} ~ 0  \cr} \right ] 
=\sum _{p_1, \cdots, p_\L \in \Lambda } '   \delta _{p,0} \prod _{r =1}^{\L-1} 
{ (\tau_2) ^{a_r} \pi ^{- \half a_r - \half b_r} \over (p_r) ^{a_r} \, (\bar p_r) ^{b_r}}
\eea
where $p=p_1+\cdots + p_\L$. The sum over the single momentum $p_\L$ may be carried out explicitly and, since $p_\L \not= 0$, the sum over $p_\L$ has the only effect of requiring that the sum over the remaining $p _r$ for $r=1,\cdots, \L-1$ cannot vanish. Hence we get,
\bea
\cC ^+ \left [ \matrix{a_1 \, \cdots \, a_{\L-1} ~ 0 \cr b_1 \, \cdots \, b_{\L-1} ~ \, 0  \cr} \right ] 
=\sum _{p_1, \cdots, p_{\L-1} \in \Lambda} '  (1 - \delta _{p',0}) \prod _{r =1}^{\L-1} 
{  (\tau_2) ^{ a_r } \pi ^{-\half a_r - \half b_r} \over (p_r )^{a_r} \, (\bar p_r )^{b_r}}
\eea
where $p' = p_1 + \cdots + p_{\L-1}$. Decomposing the sum into the two contributions arising from the terms in the parentheses on the right side, we get  (\ref{4e1}) using  (\ref{Gk}).

\subsection{Sieve algorithm for constructing modular identities}
\label{sec:46}

In this subsection,  the sieve algorithm  by which polynomial identities between modular graph functions are produced for arbitrary \weight will be made explicit. We shall explain how the fundamental  condition $(\nabla )^{w-1}F=0$ of (\ref{FF}) may once again be obtained by a sieve algorithm based on holomorphic subgraph reduction. 

\sm

The space of  polynomials in modular graph functions that are homogeneous of \weight $w$ is generated by the modular graph functions associated with irreducible graphs, and will be denoted $\cV_w$.  For high enough $w$, the space $\cV_w$ will include contributions from dihedral and trihedral graphs, as well as from graphs with larger numbers of vertices.  The operator $\nabla$ maps $\cV_w$ to modular graph forms outside of $\cV_w$. Following the nomenclature of subsection \ref{sec:23}, a general modular graph form may be represented with the help of an array of exponents, 
\bea
\label{4f1}
 \cC^+ \left [ 
\matrix{A_{12} \cr B_{12} \cr } \Bigg |  \cdots \Bigg | 
\matrix{A_{1V} \cr B_{1V} \cr } \Bigg | 
\matrix{A_{23} \cr B_{23} \cr } \Bigg |\cdots \Bigg | 
\matrix{A_{2V} \cr B_{2V} \cr } \Bigg | \cdots \Bigg | 
\matrix{A_{V-1 \, V} \cr B_{V-1 \, V} \cr }
\right ] \, .
\eea
Here $V$ is the number of vertices of the graph with valence at least equal to three, which are labeled by $i =1,\cdots , V$ and the ordered pairs of vertices $i<j$ label the decorated edges.   The exponent matrices $A_{ij}, B_{ij}$, and the total exponents $a,b$  are given by,
\bea
\label{4f2}
A_{ij} & = & [ a_{ij \, 1} ~~  a_{ij \, 2} ~~ \cdots ~~ a_{ij \, \mu_{ij}} ]
\hskip 1in a = \sum _{i<j}^V \sum _{r=1}^{\mu_{ij} } a_{ij \, \mu _{ij}}
\no \\
B_{ij} & = & [ b_{ij \, 1} ~~  \, b_{ij \, 2} ~~  \cdots ~~ b_{ij \, \mu_{ij}} ]
\hskip 1in b = \sum _{i<j}^V \sum _{r=1}^{\mu_{ij} } b_{ij \, \mu _{ij}}
\eea
where $\mu_{ij}$ is the number of decorated edges spanned between vertices $i$ and $j$. 
We shall refer to the pair $(a,b)$ as the \weight of the modular graph form of (\ref{4f1}), and extend the notion of \weight to products of modular graph forms by adding their weights. Modular graph functions in $\cV_w$ have \weight $(w,w)$, and all the exponents satisfy $a_{ij \, r} =b_{ij \, r}  \geq 1$.

\subsubsection{Holomorphic subgraph reduction}

Applying the operator $\nabla ^n$ to a function in $\cV_w$ with the help of (\ref{4c1}) gives a modular graph form of \weight $(a+n,a-n)$ with $a=w$. When $n = {\rm min}_{i,j,r}  (b_{ij \, r})$ at least one of the terms in this linear combination has one vanishing lower exponent.  This phenomenon is nicely illustrated  in the example of the $D_4$ graph where the first derivative in the second equation in (\ref{4d1}) produces one vanishing lower exponent.  

\sm

Applying one further derivative will produce one term where a lower exponent becomes $-1$, 
while all other exponents remain strictly positive (the second term on the right side of (\ref{4d2}) in the $D_4$ example), along with a number of other terms for which one or two lower exponents vanish (the first term on the right side of (\ref{4d2}) in the $D_4$ example). The $-1$ entry may be removed from the lower exponents by using the momentum conservation identities, and produces terms with either one or two vanishing lower exponents (as in (\ref{4d4}) for the $D_4$ example). If the vanishing lower exponents occur in the same block $B_{ij}$ then applying yet one further derivative will produce negative lower exponents which {\sl cannot be removed} by the momentum conservation identities. 

\sm

More generally, if all the edges on which the lower exponents vanish form a subgraph which contains a closed loop, then further application of a derivative will produce negative exponents which cannot be removed by the momentum conservation identities. However, when all the edges on which the lower exponents vanish  form a subgraph without any closed loops, then momentum conservation identities always allow one to remove all negative exponents.  Therefore, {\sl holomorphic one-loop subgraphs} pose an obstruction to removing all negative lower exponents at the subsequent order of differentiation in $\nabla$. Holomorphic subgraphs present the {\sl only} such obstruction.

\sm

To make the above discussion more concrete, we introduce the following spaces.
\begin{enumerate}
\item $\cV_{(a,b)}$ is the vector space of polynomials of \weight $(a,b)$ generated by irreducible modular graph functions  for which $a_{ij \, r} \geq 1$  and $b_{ij \, r} \geq 0$ for all $i,j,r$, such that the edges on which $b_{ij \, r}=0$ form {\sl a subgraph which contains no closed loops}.
\item $\cV_{(a,b)} ^0$  is the vector space of polynomials of \weight $(a,b)$ generated by irreducible modular graph functions  for which $a_{ij \, r} \geq 1$  and $b_{ij \, r} \geq 0$ for all $i,j,r$, such that the edges on which $b_{ij \, r}=0$ form {\sl a subgraph which contains precisely one closed loop}.
\end{enumerate}
From these definitions, it is clear that,
\bea
\cV_{(a,b)} \cap \cV^0 _{(a,b)} =0
 \hskip 1in  
 \cV_a \subset \cV_{(a,a)}
\hskip 1in 
\cV_a \cap \cV^0_{(a,a)}=0
\eea 
Implementing the arguments of the preceding paragraph,  the differential operator $\nabla$ is found to act as follows,
\bea
\label{map}
\nabla : \cV _{(a,b)} & \to & \cV_{(a+1, b-1)} \, \oplus \, \cV^0_{(a+1,b-1)} \, .
\eea
The component of the range of $\nabla$ along the subspace $\cV^0_{(a+1,b-1)}$ may or may not vanish. 
This is nicely illustrated by the action of $\nabla$ on the space of \weight 4 modular graph functions, calculated in subsection \ref{sec:44}, for which we have for example,
\bea
\nabla D_4 & \in  & \cV_{(5,3)}
\no \\
\nabla^2 D_4 & \in & \cV_{(6,2)} \oplus \, \cV^0_{(6,2)} \, .
\eea
In this case, the space $\cV^0_{(6,2)}$ contains the combination of two terms on the right side of (\ref{4d4}),
each of which contains a holomorphic one-loop subgraph. 

\sm

Holomorphic one-loop subgraphs may be reduced  to a linear combination of subgraphs without closed loops, whose coefficients are holomorphic modular forms $G_{2k}$. The phenomenon of holomorphic subgraph reduction was already observed in the $D_4$ example of (\ref{4d8}) where the holomorphic two-point subgraphs produce a term proportional to $G_4$, and  will be presented in detail for three-point subgraphs in section \ref{sec:10}. The procedure applies to holomorphic one-loop subgraphs with an arbitrary number of points, and leads to the following decomposition of an arbitrary element $v^0 \in \cV^0_{(a,b)}$, 
\bea
v^0 = v + \sum _{k=2}^{[a/2 ]} \tau_2^k G_{2k} \, v_{(k)}
\eea
where $v \in \cV_{(a,b)}$ and $v_{(k)} \in \cV_{(a-2k,b)}$. The modular graph forms $v_{(k)}$ correspond to graphs which have one fewer loop than $v^0$ does, as each holomorphic form $G_{2k}$ accounts for one loop. The modular graph form $v$ will also correspond to a graph with fewer loops than $v^0$.

\subsubsection{The sieve algorithm}

The sieve algorithm proceeds as follows. We define a sieve of subspaces of $\cV_{(a,b)}$,
\bea
\cV_{(a,b)} \supset \cV_{(a,b)}^{(1)}  \supset \cV_{(a,b)}^{(2)} \supset \cdots \supset \cV_{(a,b)}^{(b-1)}
\eea
by the following iterative procedure. Starting with $n=1$, we define $\cV^{(1)}_{(a,b)}$ to be the maximal subspace of $\cV_{(a,b)}$ whose range under $\nabla$ has vanishing component along $\cV^0_{(a,b)}$, and therefore excludes contributions with holomorphic closed loop subgraphs,
\bea
\nabla : \cV^{(1)}_{(a,b)} & \to & \cV_{(a+1,b-1)} \, .
\eea
The process may be iterated to define the maximal subspace $\cV^{(1)}_{(a+1,b-1)}$ of $\cV_{(a+1,b-1)}$, and the maximal subspace $\cV^{(2)}_{(a,b)}$ of $\cV^{(1)} _{(a,b)}$  which similarly exclude holomorphic closed loop subgraph contributions from the respective ranges of $\nabla$, 
\bea
\nabla : \cV^{(1)}_{(a+1,b-1)} & \to & \cV_{(a+2,b-2)}
\no \\
\nabla : \cV^{(2)}_{(a,b)} \hskip 0.3in & \to & \cV^{(1)} _{(a+1,b-1)}
\no \\
\nabla ^2 : \cV^{(2)}_{(a,b)} \hskip 0.3in & \to  & \cV_{(a+2,b-2)} \, .
\eea
More generally, we define recursively the space $\cV^{(n)} _{(a,b)}$ as the maximal subspace of $\cV^{(n-1)}_{(a,b)}$ which  excludes holomorphic closed loop subgraph contributions from the range of $\nabla$, 
\bea
\nabla : \cV^{(n)}_{(a,b)} & \to & \cV^{(n-1)}_{(a+1,b-1)}
\eea
and therefore, 
\bea
\nabla^n : \cV^{(n)}_{(a,b)} & \to & \cV_{(a+n,b-n)} \, .
\eea
The subspaces $\cV_w ^{(n)}$ may then be defined by the inclusion $\cV_w \subset \cV_{(w,w)}$, namely, $\cV_w^{(1)}$ is the subspace of $\cV_w$ such that,
\bea
\nabla : \cV^{(1)}_w & \to & \cV_{(w+1,w-1)}
\eea 
and so on. These spaces produce the associated sieve,
\bea
\cV_w \supset \cV_w^{(1)}  \supset \cV_w^{(2)} \supset \cdots \supset \cV_w^{(w-1)}
\eea
 The Eisenstein series $E_w$ belongs to each space in the sieve $E_w \in \cV_w ^{(n)}$ for all $0\leq n \leq w-1$. The number of linearly independent identities between modular graph functions at \weight $w$ is then given by,
\bea
\dim  \cV_w ^{(w-1)}  -1
\eea
since a further action of $\nabla$ on $\cV_w ^{(w-1)}$ will contain only one-loop holomorphic modular graph forms, which must all be proportional to $G_{2w}$.

\sm

Illustrating this iterative construction with the example of weight 4, we have,
\bea
\cV_4 \hskip 0.1in & = & \{ D_4, \,  E_2^2 , \, C_{2,1,1}, \, E_4 \}
\no \\
\cV_4 ^{(1)} & = & \{ D_4, \,  E_2^2 , \, C_{2,1,1}, \, E_4 \}
\no \\
\cV_4 ^{(2)} & = & \{ D_4-3 E_2^2 , \, C_{2,1,1}, \, E_4 \}
\no \\
\cV_4 ^{(3)} & = & \{ D_4-3 E_2^2 -24 C_{2,1,1}, \, E_4 \} \, .
\eea
Acting once again with $\nabla$ on $\nabla ^3 \cV^{(3)}_4$ produces only holomorphic modular graph functions which are one-loop, and thus must be proportional to $G_8$. Thus, there must exist a single linear combination in $\nabla ^3 \cV_4^{(3)}$ which is mapped to 0 by $\nabla$. Since this combination is a form of modular weight $(0,-6)$ it must vanish, which leads to the relation $\nabla ^3 F_4 = 0$.

\section{Holomorphic subgraph reduction in dihedral graphs}
\setcounter{equation}{0}
\label{sec:5}

In this section, we shall derive general holomorphic subgraph reduction identities for dihedral graphs, leaving the discussion for trihedral graphs to section \ref{sec:10}.  Holomorhic subgraph reduction for dihedral graphs will apply to modular graph functions of the following type, 
\bea
\cC^+ \left [ \matrix{a_+ & a_-  & A   \cr 0 & 0 & B   \cr} \right ] 
\hskip 1in
\matrix{ A  = [ \, a_1 ~ a_2 ~ \cdots ~ a_r \, ]  \cr  B = [ \, b_1 ~ b_2 ~ \cdots ~ b_r \, ]   \cr} 
\eea
where we shall assume that $a_+, a_-\geq 1$, and $a_\rho, b_\rho \geq 1$ for all $\rho=1, \cdots, r$.
To guarantee convergence of all subgraphs, we shall also assume that  $a_+ + a_- \geq 3$. 
The corresponding  lattice sum may be arranged as follows,
\bea
\label{5a2}
\cC^+ \left [ \matrix{a_+ & a_-  & A   \cr 0 & 0 & B   \cr} \right ] 
= \sum _{ p_1 , \cdots , p_r \in \Lambda} ' ~ \sum _{p_0 \in \Lambda}  
 \delta _{p_0 + p,0}  \,  (\tau _2 )^{a_0} \, \cG _{a_+,  a_-} (p_0)
\prod _{\rho =1} ^r { (\tau _2)^{a_\rho} \pi ^{-\half a_\rho - \half b_\rho} \over (p_\rho) ^{a_\rho} \, (\bar p _\rho) ^{b_\rho}}  \, .
\eea
Here, we use the notation  $p_0 = p_+ + p_- $ for the external momentum through the holomorphic subgraph, $a_0 = a_+ + a_-$ for the partial weight of the subgraph, and $ p = p_1 + \cdots + p_r$. Note that the summation range of $p_0$ is over the full lattice $\Lambda$, including $p_0=0$. 
We have factored the part of the sum that corresponds to the subgraph with only holomorphic momenta $p_\pm$ and corresponding exponents $a_\pm$, since this is the holomorphic subgraph which we wish to reduce. The summation corresponding to the holomorphic subgraph is given by,
\bea
\label{5a3}
\cG _{a_+,  a_-} (p_0) = \sum _{{p_+, \, p_- \in \Lambda \atop  p_++p_-=p_0 }} '  \, { 1 \over 2 \pi ^{\half a_0} }
\left ( {1\over  (p_+) ^{a_+} \, (p_-) ^{a_-} } + {1\over  (p_-) ^{a_+} \, (p_+) ^{a_-} } \right )
\eea
where we have symmetrized  in the momenta $p_\pm$. 
We  have the following parity relation, 
\bea
\cG_{a_+, a_-} (-p_0) = (-)^{a_0} \, \cG_{a_+, \, a_-}(p_0) \, .
\eea

\subsection{Evaluating holomorphic subgraphs}

In this subsection, we evaluate the holomorphic sums $\cG_{a_+, a_-} (p)$.  For $p_0=0$, the sum $\cG_{a_+, \, a_-}(0)$ vanishes identically when $a_0$ is odd, by  parity , while when $a_0$ is even we have,
\bea
\label{5b4}
\cG_{a_+, \, a_-} (0) = (-)^{a_+} G_{a_0}
\eea
where $G_{a_0}$ is the holomorphic Eisenstein series of modular weight $(a_0,0)$ defined in (\ref{Gs}).

\sm

To evaluate $\cG_{a_+, a_-}(p_0)$ for $p_0 \not=0$, we use the following standard algebraic identity, 
\bea
{ 1 \over (p_+) ^{a_+} \, (p_-) ^{a_-} }
& = & 
 \sum _{k=1} ^{a_+} \left ( \matrix{ a_0 -1 -k  \cr a_+ -k \cr} \right ) 
{ 1 \over (p_+)^k \, (p_0)^{a_0 - k}}
\no \\ &&
+ \sum _{k=1} ^{a_- } \left ( \matrix{ a_0  -1-k  \cr a_- -k \cr} \right ) 
{ 1 \over (p_-)^k \, (p_0)^{a_0 -k }}
\, .
\eea
To evaluate $\cG_{a_+, \, a_-}(p_0)$, we need to interchange the sums over $k$ with the sums over $p_\pm$. The sums over $p_\pm$ are absolutely convergent for $k \geq 3$, but are only conditionally convergent for $k=1,2$. To be allowed to interchange  the summations, we introduce the {\sl Eisenstein regulator} on the summation integers $m,n$ which label the lattice momenta by $p=m  + n \tau$.   We define the following Eisenstein-regularized sums, 
\bea
Q_k (p_0) = \lim _{ P \to \infty} \sum _{-P \leq m,n \leq P} '  \delta _{p_+ + p_- , p_0} 
\left ( { 1 \over 2 (p_+)^k } + { 1 \over 2 (p_-)^k} \right )
\eea
where the cut-off $P$ is required in order to define the  conditionally convergent multiple sums when $k=1,2$.   In terms of $Q_k(p_0)$, the modular form $\cG_{a_+, a_-}(p_0)$ evaluates to, 
\bea
\label{5b9}
\pi ^{\half a_0} \, \cG_{a_+, \, a_-} (p_0) =
\sum _{k=1} ^{a_+} \left ( \matrix{ a_0 -1 -k  \cr a_+ -k \cr} \right ) 
{ Q_k(p_0) \over  (p_0)^{a_0 - k}}
+ \sum _{k=1} ^{a_- } \left ( \matrix{ a_0  -1-k  \cr a_- -k \cr} \right ) 
{ Q_k(p_0) \over (p_0)^{a_0 -k }}
\, .
\eea
It remains to evaluate the  holomorphic sums $Q_k(p_0)$, which we shall do next. 

\sm

 For $k \geq 3$, the sum over $p_+$ in $Q_k(p)$ is absolutely convergent, and the Eisenstein regularization is not required for its evaluation. The two terms in the summand contribute equally. When $p_0=0$, we find
 the modular form $G_k$, 
\bea
Q_k(0) = \sum _{p_+ \not= 0 }    { 1 \over  (p_+)^k } = \pi ^{k/2} G_k
\, .
\eea
When $p_0 \not= 0$, we solve for $p_-$ in terms of $p_0$ and $p_+$ by enforcing the constraint, 
and we have $p_-=p_0-p_+$. This requires that in the remaining sum over $p_+$, 
we set $p_+ \not=0$ as well as $p_- \not = p_0$. As a result, we find, 
\bea
\label{5f1}
Q_k (p_0) = \sum _{p_+ \not= 0, p_0 }    { 1 \over  (p_+)^k } = \pi ^{k/2} G_k - { 1 \over (p_0)^k}
\, .
\eea
Note that $G_k$ is independent of $p_0$, so that all $p_0$-dependence is in the second term.

\sm

For $k=1,2$, the sum is no longer absolutely convergent, and we  use Eisenstein regularization.
The two terms in the summand still contribute equally. The calculation is standard, and will be reviewed in appendix \ref{app:B}. The results for $p_0\not=0$ are as follows, 
\bea
\label{5f2}
Q_2(p_0) & = & \pi G_2 - { 1 \over (p_0)^2} \hskip 1in Q_2(0) = - 4 \pi i \p_\tau \ln \eta (\tau)
\no \\
Q_1(p_0) & = & - i \pi { p _0 - \bar p_0 \over \tau - \bar \tau} - { 1 \over p_0}
\eea
where $\eta (\tau)$ is the Dedekind $\eta$-function, and $G_2$ the regularized Eisenstein series.

\subsection{Eliminating holomorphic subgraphs}

Using the results of the preceding subsection, we are now in a position to provide a general formula for the value of the holomorphic subgraphs, and use this evaluation to eliminate the contribution of the holomorphic subgraph from (\ref{5a2}). For $p_0=0$, the result was already given in  (\ref{5b4}). For $p_0 \not =0$, the result of (\ref{5b9}) may be recast in the following form, 
\bea
\label{5b5}
\pi ^{\half a_0} \cG_{a_+, \, a_-} (p_0) & = &
 \sum _{k=3} ^{a_+} \left ( \matrix{ a_0 -1 -k  \cr a_+ -k \cr} \right ) 
{ \pi^{k/2} G_k  \over  (p_0)^{a_0-k} }
+ \sum _{k=3} ^{a_- } \left ( \matrix{ a_0  -1-k  \cr a_- -k \cr} \right ) 
{ \pi^{k/2} G_k  \over (p_0)^{a_0-k}}
\no \\ &&
-{ 1 \over (p_0)^{a_0} }  \left ( \matrix{ a_0 \cr a_+ } \right) + \left ( \matrix{ a_0 -2  \cr a_+ -1 \cr} \right ) 
\left ( {\pi G_2 \over (p_0) ^{a_0-2} }  + { \pi \, \bar p \over \tau_2 \, (p_0)^{a_0-1} } \right )
\, .
\eea
Here, we have used the following relation between binomial coefficients to evaluate the coefficient in the first term on the the second  line, 
\bea
 \sum _{k=1} ^{a_+} \left ( \matrix{ a_0 -1-k  \cr a_+ -k  \cr} \right )  
+ \sum _{k=1} ^{a_- } \left ( \matrix{ a_0  -1-k \cr a_-  -k \cr} \right )
= \left ( \matrix{ a_0 \cr a_+ } \right)
\, .
\eea
Assembling all the contributions to the elimination of an arbitrary one-loop holomorphic subgraph, we obtain, 
\bea
\label{redux}
\cC^+ \left [ \matrix{a_+ ~ a_-  ~ A   \cr 0 ~~ 0 ~~ ~ B   \cr} \right ] 
& = &
\cC^+ \left [ \matrix{a_+ ~ a_-   \cr 0 ~~~ 0   \cr} \right ]  \, \cC^+ \left [ \matrix{A   \cr B   \cr} \right ] 
-  \left ( \matrix{ a_0 \cr a_+ } \right)\, \cC^{+} \!  \left [ \matrix{a_0  & A   \cr  0 & B   \cr} \right ] 
\\ &&
+ \sum _{k=4} ^{a_+} \left ( \matrix{ a_0 -1-k \cr a_+ -k \cr} \right )  
\cC^+ \left [ \matrix{k ~ 0   \cr 0 ~ 0   \cr} \right ]   \, \cC^+ \! \left [ \matrix{a_0 - k & A   \cr  0 & B   \cr} \right ] 
\no \\ &&
+ \sum _{k=4} ^{a_- } \left ( \matrix{ a_0  -1 -k \cr a_-  -k \cr} \right )  
\cC^+ \left [ \matrix{k ~ 0   \cr 0 ~ 0   \cr} \right ] \, \cC^+ \! \left [ \matrix{a_0 -k & A   \cr  0 & B   \cr} \right ] 
\no \\ &&
+ \left ( \matrix{ a_0  -2 \cr a_+ -1 \cr} \right ) \left \{ \tau_2^2 G_2 \, \cC^+ \! \left [ \matrix{a_0 -2 & A   \cr  0 & B   \cr} \right ]
+ \pi \tau_2  \, \cC^+ \! \left [ \matrix{a_0 -1    & A   \cr  -1 & B   \cr} \right ] \right \}
\, .
\no
\eea
The first term on the right side arises from the contribution of $p_0=0$, while the remaining terms arise from $p_0 \not=0$ with the help of (\ref{5b5}). The terms for $k=3$ on the second and third lines, which were present in (\ref{5b5}),  have been omitted as they vanish in view of $G_3=0$. Similarly, the contributions to the sum for any odd value of  $k$ will also vanish, using (\ref{Gk}) with $2k$ replaced by $k$..

\subsection{Iterative structure of the  reduction}

The starting point of any iterative procedure will be a modular graph form with {\sl positive exponents}.  The recursive procedure that will be used  throughout consists in applying the operator $\DD$ repeatedly  until one of the exponents in the lower (anti-holomorphic) row vanishes, while all other exponents in the lower row are 
strictly positive. Thus, the next application of $\DD$ is then to a modular graph form of the  following type,
\bea
\cC^+_\Gamma= \cC^+ \left [ \matrix{ a_1 ~ a_2 ~ \cdots ~ a_\ell & a_{\ell+1} ~ \cdots ~ a_r   \cr 
0 ~~ 1 ~~  \cdots ~ 1 & b_{\ell+1} ~ \cdots ~ b_r   \cr} \right ] 
\hskip 1in 1 \leq \ell \leq r
\, .
\eea
Here, we have used the permutation symmetry of dihedral graphs on  pairs  to order the lower exponents so that $2 \leq b_{\ell+1}, \cdots, b_r$,   with the understanding that for $\ell=1$ there are no entries in the lower  row which equal 1. Working out the derivative gives,
\bea
\DD \, \cC^+ _\Gamma
& = & 
a_1 \, \cC^+ \left [ \matrix{ a_1+1  & a_2 ~ \cdots ~ a_\ell & a_{\ell+1} ~ \cdots ~ a_r   \cr 
-1 & 1 ~~  \cdots ~ 1 & b_{\ell+1} ~ \cdots ~ b_r   \cr} \right ] 
\no \\ && 
+ \sum _{j=2}^\ell 
a_j \, \cC^+ \left [ \matrix{ a_1  ~ a_2 ~ \cdots & a_j+1 & \cdots  a_\ell & a_{\ell+1} ~ \cdots ~ a_r   \cr 
0 ~~ 1 ~~  \cdots  & 0 & \cdots  ~ 1 & b_{\ell+1} ~ \cdots ~ b_r   \cr} \right ] 
\no \\ && 
+ \sum _{j=\ell +1}^r
a_j \, \cC^+ \left [ \matrix{ a_1  ~ a_2 ~ \cdots   a_\ell & a_{\ell+1} ~ \cdots & a_j+1 & \cdots ~ a_r   \cr 
0 ~~ 1 ~~  \cdots  ~ 1 & b_{\ell+1} \cdots  & b_j -1 & \cdots ~ b_r   \cr} \right ] 
\, .
\eea
The terms on the third line above have only one zero on the bottom row, so they are ready to proceed to the next iterative step. To the term on the first line we may apply the momentum conservation identity. This will produce $r-\ell$ terms  that have only one zero on the bottom row, and are ready for the next iterative step. It will also produce $\ell-1$  terms with two zeros on the bottom row of exponents, which together with the terms on the second line give, 
\bea
\DD \, \cC^+ _\Gamma
& = &
 \sum _{j=2}^\ell 
a_j \, \cC^+ \left [ \matrix{ a_1  ~ a_2 ~ \cdots & a_j+1 & \cdots  a_\ell & a_{\ell+1} ~ \cdots ~ a_r   \cr 
0 ~~ 1 ~~  \cdots  & 0 & \cdots  ~ 1 & b_{\ell+1} ~ \cdots ~ b_r   \cr} \right ] 
\no \\ && 
- \sum _{j=2}^\ell 
a_1 \, \cC^+ \left [ \matrix{ a_1+1 &  a_2 ~ \cdots & a_j & \cdots  a_\ell & a_{\ell+1} ~ \cdots ~ a_r   \cr 
0 & 1 ~~  \cdots  & 0 & \cdots  ~ 1 & b_{\ell+1} ~ \cdots ~ b_r   \cr} \right ] + \cdots 
\eea
where the ellipsis indicates terms with only a single zero on the bottom row of exponents.

\sm

We shall now show that the Eisenstein-regularized terms in the holomorphic reduction identities  cancel.
In the general expression  given in (\ref{redux}), these terms are on the last line on the right side. We specialize to the two terms with label $j$, for which $a_0=a_1+a_j+1$. Using the results of (\ref{redux}), we find a common factor which only depends on $j$, and from the terms on the second and third lines above, we find the following contribution for given~$j$,
\bea
a_j \left ( \matrix{ a_1 + a_j -1 \cr  a_j \cr } \right ) - a_1 \left ( \matrix{ a_1 + a_j -1 \cr  a_1 \cr } \right ) 
= 0
\eea
which cancels for all $j$. The remaining terms in the holomorphic  reduction formula correspond to the first three lines of (\ref{redux}).

\subsection{Holomorphic subgraph reduction for low weight $a_0$}
\label{sec:54}

We shall now list the  combinations of holomorphic subgraphs, discussed in the previous section,  up to weight $a_0 \leq 8$, and for any value $r \geq 1$. For $r=1$ and even weight $a_0$, the formulas below  hold provided their last term on the right side,  proportional to $\cC^+ [ A ~  B ]^t$,  is set to zero, in accord with (\ref{3a9}). 
 
\sm

\noindent
$\bullet$ For $a_0=4$  we have the combination, 
\bea
\cC^+ \! \left [ \matrix{2 \, 2 ~ A  \cr 0 \, 0 ~ B  \cr} \right ] 
- 2 \, \cC^+ \! \left [ \matrix{3 \, 1 ~ A \cr 0 \, 0 ~ B   \cr} \right ] 
=
 2 \, \cC^+ \! \left [ \matrix{4 ~ A  \cr 0  ~ B  \cr} \right ] 
+ 3 \tau_2^4 G_4 \, \cC^+ \! \left [ \matrix{A  \cr  B  \cr} \right ] 
\, .
\eea

\sm

\noindent
$\bullet$ For $a_0=5$   we have the combination, 
\bea
 \cC^+ \! \left [ \matrix{3 \, 2 ~ A \cr 0 \, 0 ~ B  \cr} \right ] 
- 3 \, \cC^+ \! \left [ \matrix{4 \, 1 ~ A  \cr 0 \, 0 ~ B   \cr} \right ] 
= 
 5 \, \cC^+ \! \left [ \matrix{5 ~ A \cr 0  ~ B  \cr} \right ] 
- 3 \tau_2^4 G_4 \, \cC^+ \! \left [ \matrix{1 ~ A  \cr 0  ~ B   \cr} \right ] 
\, .
\eea

\sm

\noindent
$\bullet$ For $a_0=6$  we have the combinations,
\bea
\label{id8}
\cC^+ \! \left [ \matrix{3 \, 3 ~ A \cr 0 \, 0 ~ B  \cr} \right ] 
- 6 \, \cC^+ \! \left [ \matrix{5 \, 1 ~ A  \cr 0 \, 0 ~ B   \cr} \right ] 
& = & 
 16 \, \cC^+ \! \left [ \matrix{6 ~ A \cr 0  ~ B  \cr} \right ]
- 6 \tau_2^4 G_4 \, \cC^+ \! \left [ \matrix{2 ~ A  \cr 0  ~ B   \cr} \right ] 
+ 5 \tau_2^6 G_6 \, \cC^+ \! \left [ \matrix{ A  \cr  B   \cr} \right ] 
\no \\
\cC^+ \! \left [ \matrix{4 \, 2 ~ A \cr 0 \, 0 ~ B  \cr} \right ] 
- 4 \, \cC^+ \! \left [ \matrix{5 \, 1 ~ A  \cr 0 \, 0 ~ B   \cr} \right ] 
& = & 
 9 \, \cC^+ \! \left [ \matrix{6 ~ A \cr 0  ~ B  \cr} \right ]
- 3 \tau_2^4 G_4 \, \cC^+ \! \left [ \matrix{2 ~ A  \cr 0  ~ B   \cr} \right ] 
+ 5 \tau_2^6 G_6 \, \cC^+ \! \left [ \matrix{ A  \cr  B   \cr} \right ] 
\, .
\qquad
\eea

\sm

\noindent
$\bullet$ For $a_0=7$  we have the combinations, 
\bea
\cC^+ \! \left [ \matrix{4 \, 3 ~ A \cr 0 \, 0 ~ B  \cr} \right ] 
- 2 \, \cC^+ \! \left [ \matrix{5 \, 2 ~ A  \cr 0 \, 0 ~ B   \cr} \right ] 
& = & 
 7 \, \cC^+ \! \left [ \matrix{7 ~ A \cr 0  ~ B  \cr} \right ]
- 3 \tau_2^4 G_4 \, \cC^+ \! \left [ \matrix{3 ~ A  \cr 0  ~ B   \cr} \right ] 
 \\
\cC^+ \! \left [ \matrix{5 \, 2 ~ A \cr 0 \, 0 ~ B  \cr} \right ] 
- 5 \, \cC^+ \! \left [ \matrix{6 \, 1 ~ A  \cr 0 \, 0 ~ B   \cr} \right ] 
& = & 
 14 \, \cC^+ \! \left [ \matrix{6 ~ A \cr 0  ~ B  \cr} \right ]
- 3 \tau_2^4 G_4 \, \cC^+ \! \left [ \matrix{3 ~ A  \cr 0  ~ B   \cr} \right ] 
- 5 \tau_2^6 G_6 \, \cC^+ \! \left [ \matrix{1 ~ A  \cr  0 ~B   \cr} \right ] 
\no
\eea

\sm

\noindent
$\bullet$ For $a_0=8$  we have the combinations,
\bea
3 \cC^+ \! \left [ \matrix{4 \, 4 ~ A \cr 0 \, 0 ~ B  \cr} \right ] 
- 4 \, \cC^+ \! \left [ \matrix{5 \, 3 ~ A  \cr 0 \, 0 ~ B   \cr} \right ] 
& = & 
 14 \, \cC^+ \! \left [ \matrix{8 ~ A \cr 0  ~ B  \cr} \right ]
- 6  \tau_2^4 G_4 \, \cC^+ \! \left [ \matrix{4 ~ A  \cr 0  ~ B   \cr} \right ] 
+ 7 \tau_2^8 G_8 \, \cC^+ \! \left [ \matrix{ A  \cr  B   \cr} \right ] 
\no \\
2 \cC^+ \! \left [ \matrix{5 \, 3 ~ A \cr 0 \, 0 ~ B  \cr} \right ] 
- 5 \, \cC^+ \! \left [ \matrix{6 \, 2 ~ A  \cr 0 \, 0 ~ B   \cr} \right ] 
& = & 
 28 \, \cC^+ \! \left [ \matrix{8 ~ A \cr 0  ~ B  \cr} \right ]
- 9 \tau_2^4 G_4 \, \cC^+ \! \left [ \matrix{4 ~ A  \cr 0  ~ B   \cr} \right ] 
\no \\ && 
- 5 \tau_2^6 G_6 \, \cC^+ \! \left [ \matrix{ 2 ~ A  \cr 0 ~ B   \cr} \right ]  
- 7 \tau_2^8 G_8 \, \cC^+ \! \left [ \matrix{ A  \cr B   \cr} \right ] 
\no \\
\cC^+ \! \left [ \matrix{6 \, 2 ~ A \cr 0 \, 0 ~ B  \cr} \right ] 
- 6 \, \cC^+ \! \left [ \matrix{7 \, 1 ~ A  \cr 0 \, 0 ~ B   \cr} \right ] 
& = & 
 20 \, \cC^+ \! \left [ \matrix{8 ~ A \cr 0  ~ B  \cr} \right ]
- 3 \tau_2^4 G_4 \, \cC^+ \! \left [ \matrix{4 ~ A  \cr 0  ~ B   \cr} \right ] 
\no \\ && 
- 5 \tau_2^6 G_6 \, \cC^+ \! \left [ \matrix{ 2 ~ A  \cr 0 ~ B   \cr} \right ]  
+ 7 \tau_2^8 G_8 \, \cC^+ \! \left [ \matrix{ A  \cr B   \cr} \right ] 
\, .
\qquad
\eea
The holomorphic subgraph reduction formula obtained more informally in (\ref{4d8}) is seen to be a special case of the identity given above for $a_0=4$ and $A=B=[1 ~ 1]$.

\section{Proving dihedral graph $D_5$ and $D_{3,1,1}$ conjectures}
\setcounter{equation}{0}
\label{sec:6}

In this section, we shall prove the $D_5$ and $D_{3,1,1}$ parts of Theorem 1 in (\ref{4a1}).  They are given by the vanishing of the modular functions  $F_5$ and $F_{3,1,1}$ defined as follows,
\bea
\label{6a0}
F_5 & = & D_5 - 60 C_{3,1,1} - 10 E_2 C_{1,1,1} + 48 E_5 - 16 \zeta (5)
\no \\
40 F_{3,1,1} & = & 40 D_{3,1,1} - 300  C_{3,1,1} - 120 E_2 E_3 +276 E_5 - 7 \zeta (5)
\, .
\eea

\subsection{Preliminary derivative formulas}

Given the identifications $D_5=C_{1,1,1,1,1}$, $D_{3,1,1}= C_{2,1,1,1}$, and $C_{3,1,1}$,  it is clear that these modular graph functions  are of the type $C_{a_1, \cdots, a_\L}$, and are special cases of the following general classes of dihedral modular graph functions,
\bea
D_\ell = C_{1_\ell} = \cC \left [ \matrix{ 1_\ell  \cr 1_\ell  \cr } \right ] 
\hskip 1in 
C_{k, 1_{\ell}} = \cC \left [ \matrix{ k & 1_\ell  \cr k & 1_\ell  \cr } \right ]
\eea
where $k \geq 2$ and we use the abbreviation $1_\ell = [ 1 ~ 1 \, \cdots \,  1]$ for a row vector of length $\ell$ whose entries are all equal to one. Their first derivatives may be evaluated with the help of the algebraic representation of (\ref{3b6}) and (\ref{3b8}), and we have,
\bea
\DD \, \cC \left [ \matrix{ 1_\ell  \cr 1_\ell  \cr } \right ]  
& = & \ell \, \cC^+ \left [ \matrix{ 2 & 1_{\ell-1}  \cr 0 & 1_{\ell-1}  \cr } \right ] 
\no \\
\DD \, \cC \left [ \matrix{ k & 1_\ell  \cr  k & 1_\ell  \cr } \right ]  
& = & 
k \, \cC^+ \left [ \matrix{ k+1  & 1_\ell  \cr  k-1  & 1_\ell  \cr } \right ]  
+ \ell \, \cC^+ \left [ \matrix{ k & 2 & 1_{\ell-1}  \cr k & 0 & 1_{\ell-1}  \cr } \right ] 
\, .
\eea
Their second derivatives may be evaluated using (\ref{3b6}) and (\ref{3b8}), and simplified with the help of the holomorphic subgraph reduction identities, and we find, 
\bea
\label{secder}
\DD^2 \, \cC \left [ \matrix{ 1_\ell  \cr 1_\ell  \cr } \right ]  
& = & 2 \ell (\ell-1) \, \cC^+ \left [ \matrix{ 4 & 1_{\ell-2}  \cr 0 & 1_{\ell-2}  \cr } \right ] 
+ 3 \ell (\ell-1) \tau_2^4 G_4  \, 
\cC^+ \left [ \matrix{  1_{\ell-2}  \cr  1_{\ell-2}  \cr } \right ] 
\\
\DD^2 \, \cC \left [ \matrix{ k & 1_\ell  \cr  k & 1_\ell  \cr } \right ]  
& = & 
k(k+1) \, \cC^+ \left [ \matrix{ k+2  & 1_\ell  \cr  k-2  & 1_\ell  \cr } \right ]  
+ 2 k \ell \, \cC^+ \left [ \matrix{ k+1 & 2 & 1_{\ell-1}  \cr k-1 & 0 & 1_{\ell-1}  \cr } \right ] 
\no \\ &&
+ 2 \ell (\ell-1) \, \cC^+ \left [ \matrix{ k & 4 & 1_{\ell-2}  \cr k & 0 & 1_{\ell-2}  \cr } \right ] 
+ 3 \ell (\ell-1) \tau_2^4 G_4 \, 
\cC^+ \left [ \matrix{  k & 1_{\ell-2}  \cr  k & 1_{\ell-2}  \cr } \right ] 
\, .
\no
\eea
These relations will suffice to evaluate all the second order derivatives needed to prove the conjectured identities for dihedral modular graph functions. To evaluate higher derivatives, it will be convenient to do so graph by graph, which we shall do next.

\subsection{Proof of the $D_5$ conjecture}

The proof of the $D_5$ conjecture proceeds parallel to the proof for $D_4$. It will be helpful to use the following relation in equation in (\ref{6a0}), $C_{1,1,1} =  E_3 + \zeta (3)$, established in \cite{Zagier:2014,D'Hoker:2015foa}. We begin by computing the second derivatives of $D_5$ and $C_{3,1,1}$ using the formulas of (\ref{secder}), 
\bea
\label{6c1}
\DD ^2  D_5 & = & 
40 \, \cC^+ \!  \left [ \matrix{ 4 \, 1 \, 1 \, 1 \cr 0  \, 1 \, 1 \, 1 \cr } \right ] 
+ 60 \tau_2^4 G_4 \, C_{1,1,1}
\no \\
\DD^2  C_{3,1,1} & = & 
12 \, \cC^+ \!  \left [ \matrix{ 5 \, 1 \, 1  \cr 1  \, 1 \, 1  \cr } \right ] 
+ 12 \, \cC^+ \!  \left [ \matrix{ 4 \, 2 \, 1  \cr 2  \, 0 \, 1  \cr } \right ] 
- 4 \, \cC^+ \!  \left [ \matrix{ 3 \, 3 \, 1  \cr 2  \, 0 \, 1  \cr } \right ] 
+ 4 \, \cC^+ \!  \left [ \matrix{ 7 \, 0  \cr 3  \, 0  \cr } \right ] 
\, .
\eea
Since the holomorphic form $G_4$ cannot be differentiated further without producing non-holomorphic terms, we shall  re-express it in terms of $E_2$ by the relation $\DD ^2 E_2 = 6 \tau_2^4 G_4$ which is already familiar from the case of the proof for $D_4$.  The holomorphic factor may then be eliminated from the right side the first line in (\ref{6c1}) by subtracting the term $\DD ^2 (E_2 C_{1,1,1})$ from both sides of the first line in (\ref{6c1}), so that we obtain, 
\bea
\DD ^2 \Big ( D_5 - 10 E_2 C_{1,1,1}  \Big ) = 
40 \, \cC^+ \!  \left [ \matrix{ 4 \, 1 \, 1 \, 1 \cr 0  \, 1 \, 1 \, 1 \cr } \right ] 
- 20  (\DD E_2) \,  (\DD   E_3)   - 10 E_2 (\DD ^2 E_3)  
\, .
\eea
Taking one further derivative with the help of  (\ref{id8}) of this combination and $\DD ^2 C_{3,1,1}$ gives,
\bea
\label{6c6}
 \DD ^3 \Big ( D_5 - 10 E_2 C_{1,1,1}  \Big ) & = & 
1080 \, \cC^+ \!  \left [ \matrix{ 6 \, 1 \, 1  \cr 0  \, 1 \, 1  \cr } \right ] 
- 30 (\DD E_2) (\DD ^2 E_3)  
- 240 \tau_2^4 G_4 \,  (\DD   E_3)   
\no \\
\DD ^3 \, C_{3,1,1} & = & 
60 \, \cC^+ \!  \left [ \matrix{ 6 \, 1 \, 1  \cr 0  \, 1 \, 1  \cr } \right ] 
+ 72 \, \cC^+ \!  \left [ \matrix{ 5 \, 2 \, 1  \cr 1  \, 0 \, 1  \cr } \right ] 
- 36 \, \cC^+ \!  \left [ \matrix{ 4 \, 3 \, 1  \cr 1  \, 0 \, 1  \cr } \right ] 
\no \\ &&
+ 12 \, \cC^+ \!  \left [ \matrix{ 3 \, 4 \, 1  \cr 1  \, 0 \, 1  \cr } \right ] 
+ 72  \, \cC^+ \!  \left [ \matrix{ 8 \, 0  \cr 2  \, 0  \cr } \right ] 
- 4 \tau_2^4 G_4 \,  (\DD E_3)
\, .
\eea
Combining the two lines of (\ref{6c6}) into $F_5$, the pre-factor $-60$ of $C_{3,1,1}$ which occurs  in the
combination $F_5$ is seen to be responsible for precisely canceling against one another  the terms proportional to the holomorphic form $G_4$.  Collecting all contributions, the third derivative of $F_5$ is found with the help of (\ref{Es1}) to be given by,
\bea
\DD^3  F_5 & = & 
- 2520  \, \cC^+ \!  \left [ \matrix{ 6 \, 1 \, 1  \cr 0  \, 1 \, 1  \cr } \right ] 
- 4320 \, \cC^+ \!  \left [ \matrix{ 5 \, 2 \, 1  \cr 1  \, 0 \, 1  \cr } \right ] 
+  2160 \, \cC^+ \!  \left [ \matrix{ 4 \, 3 \, 1  \cr 1  \, 0 \, 1  \cr } \right ] 
\no \\ &&
- 720 \, \cC^+ \!  \left [ \matrix{ 3 \, 4 \, 1  \cr 1  \, 0 \, 1  \cr } \right ] 
+ 5760  \, \cC^+ \!  \left [ \matrix{ 8 \, 0  \cr 2  \, 0  \cr } \right ] 
- 30 (\DD E_2) (\DD ^2 E_3)
\, .
\eea
Applying a further derivative, and using the auxiliary formulas of appendix \ref{app:D}, we find,
\bea
\DD^ 4  F_5 = 0
\, .
\eea
Therefore, the conditions of Lemma 1 are satisfied, and we conclude that $F_5$ is constant.  The proof is completed using the fact that  the Laurent polynomial  of the asymptotic expansion of $F_5$ near the cusp was already shown to vanish  in \cite{D'Hoker:2015foa}, so that $F_5=0$.

\subsection{Proof of the $D_{3,1,1}$ conjecture}

The derivatives of $C_{3,1,1}$ have already been computed in (\ref{6c1}) and (\ref{6c6}). In addition, we need the derivatives of $D_{3,1,1}= C_{2,1,1,1}$, which by (\ref{secder})  are given as follows,  
\bea
\label{6d1}
\DD^2 \, D_{3,1,1} & = & 
6 \, \cC ^+ \! \left [ \matrix{ 4 \, 1 \, 1 \, 1 \cr 0 \, 1 \, 1 \, 1 \cr } \right ]
+ 12 \, \cC^+ \!  \left [ \matrix{ 3 \, 2 \, 1 \, 1 \cr 1 \, 0 \, 1 \, 1 \cr } \right ]
-6 \, \cC^+ \!  \left [ \matrix{ 2 \, 3 \, 1 \, 1 \cr 1 \, 0 \, 1 \, 1 \cr } \right ]
\no \\ &&
+ 12 \, \cC^+ \!  \left [ \matrix{ 4 \, 2 \, 1 \cr  0 \, 2 \, 1 \cr } \right ]
+ 18 \tau_2^4 G_4 E_3
\, .
\eea
The dependence on the holomorphic form $G_4$ may be eliminated from the right side by subtracting the combination $\DD ^2 (3 E_2E_3) $ from both sides, and doing so we find, 
\bea
\DD^2 ( D_{3,1,1} - 3 E_2 E_3) & = & 
6 \, \cC ^+ \! \left [ \matrix{ 4 \, 1 \, 1 \, 1 \cr 0 \, 1 \, 1 \, 1 \cr } \right ]
+ 12 \, \cC^+ \!  \left [ \matrix{ 3 \, 2 \, 1 \, 1 \cr 1 \, 0 \, 1 \, 1 \cr } \right ]
-6 \, \cC^+ \!  \left [ \matrix{ 2 \, 3 \, 1 \, 1 \cr 1 \, 0 \, 1 \, 1 \cr } \right ]
\no \\ &&
+ 12 \, \cC^+ \!  \left [ \matrix{ 4 \, 2 \, 1 \cr  0 \, 2 \, 1 \cr } \right ]
- 6 (\DD E_2) (\DD E_3) - 3 E_2 (\DD ^2 E_3) 
\, .
\eea
Applying one further derivative, and using the auxiliary formulas of appendix \ref{app:D}, we find,  
\bea
\label{6d2}
\DD^3 ( D_{3,1,1} - 3 E_2 E_3) & = & 
72 \, \cC ^+ \! \left [ \matrix{ 6 \, 1 \, 1  \cr 0 \, 1 \, 1  \cr } \right ]
-108 \, \cC^+ \!  \left [ \matrix{ 5 \, 2 \, 1  \cr 0 \, 1 \, 1 \cr } \right ]
+ 72 \, \cC^+ \!  \left [ \matrix{ 4 \, 3 \, 1  \cr 0 \, 1 \, 1 \cr } \right ]
\no \\ &&
+ 108 \, \cC^+ \!  \left [ \matrix{ 8 \, 0  \cr  2 \, 0 \cr } \right ]
- 9 (\DD E_2) (\DD^2 E_3) - 5 (\DD^2 E_2) (\DD  E_3) 
\, .
\eea
Combining the expression for the third order derivative of $C_{3,1,1}$ from the second line of (\ref{6c6}),
with the result from (\ref{6d2}), we obtain the third derivative of $F_{3,1,1}$, 
\bea
\DD^3 \,  F_{3,1,1} & = & 
-378 \, \cC ^+ \! \left [ \matrix{ 6 \, 1 \, 1  \cr 0 \, 1 \, 1  \cr } \right ]
-108 \, \cC^+ \!  \left [ \matrix{ 5 \, 2 \, 1  \cr 0 \, 1 \, 1 \cr } \right ]
-540 \, \cC^+ \!  \left [ \matrix{ 5 \, 2 \, 1  \cr 1 \, 0 \, 1 \cr } \right ]
-18 \, \cC^+ \!  \left [ \matrix{ 4 \, 3 \, 1  \cr 0 \, 1 \, 1 \cr } \right ]
\no \\ &&
+ 270  \, \cC^+ \!  \left [ \matrix{ 4 \, 3 \, 1  \cr 1 \, 0 \, 1 \cr } \right ]
+ 1017 \, \cC^+ \!  \left [ \matrix{ 8 \, 0  \cr  2 \, 0 \cr } \right ]
- 9 (\DD E_2) (\DD^2 E_3)  
\, .
\eea
Applying one further derivative, and using the auxiliary formulas of appendix \ref{app:D}, one finds,
\bea
\DD^4 \, F_{3,1,1}=0
\, .
\eea
Therefore, the conditions of Lemma 1 are satisfied and $F_{3,1,1}$ must be constant.  Since the Laurent polynomial  of its asymptotic expansion  near the cusp was already shown to vanish  in \cite{D'Hoker:2015foa}, we conclude that of $F_{3,1,1}=0$.

\section{Trihedral graphs}
\setcounter{equation}{0}
\label{sec:10}

{\sl Trihedral graphs} have three vertices, so $V=3$.  The total number of  edges $\L$ is partitioned into three sets,  $\L=\L_1+\L_2+\L_3$,  where $\L_i$ for $i=1,2,3$  denote the numbers of edges  connecting pairs of vertices.  The labeling of the exponents is shown in the following diagram using the general notation of subsection \ref{sec:23}.
\bea
\tikzpicture[scale=1.2]
\scope[xshift=-5cm,yshift=-0.4cm]
\draw[thick] (0,0.035) -- (3,0.035);
\draw[thick] (0,-0.035) -- (3,-0.035);
\draw[thick] (0,0.05) -- (1.5,2.05);
\draw[thick] (0,-0.05) -- (1.5,1.95);
\draw[thick] (3,0.05) -- (1.5,2.05);
\draw[thick] (3,-0.05) -- (1.5,1.95);
\draw[fill=black] (0,0)  circle [radius=.07] ;
\draw[fill=black] (3,0)  circle [radius=.07] ;
\draw[fill=black] (1.5,2)  circle [radius=.07] ;
\draw (1.5,-0.3) node{$ A^3, B^3 $};
\draw (0.5, 1.1) node [rotate=50] {$A^2,B^2$};
\draw (2.5, 1.1) node [rotate=-50] {$A^1,B^1$};
\endscope
\endtikzpicture
\eea
Each line represents a collection of labels,  and the labeling of the exponents is shorthand for the  following sets of  labels
\bea
\label{7a1}
\left [ \matrix{ A \cr B \cr } \right ] = 
 \left [ \matrix{ A^1 \cr B^1 \cr} \Bigg |  \matrix{ A^2 \cr B^2 \cr} \Bigg | \matrix{ A^3 \cr B^3 \cr} \right ] 
\hskip 1in 
\matrix{ A^i \, = \, [ \, a^i _1 ~  a^i _2 ~  \cdots ~  a^i _{\L_i} \, ]  \cr 
B^i \, = \, [ \,  b^i _2 ~ b^i _2 ~ \cdots ~ b^i _{\L_i} \, ] \cr } 
\, .
\eea
The graphical representation is based on the  double edge notation of subsection \ref{sec:23}.
The associated modular graph form is given by specializing (\ref{3b5}) to the case of trihedral graphs,
\bea
\label{7a2}
\cC \left [ \matrix{ A \cr B \cr}  \right ] (\tau)
=
\sum _{  p^i_{k_i} \in \Lambda  } \delta _{\mpp^1,\mpp^2} \delta _{\mpp^2,\mpp^3} \delta _{\mpp^3,\mpp^1} 
\prod _{i=1}^3 \prod _{k_i =1}^{\L_i} 
{ (\tau_2/\pi) ^{\half a^i _{k_i} + \half b^i _{k_i}} \over (p^i _{k_i})^{a^i_{k_i}} \, (\bar p^i _{k_i})^{b^i_{k_i}} }
\, .
\eea
Corresponding modular forms $\cC^\pm$ may be defined by (\ref{3b6}).
The total momentum $\mpp^i$ passing through the double line  of the trihedral graph labeled $i$ is given by,
\bea
\label{7a3}
\mpp^i = \sum _{k_i=1}^{\L _i} p^i _{k_i}
\, .
\eea
The Kronecker $\delta$-functions require  all $\mpp^i$ to be equal to one another.  By overall momentum conservation, one of the  Kronecker  $\delta$-functions is redundant, and may be omitted.

\subsection{Symmetries of trihedral modular graph forms}

Trihedral modular graph forms are manifestly  invariant under the following permutations,
\bea
(\sigma _0,  \sigma _1, \sigma _2, \sigma _3) \in \mS_3 \times \mS_{\L_1} \times \mS_{\L_2} \times \mS_{\L_3} 
\eea
which act upon the momenta by $\sigma _0(p^i _{k_i}) = p^{\sigma _0(i)} _{k_{\sigma _0(i)}}$,
$\sigma _i (p^i _{k_i}) = p^i _{\sigma (k_i)}$ and on the exponents by, 
\bea
\label{7b1}
\sigma _0 \left [ \matrix{ A \cr B \cr } \right ] =  
 \left [ \matrix{ A^{\sigma_0 (1)}  \cr B^{\sigma _0 (1)} \cr} \Bigg |  
 \matrix{ A^{\sigma _0(2)} \cr B^{\sigma _0 (2)}  \cr} \Bigg | \matrix{ A^{\sigma _0(3)}  \cr B^{\sigma _0 (3)}  \cr} \right ] 
\hskip 0.6in 
\sigma _i \left [ \matrix{ A^i \cr B^i \cr } \right ]
= \left [ \matrix{ a^i _{\sigma _i(1)}  & a^i _{\sigma _i(2)} & \cdots & a^i _{\sigma _i(\L_i)} \cr 
b^i _{\sigma _i(1)} & b^i _{\sigma _i(2)} & \cdots & b^i _{\sigma _i(\L_i )} \cr } \right ] 
\eea
while the action of the permutations $\sigma _i$ on $A^j, B^j$ is trivial for $j \not= i$.

\subsection{Momentum conservation identities}

Momentum conservation relations on trihedral modular graph forms follow from inserting $\mpp_1 - \mpp_2$ and $\mpp_2 - \mpp_3$ into the summands of (\ref{7a2}). The resulting relations may be written out explicitly as follows, 
\bea
\sum _{k_1=1}^{\L_1} \cC  \left [ \matrix{ A^1 - S_{k_1}^{(1)} \cr B^1 \cr} \Bigg |  \matrix{ A^2 \cr B^2 \cr} \Bigg | \matrix{ A^3 \cr B^3 \cr} \right ] 
- \sum _{k_2=1}^{\L_2} \cC  \left [ \matrix{ A^1  \cr B^1 \cr} \Bigg |  \matrix{ A^2 - S_{k_2}^{(2)} \cr B^2 \cr} \Bigg | \matrix{ A^3 \cr B^3 \cr} \right ] & = & 0
\no \\
\sum _{k_2=1}^{\L_2} \cC  \left [ \matrix{ A^1  \cr B^1 \cr} \Bigg |  \matrix{ A^2 - S_{k_2}^{(2)} \cr B^2 \cr} \Bigg | \matrix{ A^3 \cr B^3 \cr} \right ] 
- \sum _{k_3=1}^{\L_3} \cC  \left [ \matrix{ A^1  \cr B^1 \cr} \Bigg |  \matrix{ A^2  \cr B^2 \cr} \Bigg | \matrix{ A^3 - S_{k_3}^{(3)} \cr B^3 \cr} \right ] & = & 0
\, .
\eea
Here, $S^{i}_{k_i}$ is a row vector of length $\L_i$, where the index runs over the range $k_i =1, \cdots \L_i$, and is defined as in (\ref{3b9}) with $\L=\L_i$. In addition, one has two complex conjugate relations in which the $S^{(i)}_{k_i}$ subtractions are being performed on the lower rows.

\subsection{Specializing to graphs with $\L_1=\L_2=2$ and $\L_3=1$}

The simplest non-trivial trihedral graph has two trivalent and one quadrivalent vertex, $\L_1=\L_2=2\L_3=2$.  In the following we shall spell out momentum conservation and permutation symmetry properties, and the algebraic and holomorphic subgraph reduction identities needed for this specific case.

\sm

A first issue concerns the momentum labeling. Given the reduced symmetry of this graph, it will be convenient to flip the sign of the last momentum when deriving the reduction identities, so that these trihedral modular graph forms  are defined by,
\bea
\label{7d1}
\cC ^+ \! \left [ \matrix{ a_1 ~ a_2 \cr b_1 ~ b_2 \cr}  \Bigg | \matrix{ a_3 ~ a_4 \cr b_3 ~ b_4 \cr} 
\Bigg |  \matrix{ a_5 \cr b_5 \cr} \right ]
= 
\sum _{p_1, \cdots, p_5 \in \Lambda} ' \delta _{p_1+p_2+p_5,0} \, \delta _{p_3+p_4+p_5,0} \, 
\prod _{r=1}^5 { \tau _2 ^{a_r} \pi ^{- \half a_r - \half b_r} \over (p_r) ^{a_r} (\bar p_r) ^{b_r}}
\, .
\eea
To economize on notation, we have labeled the exponents $a_r$ and $b_r$ sequentially with a single index~$r$. The above momentum conservation identities, with the reversed sign of $p_5$, may then be expressed in the following form,
\bea
\label{7d2}
\cC ^+ \! \left [ \matrix{ a_1 - 1 & \! a_2 \cr b_1 & \! b_2 \cr}  \Bigg | \matrix{ a_3 \, a_4 \cr b_3 \, \, b_4 \cr} 
\Bigg |  \matrix{ a_5 \cr b_5 \cr} \right ]
+ \cC ^+ \! \left [ \matrix{ a_1 \! & a_2-1 \cr b_1 \! & b_2 \cr}  \Bigg | \matrix{ a_3 \, a_4 \cr b_3 \, \, b_4 \cr} 
\Bigg |  \matrix{ a_5 \cr b_5 \cr} \right ]
+ \cC ^+ \! \left [ \matrix{ a_1 ~ a_2 \cr b_1 ~ b_2 \cr}  \Bigg | \matrix{ a_3 \, a_4 \cr b_3 \, \, b_4 \cr} 
\Bigg |  \matrix{ a_5-1 \cr b_5 \cr} \right ] & = & 0
\no \\ &&
\\
\cC ^+ \! \left [ \matrix{ a_1 \, a_2 \cr b_1 \, \, b_2 \cr}  \Bigg | \matrix{ a_3-1 & a_4 \cr b_3 & b_4 \cr} 
\Bigg |  \matrix{ a_5 \cr b_5 \cr} \right ]
+ \cC ^+ \! \left [ \matrix{ a_1 \, a_2 \cr b_1 \, \, b_2 \cr}  \Bigg | \matrix{ a_3 & a_4-1 \cr b_3 & b_4 \cr} 
\Bigg |  \matrix{ a_5 \cr b_5 \cr} \right ]
+ \cC ^+ \! \left [ \matrix{ a_1 \, a_2 \cr b_1 \, \, b_2 \cr}  \Bigg | \matrix{ a_3 \, a_4 \cr b_3 \, \, b_4 \cr} 
\Bigg |  \matrix{ a_5-1 \cr b_5 \cr} \right ] & = & 0
\, .
\no
\eea
Next, we shall derive algebraic and holomorphic subgraph reduction identities which extend to trihedral graphs the  identities of subsections \ref{sec:44} and  \ref{sec:5} derived for dihedral graphs.

\subsection{Algebraic reduction}

$\bullet$ The first algebraic reduction formula is given by,
\bea
\label{7d3}
\cC ^+ \! \left [ \matrix{ a_1 ~ a_2 \cr b_1 ~ b_2 \cr}  \Bigg | \matrix{ a_3 ~ a_4 \cr b_3 ~ b_4 \cr} \Bigg |  \matrix{ 0 \cr 0 \cr} \right ]
& = &
(-)^{a_3+b_3+a_4+b_4} \, 
\cC ^+ \! \left [ \matrix{ a_1 ~ a_2  ~ a_3 ~ a_4 \cr b_1 ~ b_2  ~ b_3 ~ \, b_4 \cr}  \right ]
\no \\ &&
- (-)^{a_2+b_2+a_3+b_3} \, 
\cC ^+ \! \left [ \matrix{ a_1 + a_2  ~\, 0 \cr b_1 + b_2  ~\, \, 0 \cr}  \right ] \, 
\cC^+ \! \left [ \matrix{ a_3 + a_4 ~\,  0 \cr b_3 + b_4 \, ~ \, 0 \cr}  \right ]
\, .
\eea
To prove this formula, we set $a_5=b_5=0$ in (\ref{7d1}), so that the summand depends on $p_5$
only through the two Kronecker $\delta$-functions. This summation may be carried out and results in,
\bea
\label{7d4}
\sum _{p_5 \in \Lambda} ' \delta _{p_1+p_2+p_5,0} \, \delta _{p_3+p_4+p_5,0}
= \delta _{p_1+p_2,p_3+p_4} \, (1 - \delta _{p_1+p_2,0})
\, .
\eea
We now evaluate the contribution to the sum of each term in turn. The first term on the right side of (\ref{7d4})
produces the first term on the right side of (\ref{7d3}), the sign factor arising from the reversal of the signs of 
$p_3$ and $p_4$ needed to match the definition of the $\cC$-function with four indices. The second term on the right side of (\ref{7d4}) produces the second term on the right side of (\ref{7d3}), the sign factor again arising from the reversal of the signs of $p_2, p_3$.

\sm

\noindent
$\bullet$ The second algebraic reduction formula is given by, 
\bea
\label{7d5}
\cC ^+ \! \left [ \matrix{ a_1 ~ a_2 \cr b_1 ~ b_2 \cr}  \Bigg | \matrix{ a_3 ~ 0 \cr b_3 ~ 0 \cr} \Bigg |  \matrix{ a_5 \cr b_5 \cr} \right ]
=
\cC ^+ \! \left [ \matrix{ a_3 ~ 0 \cr b_3 ~ 0 \cr} \right ] \, \cC ^+ \! \left [ \matrix{ a_1 ~ a_2 ~ a_5  \cr b_1 ~ b_2 ~ b_5 \cr} \right ] 
- (-)^{a_3+b_3} \cC ^+ \! \left [ \matrix{ a_1 ~ a_2 & a_3+a_5  \cr b_1 ~ b_2 & b_3+ b_5 \cr} \right ] 
\, .
\eea
An analogous formula arises upon interchange of the indices $(1,2)$ and $(3,4)$.
To prove this formula, we set $a_4=b_4=0$ in (\ref{7d1}), so that the summand depends
on $p_4$ only through the single Kronecker $\delta$-function $\delta _{p_3+p_4+p_5}$. This summation may be 
carried out and results in,
\bea
\label{7d6}
\sum _{p_4 \in \Lambda } '  \delta _{p_3+p_4+p_5,0} =  1 - \delta _{p_3+p_5,0}
\, .
\eea
The first term on the right side of (\ref{7d6}) produces the first term on the right side of (\ref{7d5}),
since the summation over $p_3$ is decoupled from the summations over $p_1, p_2, p_5$.  The second term on the  right side of (\ref{7d6}) produces the second term on the right side of (\ref{7d5}), the sign factor again arising from the reversal of the sign of $p_3$.

\subsection{Holomorphic reduction for 2-point subgraphs}

In a dihedral graph, the only one-loop subgraph with vanishing lower exponents  contains the two  vertices of valence at least equal to three.  For a trihedral graph, there are actually two possibilities: the subgraph can contain either two or three vertices with valence at least equal to three. With two higher valence vertices, we again have a two-point subgraph, while with three higher valence vertices, the holomorphic subgraph is a three-point subgraph.  In this section, we give the simplest holomorphic subgraph reduction for two-point subgraphs, leaving the three-point subgraph case for the next subsection. 

\sm

The first holomorphic subgraph reduction identity, which corresponds to a two-point function of holomorphic weight 4,  is given by,
\bea
\label{11b3}
\cC ^+ \! \left [ \matrix{ a_1 ~ a_2 \cr b_1 ~ b_2 \cr}  \Bigg | \matrix{ 2 ~ 2 \cr 0 ~ 0 \cr} \Bigg |  \matrix{ a_5 \cr b_5 \cr} \right ]
- 2 \, \cC ^+ \! \left [ \matrix{ a_1 ~ a_2 \cr b_1 ~ b_2 \cr}  \Bigg | \matrix{ 3 ~ 1 \cr 0 ~ 0 \cr} \Bigg |  \matrix{ a_5 \cr b_5 \cr} \right ]
= 2 \, \cC ^+ \! \left [ \matrix{ a_1 ~ a_2 & 4+a_5  \cr b_1 ~ b_2 & b_5 \cr} \right ] 
\, .
\eea
To prove this formula, we carry out the sum in (\ref{7d1}) over $p_3$ and $p_4$ first,
\bea
\sum _{p_3, p_4 \in \Lambda } ' \delta _{p_3+p_4+p_5,0} \, \left ( { 1 \over p_3^2 \, p_4^2} - { 2  \over p_3^3 \, p_4} \right )
= \pi^2 \cG_{2,2}(-p_5) - 2 \pi^2 \cG_{3,1}(-p_5)
\eea
using the general expression for $\cG$ introduced in (\ref{5a3}). Here $\cG$ is evaluated on a non-vanishing argument, $-p_5$, and we find, 
\bea
\cG_{2,2}(-p_5) - 2 \cG_{3,1}(-p_5) = { 2 \over \pi^2 p_5^4}
\eea
which immediately results in formula (\ref{11b3}). This completes the proof of the basic  reduction identities for two-point holomorphic subgraphs that we shall need to prove the $D_{2,2,1}$ conjecture in section \ref{sec:11}.

\subsection{Holomorphic reduction for 3-point subgraphs}
\label{sec:76}

The second holomorphic subgraph identity we shall need pertains to a three-point function of holomorphic weight 6,
\bea
\label{7f1}
\cL & = & 
\cC ^+ \! \left [ \matrix{ 2 ~ a_2 \cr 0 ~ b_2 \cr}  \Bigg | \matrix{ 2 ~ a_4 \cr 0 ~ b_4 \cr} \Bigg |  \matrix{ 2 \cr 0 \cr} \right ]
- 2 \, \cC ^+ \! \left [ \matrix{ 3 ~ a_2 \cr 0 ~ b_2 \cr}  \Bigg | \matrix{ 2 ~ a_4 \cr 0 ~ b_4 \cr} \Bigg |  \matrix{ 1 \cr 0 \cr} \right ]
- 2 \, \cC ^+ \! \left [ \matrix{ 2 ~ a_2 \cr 0 ~ b_2 \cr}  \Bigg | \matrix{ 3 ~ a_4 \cr 0 ~ b_4 \cr} \Bigg |  \matrix{ 1 \cr 0 \cr} \right ]
\, .
\eea
We shall prove below that the expression reduces to,
\bea
\label{7f2}
\cL & = & 9 \,  \cC ^+ \! \left [ \matrix{ a_2 + a_4 + 6 & 0  \cr b_2+ b_4 & 0 \cr} \right ] 
- 3 \tau_2^4 \, G_4 \,   \cC ^+ \! \left [ \matrix{ a_2 + a_4 + 2 & 0  \cr b_2+ b_4 & 0 \cr} \right ] + (-)^{a_4+b_4} \cL'
\eea
where $\cL'$ is given by,
\bea
\cL' & = & 
2 \, \cC ^+ \! \left [ \matrix{ a_2+4  & a_4 & 2  \cr b_2 & b_4 & 0 \cr} \right ] 
+ 2 \, \cC ^+ \! \left [ \matrix{ a_2  & a_4+4 & 2  \cr b_2 & b_4 & 0 \cr} \right ] 
\no \\ &&
-3 \, \cC ^+ \! \left [ \matrix{ a_2+2  & a_4+2 & 2  \cr b_2 & b_4 & 0 \cr} \right ] 
+ 2 \, \cC ^+ \! \left [ \matrix{ a_2+1  & a_4+1 & 4  \cr b_2 & b_4 & 0 \cr} \right ] 
\, .
\eea
To prove this identity, we write out $\cL$ as a sum,
\bea
\label{7f3}
\cL  =  \sum _{p_1, \cdots, p_5 \in \Lambda } '  \delta _{p_1+p_2+p_5,0} \, \delta _{p_3+p_4+p_5,0} \, 
{ (\tau_2)^{6+a_2+a_4} \over \pi^c \, p_2^{a_2} \, \bar p_2^{b_2} \, p_4^{a_4} \, \bar p_4^{b_4}}
\left ( { 1 \over p_1^2 p_3^2 p_5^2} - { 2 \over p_1^3 p_3^2 p_5 } - { 2 \over p_1^2 p_3^3 p_5} \right )
\eea
and we have used the abbreviation $c=3 + \half ( a_2+a_4+b_2+b_4)$. 
Next, we solve the Kronecker $\delta$-function constraints for $p_1$ and $p_3$ in terms of $p_2, p_4, p_5$,
which we retain as independent summation variables, subject to the following constraints, 
\bea
p_1 = -p_2 -p_5 & \hskip 0.8in & p_5 \not= - p_2
\no \\ 
p_3 = -p_4 -p_5 & \hskip 0.8in & p_5 \not= - p_4
\, .
\eea
The loop momentum through the holomorphic three-point subgraph is then $p_5$, so that we split up the summation as follows,
\bea
\label{7f5}
\cL = \sum _{p_2,p_4 \in \Lambda } ' 
{ (\tau_2 )^{6+a_2+a_4} \over \pi^c \, p_2^{a_2} \, \bar p_2^{b_2} \, p_4^{a_4} \, \bar p_4^{b_4}} \, \cH(p_2,p_4)
\eea
where the three-point holomorphic subgraph sum is given as follows,
\bea
\label{7f6}
\cH(p_2,p_4) = \! 
\sum _{p_5 \not= -p_2, -p_4 \in \Lambda } ' \! { 1 \over (p_4+p_5)^2}
\left ( { {1 \over 2}  \over (p_2+p_5)^2  p_5^2} + { 2 \over (p_2+p_5)^3  p_5 } 
\right ) + \, (p_2 \leftrightarrow p_4)
\, .
\quad
\eea
Here, and below, the term $(p_2 \leftrightarrow p_4)$ stands for the preceding sum in which the arguments $p_2$ and $p_4$ have been swapped.
We partition the summation over $p_2$ and $p_4$ into the contributions from $p_2=p_4$ and from $p_2 \not= p_4$. The first reduces to a sum of two-point holomorphic subgraph contributions, and may be evaluated using the methods of section \ref{sec:5}. Although an Eisenstein regularization is required at intermediate stages of the calculation, all contributions in $G_2$ and $Q_1(p)$ cancel, and we are left with,
\bea
\cH(p_2,p_2) = { 9 \over p_2^6 } - { 3 \pi^2 G_4 \over p_2^2} 
\, .
\eea
Substituting this result into the part of (\ref{7f5}) with $p_2=p_4$, we find the contributions of the first two terms on the right side of (\ref{7f2}). To compute the contribution from  $p_2 \not= p_4$, we decompose the summand in (\ref{7f6}) into partial fractions with respect to the momentum summation variable $p_5$. All summations over $p_5$ may be performed with the help of the formulas (\ref{5f1}) and (\ref{5f2}). All terms in $G_2$ and $Q_1$ cancel, and we are left with,
\bea
\cH(p_2,p_4) = 
{ 2 \over p_2^4 (p_2-p_4)^2 } + { 2 \over p_4^4 (p_2-p_4)^2 }  - { 3 \over p_2^2p_4^2 (p_2-p_4)^2 }  - { 2 \over p_2 p_4 (p_2-p_4)^4 } 
\eea
for $p_2 \not=p_4$. Substituting this result into the part of (\ref{7f5}) with $p_2\not= p_4$, we find the terms in $\cL'$, where the sign factor in front of $\cL'$ in (\ref{7f2}) arises from reversing the sign of $p_4$.

\section{Proving the trihedral $D_{2,2,1}$ conjecture}
\setcounter{equation}{0}
\label{sec:11}

The trihedral modular graph function $D_{2,2,1}$ was investigated in \cite{D'Hoker:2015foa}. It may be represented using the notation developed in the preceding section by, 
\bea
D_{2,2,1}= 
\cC \left [ \matrix{ 1 \, 1 \cr 1 \, 1 \cr}  \Bigg | \matrix{ 1 \, 1 \cr 1 \, 1 \cr} \Bigg |  \matrix{ 1 \cr 1 \cr} \right ]
\, .
\eea
In this section, we shall use the  formalism of this paper to prove the $D_{2,2,1}$ conjecture, which states
the vanishing of the following modular function, 
\bea
\label{11a1}
F_{2,2,1} = D_{2,2,1} - 2 C_{3,1,1} + { 2 \over 5 } E_5 -{ 3 \over 10 } \zeta (5)
\, .
\eea
The $\DD$ derivatives of $C_{3,1,1}$ have already been obtained in (\ref{6c1}) and (\ref{6c6}) up to third order. It remains to compute the $\DD$ derivatives of $D_{2,2,1}$, which we do in the next subsection.

\subsection{Second order derivative of $D_{2,2,1}$}

We begin by computing the first and second order derivatives of $D_{2,2,1}$,
\bea
\label{11a2}
\DD  D_{2,2,1} & = & 
4 \cC ^+ \! \left [ \matrix{ 2 \, 1 \cr 0 \, 1 \cr}  \Bigg | \matrix{ 1 \, 1 \cr 1 \, 1 \cr} \Bigg |  \matrix{ 1 \cr 1 \cr} \right ]
+  \cC ^+ \! \left [ \matrix{ 1 \, 1 \cr 1 \, 1 \cr}  \Bigg | \matrix{ 1 \, 1 \cr 1 \, 1 \cr} \Bigg |  \matrix{ 2 \cr 0 \cr} \right ]
\no \\
\DD^2  D_{2,2,1} & = & 
8 \, \cC ^+ \! \left [ \matrix{ 2 \, 1 \cr 0 \, 1 \cr}  \Bigg | \matrix{ 2 \, 1 \cr 0 \, 1 \cr} \Bigg |  \matrix{ 1 \cr 1 \cr} \right ]
+ \cL_2
\eea
where we have collected in $\cL_2$ the following sum of terms,
\bea
\label{11a3}
\cL_2 = 
8 \, \cC^+ \! \left [ \matrix{ 5 \, 1 \,  1 \cr 0 \, 1 \, 1 \cr } \right ]
- 8 \, \cC ^+ \! \left [ \matrix{ 3 \, 1 \cr 0 \, 1 \cr}  \Bigg | \matrix{ 1 \, 1 \cr 1 \, 1 \cr} \Bigg |  \matrix{ 1 \cr 0 \cr} \right ]
+ 8 \, \cC ^+ \! \left [ \matrix{ 2 \, 1 \cr 0 \, 1 \cr}  \Bigg | \matrix{ 1 \, 1 \cr 1 \, 1 \cr} \Bigg |  \matrix{ 2 \cr 0 \cr} \right ]
- 4 \, \cC ^+ \! \left [ \matrix{ 1 \, 1 \cr 0 \, 1 \cr}  \Bigg | \matrix{ 1 \, 1 \cr 1 \, 1 \cr} \Bigg |  \matrix{ 3 \cr 0 \cr} \right ]
\, .
\eea
To obtain the first term on the right side above, we have made use of the  holomorphic subgraph reduction identity derived from (\ref{11b3}) for $a_1=a_2=a_5=b_1=b_2=b_5=1$.

\sm

The remaining three terms on the right side of (\ref{11a3}) may be further simplified with the help of the momentum conservation identities of (\ref{7d2}) combined with the algebraic reduction identities of (\ref{7d3}) and (\ref{7d5}). We shall illustrate how this works on the first term, to which we first apply the momentum conservation identity on the first line of (\ref{7d2}), and we obtain, 
\bea
\label{11b1}
\cC ^+ \! \left [ \matrix{ 3 \, 1 \cr 0 \, 1 \cr}  \Bigg | \matrix{ 1 \, 1 \cr 1 \, 1 \cr} \Bigg |  \matrix{ 1 \cr 0 \cr} \right ]
= 
- \cC ^+ \! \left [ \matrix{ 3 \, 2 \cr 0 \, 1 \cr}  \Bigg | \matrix{ 1 \, 1 \cr 1 \, 1 \cr} \Bigg |  \matrix{ 0 \cr 0 \cr} \right ]
- \cC ^+ \! \left [ \matrix{ 2 \, 2 \cr 0 \, 1 \cr}  \Bigg | \matrix{ 1 \, 1 \cr 1 \, 1 \cr} \Bigg |  \matrix{ 1 \cr 0 \cr} \right ]
\, .
\eea
To the first term on the right side of (\ref{11b1}) is now ready for the application of the algebraic reduction formula (\ref{7d3}),
but the second is not. To the second term, we again apply the momentum conservation identities of (\ref{7d2}), and we find, 
\bea
\cC ^+ \! \left [ \matrix{ 2 \,  2 \cr 0 \, 1 \cr}  \Bigg | \matrix{ 1 \, 1 \cr 1 \, 1 \cr} \Bigg |  \matrix{ 1 \cr 0 \cr} \right ]
= 
- \cC ^+ \! \left [ \matrix{ 2 \, 3 \cr 0 \, 1 \cr}  \Bigg | \matrix{ 1 \, 1 \cr 1 \, 1 \cr} \Bigg |  \matrix{ 0 \cr 0 \cr} \right ]
+ \cC ^+ \! \left [ \matrix{ 1 \, 4 \cr 0 \, 1 \cr}  \Bigg | \matrix{ 1 \, 1 \cr 1 \, 1 \cr} \Bigg |  \matrix{ 0 \cr 0 \cr} \right ]
+ \cC ^+ \! \left [ \matrix{ 0 \, 4 \cr 0 \, 1 \cr}  \Bigg | \matrix{ 1 \, 1 \cr 1 \, 1 \cr} \Bigg |  \matrix{ 1 \cr 0 \cr} \right ]
\, .
\eea
Proceeding analogously for the last two  terms on the right side of (\ref{11a3}), and collecting all contributions, we obtain,  
\bea
\cL_2 & = & 8 \, \cC^+ \! \left [ \matrix{ 5 ~ 1 ~  1 \cr 0 ~ 1 ~ 1 \cr } \right ]
+ 4 \, \cC ^+ \! \left [ \matrix{ 0 ~ 2 \cr 0 ~ 1 \cr}  \Bigg | \matrix{ 1 ~ 1 \cr 1 ~ 1 \cr} \Bigg |  \matrix{ 3 \cr 0 \cr} \right ]
+ 4 \, \cC ^+ \! \left [ \matrix{ 0 ~ 3 \cr 0 ~ 1 \cr}  \Bigg | \matrix{ 1 ~ 1 \cr 1 ~ 1 \cr} \Bigg |  \matrix{ 2 \cr 0 \cr} \right ]
\no \\ &&
- 4 \, \cC ^+ \! \left [ \matrix{ 0 ~ 4 \cr 0 ~ 1 \cr}  \Bigg | \matrix{ 1 ~ 1 \cr 1 ~ 1 \cr} \Bigg |  \matrix{ 1 \cr 0 \cr} \right ]
- 4 \,  \cC ^+ \! \left [ \matrix{ 1 ~ 4 \cr 0 ~ 1 \cr}  \Bigg | \matrix{ 1 ~ 1 \cr 1 ~ 1 \cr} \Bigg |  \matrix{ 0 \cr 0 \cr} \right ]
+ 8 \, \cC ^+ \! \left [ \matrix{ 3 ~ 2 \cr 0 ~ 1 \cr}  \Bigg | \matrix{ 1 ~ 1 \cr 1 ~ 1 \cr} \Bigg |  \matrix{ 0 \cr 0 \cr} \right ]
\, .
\eea
Finally, each term on the right side, except the first one, may be simplified with the help of one of the algebraic reductions identities, to obtain,
\bea
\cL_2 & = & 
4 \, \cC^+ \! \left [ \matrix{ 5 ~ 1 ~  1 \cr 1 ~ 1 ~ 1 \cr } \right ]
- 4 \, \cC^+ \! \left [ \matrix{ 4 ~ 1 ~  1 ~ 1 \cr 1 ~ 0 ~ 1 ~ 1 \cr } \right ]
+ 8 \, \cC^+ \! \left [ \matrix{ 3 ~ 2 ~  1 ~ 1 \cr 0 ~ 1 ~ 1 ~ 1 \cr } \right ]
\no \\ &&
+ 4 \, \cC^+ \! \left [ \matrix{ 5 ~ 0 \cr 1 ~ 0  \cr } \right ] \cC^+ \! \left [ \matrix{ 2 ~ 0  \cr 2 ~ 0 \cr } \right ]
+ 4 \,  \cC^+ \! \left [ \matrix{ 3 ~ 0 \cr 1 ~ 0  \cr } \right ] \cC^+ \! \left [ \matrix{ 2 ~ 1 ~  1 \cr 0 ~ 1 ~ 1 \cr } \right ]
\, .
\eea
Thus, the entire contribution of $\cL_2$ has now been reduced to modular graph forms associated with dihedral graphs.

\subsection{Third order derivatives of $D_{2,2,1}$ and $F_{2,2,1}$}

Next, we evaluate the third order derivative of $D_{2,2,1}$, by applying $\DD$ to the second line in (\ref{11a2}). The calculation of $\DD \cL_2$ is uneventful, and may be simplified with the help of algebraic and holomorphic subgraph reduction identities for dihedral graphs, and we find, 
\bea
\label{derL2}
\DD \cL_2 =
96 \, \cC^+ \! \left [ \matrix{ 6 \, 1 \,  1 \cr 0 \, 1 \, 1 \cr } \right ] 
+ 80 \, \cC^+ \! \left [ \matrix{ 5 \, 2 \,  1 \cr 0 \, 1 \, 1 \cr } \right ] 
+ 8 \, \cC^+ \! \left [ \matrix{ 5 \, 2 \,  1 \cr 1 \, 0 \, 1 \cr } \right ]
+ 24 \,  \cC^+ \! \left [ \matrix{ 5 \, 0 \cr 1 \, 0  \cr } \right ] \cC^+ \! \left [ \matrix{ 3 \, 0  \cr 1 \, 0 \cr } \right ]
\, .
\eea
The calculation  of the remaining derivative will involve the holomorphic subgraph reduction formula of (\ref{7f1}) and (\ref{7f2}). The starting point is,
\bea
\label{8f1}
\DD \, \cC ^+ \! \left [ \matrix{ 2 \, 1 \cr 0 \, 1 \cr}  \Bigg | \matrix{ 2 \, 1 \cr 0 \, 1 \cr} \Bigg |  \matrix{ 1 \cr 1 \cr} \right ]
=
4 \, \cC^+ \! \left [ \matrix{ 5 \, 2 \,  1 \cr 1 \, 0 \, 1 \cr } \right ] 
+ \cC ^+ \! \left [ \matrix{ 2 \, 1 \cr 0 \, 1 \cr}  \Bigg | \matrix{ 2 \, 1 \cr 0 \, 1 \cr} \Bigg |  \matrix{ 2 \cr 0 \cr} \right ]
- 4 \, \cC ^+ \! \left [ \matrix{ 3 \, 1 \cr 0 \, 1 \cr}  \Bigg | \matrix{ 2 \, 1 \cr 0 \, 1 \cr} \Bigg |  \matrix{ 1 \cr 0 \cr} \right ]
\eea
where the first term is the result of a simplification obtained using the holomorphic subgraph reduction formula of (\ref{11b3}).  The sum of the second and third terms on the right of (\ref{8f1}) may be reduced using (\ref{7f1}) and (\ref{7f2}) with $a_2=a_4=b_2=b_4=1$, and we find,
\bea
\label{8f1a}
\DD \, \cC ^+ \! \left [ \matrix{ 2 \, 1 \cr 0 \, 1 \cr}  \Bigg | \matrix{ 2 \, 1 \cr 0 \, 1 \cr} \Bigg |  \matrix{ 1 \cr 1 \cr} \right ]
& = &
8 \, \cC^+ \! \left [ \matrix{ 5 \, 2 \,  1 \cr 1 \, 0 \, 1 \cr } \right ] 
- 3 \,  \cC^+ \! \left [ \matrix{ 3 \, 3 \,  2 \cr 1 \, 1 \, 0 \cr } \right ] 
+ 2 \,  \cC^+ \! \left [ \matrix{ 4 \, 2 \,  2 \cr 0 \, 1 \, 1 \cr } \right ] 
\no \\ &&
+ 9 \,  \cC^+ \! \left [ \matrix{ 8 \, 0 \cr 2 \, 0 \cr } \right ] 
- 3 \,  \cC^+ \! \left [ \matrix{ 4 \, 0 \cr  0 \, 0 \cr } \right ]  \,  \cC^+ \! \left [ \matrix{ 4 \, 0 \cr 2 \, 0 \cr } \right ] 
\, .
\eea
Assembling the third order derivatives of $F_{2,2,1}$ by using the third order derivative $\DD^3 C_{3,1,1}$ already obtained in (\ref{6c6}), we find after some mild simplifications, 
\bea
\DD ^3 F_{2,2,1} & = & 
-24 \, \cC^+ \! \left [ \matrix{ 6 \, 1 \,  1 \cr 0 \, 1 \, 1 \cr } \right ] 
+ 72 \, \cC^+ \! \left [ \matrix{ 5 \, 2 \,  1 \cr 0 \, 1 \, 1 \cr } \right ] 
- 72  \, \cC^+ \! \left [ \matrix{ 5 \, 2 \,  1 \cr 1 \, 0 \, 1 \cr } \right ]
- 48  \,  \cC^+ \! \left [ \matrix{ 4 \, 3 \,  1 \cr 0 \, 1 \, 1 \cr } \right ] 
\no \\ &&
+ 48 \,  \cC^+ \! \left [ \matrix{ 4 \, 3 \,  1 \cr 1 \, 0 \, 1 \cr } \right ] 
+ 24 \,  \cC^+ \! \left [ \matrix{ 5 \, 0 \cr 1 \, 0  \cr } \right ] \cC^+ \! \left [ \matrix{ 3 \, 0  \cr 1 \, 0 \cr } \right ]
+ 12 \,  \cC^+ \! \left [ \matrix{ 8 \, 0 \cr 2 \, 0 \cr } \right ]  
\, .
\eea
Applying one more derivative, and extensively using the holomorphic subgraph reduction identities of subsection \ref{sec:54} produces the identities summarized in appendix \ref{app:D}, and we find, 
\bea
\DD^4 F_{2,2,1}=0
\, .
\eea
This result, along with the matching asymptotic Laurent series, proves the $D_{2,2,1}$ conjecture.

\section{Discussion}
\setcounter{equation}{0}
\label{sec:12}

Modular graph forms considered in this paper generalize the  modular graph functions that  were originally motivated by the expansion of genus-one superstring  graviton scattering amplitudes, and are the subject of earlier papers \cite{D'Hoker:2015zfa,D'Hoker:2015foa,D'Hoker:2015qmf}.  We have found this to be a useful generalization for formulating streamlined  proofs of the relationships between modular graph functions conjectured in \cite{D'Hoker:2015zfa}.  It is also clear that this  enlarged space is essential for developing a more complete understanding of the basis of these functions.  As stressed in  \cite{D'Hoker:2015qmf} modular graph functions may be viewed as special values of single-valued elliptic multiple polylogarithms and are elliptic generalisations of the single-valued elliptic polylogarithms introduced by Zagier \cite{Zagier:1990},  and denoted $D_{a,b}(q;e^{2i \pi z})$, in the special cases for which $a=b$.  The modular graph forms discussed in this paper are similarly related to generalizations  in which $a\ne b$. 

\sm

The polynomial relations between modular graph functions conjectured in  \cite{D'Hoker:2015zfa}, which have now been proved, are simple low-weight examples of what must surely be a web  of similar relations at arbitrary weight.  The relations that have been considered so far at a given weight $w$ (such as those  with $w= 4$ and $5$ in (\ref{6a0})) imply the vanishing of certain polynomials in modular graph functions where the terms in each polynomial have weight $w$ and rational coefficients. Recently, an interesting identity was obtained for the simplest tetrahedral modular graph function of weight six  in \cite{Basu:2015ayg} using the methods of \cite{D'Hoker:2015zfa}.  The functions that enter in these relations are all of the form (\ref{3b5})  with the array $A$ of holomorphic exponents equal to the array  $B$  of anti-holomorphic exponents.  

\sm

However, we expect that more general relations apply at higher weight.  The generalizations involved should be of two kinds:
\begin{description}
\item[$(i)$] Modular graph functions, i.e. modular forms of weight  $(0,0)$, should be included for which the arrays of exponents $A$ and $B$ that enter the definition in (\ref{3b5}) are different, even though their net holomorphic and anti-holomorphic weights vanish.  Such functions enter, for example,  in the low energy expansion of superstring $N$-graviton one-loop scattering amplitudes  when $N>4$, as was shown in  \cite{Green:2013bza}.
\item[$(ii)$] We also expect that modular forms with non-zero modular weights will enter into more general polynomial relations.  In other words, the general relations  should  be polynomial in modular graph forms and all the terms in the polynomial not only have the same weight $w$, but also the same modular weight, which may be non-zero. Rather trivial examples of this kind can be obtained simply by applying modular covariant derivatives to any of the relations conjectured in \cite{D'Hoker:2015zfa}.  Less trivial examples are provided by the holomorphic subgraph reduction identities of Section \ref{sec:5}, and in particular the formulas of subsection \ref{sec:54}.  More generally we would expect less obvious polynomial relations of non-zero modular weight to appear.
\end{description}

\sm
 
The results of this paper may be expected to have important relations with the single-valued multiple zeta values (svMZV),  which arise as coefficients in the expansion of tree-level closed-string amplitudes~\cite{Stieberger:2013wea,Stieberger:2014hba} and are special values of single-valued multiple polylogarithms  \cite{brownCRAS, Schnetz:2013hqa, Brown:2013gia}.  Indeed, it was as argued in \cite{D'Hoker:2015qmf}, and demonstrated explicitly in \cite{Zerbini}, that the Laurent series of a modular graph form of high enough weight has coefficients that are irreducible svMZVs.  Therefore the condition that the Laurent series for a polynomial of modular graph forms be zero must involve the subtle algebraic relations between svMZVs.  These relations may be viewed as an elliptic extension of the polynomial relations between svMZVs. 
 
 \sm
 
A possibly related and very interesting question is how modular graph forms may be connected to the elliptic multiple polylogarithms of Brown and Levin \cite{Brown:2011}.  These arise in the expansion of one-loop amplitudes in open string theory as iterated integrals on an elliptic curve \cite{Broedel:2014vla, Broedel:2015hia}.
 
 \sm
 
Finally, the existence of polynomial identities between modular graph functions may now be approached in two different ways. The first is by matching their Laurent polynomial at the cusp $\tau_2 \to \infty$, while the second is by the use of Lemma 1 along with algebraic and holomorphic subgraph identities developed in this paper. The method of matching Laurent polynomials was used in \cite{D'Hoker:2015zfa} to conjecture the identities which are ultimately proven in this paper. In the absence of a general theorem which allows one to conclude the vanishing of a modular graph function whose Laurent polynomial vanishes, the method based on matching Laurent polynomials cannot produce results beyond the conjectural level. 

\sm

At a practical level, the calculation of the Laurent polynomial for a given modular graph function becomes quite difficult to carry out beyond the simplest cases, although in the very special case of $D_\ell$ dihedral modular graph functions a general formula is available \cite{Green:2008uj}. The method based on the sieve algorithm,  holomorphic subgraph reduction, and Lemma 1, however, allows for a relatively simple and completely systematic derivation of polynomial identities at any given weight $w$, and automatically provides proofs of the identities.  The search for such identities could be rather easily automatized with the help of a computer program. 

\sm

It would therefore be very interesting to discover if it is a general property that the Laurent series of a modular graph form determines the complete function.   Elucidating the relation between the holomorphic approach of the present paper and the approach of matching the Laurent series at the cusp  is a goal that will be reserved for future work.

\bigskip

\subsection*{Acknowledgments}

We are happy to acknowledge the long-time collaboration with Pierre Vanhove on work that led to the present results. We have also benefited from conversations and correspondence with Pierre Deligne, William Duke, Stephen Miller, and Don Zagier over the course of this work. This research was supported in part by National Science Foundation grants PHY-13-13986 and PHY-16-19926 and by the European Research Council under the European Community's Seventh Framework Programme (FP7/2007-2013)/ERC grant agreement no. [247252]. ED is pleased to acknowledge support by a Fellowship from the Simons Foundation. Finally, we are very grateful to the referee for an extremely careful reading of the manuscript and for many helpful comments, all of which we have incorporated.

\newpage

\appendix

\section{Modular covariant derivatives}
\setcounter{equation}{0}
\label{app:AA}

In this appendix, we shall  compare various customary normalizations for the modular covariant derivatives on the Poincar\'e upper half plane $\mH$. We parametrize $\mH$ by a complex parameter $\tau = \tau_1 + i \tau_2$ for $\tau_1, \tau_2 \in \RR$ and $\tau_2 >0$, and endow $\mH$ with the Poincar\'e metric of unit negative constant curvature,
\bea
\label{A1}
 ds^2 = { d \tau \, d\bar \tau \over \tau _2^2}
 \, .
\eea
The group $SL(2,\RR)$ on $\mH$ by M\"obius transformations,
\bea
\label{A2}
\tau \to \tau' = { \alpha \tau + \beta \over \gamma \tau + \delta} 
\hskip 1in 
\left ( \matrix{ \alpha & \beta \cr \gamma & \delta \cr} \right ) \in SL(2,\RR)
\eea
and maps $\mH$ into itself while leaving the Poincar\'e metric invariant. The modular group $SL(2,\ZZ)$ is the discrete subgroup of $SL(2,\RR)$ which acts on $\mH$ by the transformations given in (\ref{A2}) with $\alpha, \beta, \gamma, \delta \in \ZZ$. The element $-I \in SL(2,\ZZ)$ does not transform $\tau$, and we define $PSL(2,\ZZ) = SL(2,\ZZ)/\{ \pm 1\}$. Since $PSL(2,\ZZ)$ leaves the Poincar\'e metric invariant, it descends to a metric on the quotient $\cM_1=\mH/PSL(2,\ZZ)$, which is the moduli space of complex structures of oriented genus-one Riemann surfaces, i.e. tori.

\sm

A form $f^{(\w,\w')}$ has modular weight $(\w,\w')$ if the combination $f^{(\w,\w')} (\tau, \bar \tau)d\tau^{{\w \over 2}} d\bar \tau^{{\w' \over 2}}$ is invariant under $SL(2,\ZZ)$. Equivalently, its components transform as follows,
\bea
\label{A3}
f^{(\w,\w')} (\tau', \bar \tau') = (\gamma \tau + \delta )^\w (\gamma \bar \tau + \delta)^{\w'} \, f^{(\w,\w')} (\tau, \bar \tau)
\, .
\eea
We shall denote the space of all such forms by $\cF^{(\w,\w')}$. Note that complex conjugation interchanges $\w$ and $\w'$, so that a reality condition may be imposed on forms with $\w'=\w$.

\sm

One may regard $\mH$ as a Riemann surface, specifically a non-compact sphere with one puncture and two orbifold points where the action of $PSL(2,\ZZ)$ on $\mH$ is not transitive.  The transformation property of a form of modular weight $(\w,\w')$ of (\ref{A3}) then simply corresponds to the transformation property of a conformal tensor  of conformal weight $\left ( { \w\over 2}, {\w' \over 2} \right )$.

\sm

There are three natural customary choices for covariant derivatives (see for example \cite{D'Hoker:1988ta}) acting on the space $\cF^{(\w,\w')}$ of forms of modular weight $(\w,\w')$, 
\bea
\nabla _\tau : \cF^{(\w,\w')} & \to & \cF^{(\w+2,\w')}
\no \\
D_{\hat \tau} : \cF^{(\w,\w')} & \to & \cF^{(\w+1,\w'-1)}
\no \\
\nabla ^{\bar \tau} : \cF^{(\w,\w')} & \to & \cF^{(\w,\w'-2)}
\eea
along with their complex conjugates. In differential geometry, the first and the last correspond to covariant derivatives with respect to an affine connection acting on conformal tensors, expressed with Einstein indices $\tau$ and $\bar \tau$, while the middle case corresponds to the covariant derivative with respect to a $U(1)$ spin connection, expressed with orthonormal frame indices $\hat \tau$ and $\hat {\bar \tau}$. The Poincar\'e metric of (\ref{A1}) is expressed in conformal coordinates $\tau$ and $ \bar \tau$ with metric components $g_{\tau \bar \tau}=1/(2 \tau_2^2)$ and $g_{\tau \tau} = g _{\bar \tau \bar \tau}=0$, and frame components $e_\tau {}^{\hat \tau} = e_{\bar \tau} {}^{\hat {\bar \tau}} =1/\tau_2$ and $e_\tau {}^{\hat {\bar \tau}} =e_{\bar \tau} {}^{\hat \tau} =0$. The affine and $U(1)$ connections may then be evaluated with the help of standard differential geometry formulas, and we find,
\bea
\nabla _\tau f^{(\w,\w')} & = &  \left(  \p_\tau + {i  \w \over 2 \tau_2} \right)\, f^{(\w,\w')}(\tau, \bar \tau)
\no \\
D_{\hat \tau} f^{(\w,\w')} & = &  \left( \tau_2 \p_\tau + {i  \w \over 2} \right)\, f^{(\w,\w')}(\tau, \bar \tau)
\no \\
\nabla ^{\bar \tau} f^{(\w,\w')} & = &  \left(  2\tau_2^2 \p_\tau + i  \w \tau_2  \right)\, f^{(\w,\w')}(\tau, \bar \tau)
\, .
\eea
Note that $\nabla _\tau$ is known as the Maass weight-changing differential operator, and the other operators are related to $\nabla _\tau$ by $D_{\hat \tau} = \tau _2 \nabla _\tau$ and $\nabla ^\tau = 2 \tau_2^2 \nabla _\tau$.  For the purpose of this paper, it is the last derivative, and its complex conjugate, that will be particularly useful.  It will be convenient to introduce the following notation,
\bea
\label{nabla}
\nabla = i \, \nabla ^{\bar \tau}
\hskip 1in 
\nabla f^{(0,\w')} = 2 i \tau_2^2 \p_\tau f^{(0,\w')}
\, .
\eea
More generally, upon multiplying a form $f^{(\w,\w')}$ of arbitrary modular weight $(\w,\w')$ by a factor  of $\tau_2^\w$ will reduce its holomorphic weight to zero, in which case no connection is required in the covariant derivative $\nabla$, as shown in the second formula of (\ref{nabla}).

\sm

The ``$U(1)$ covariant derivative" $D_{\hat \tau}$  that maps modular weight $(\w,\w')$ into $(\w+1,\w'-1)$ is particularly useful when considering forms in which $\w'=-\w$, since the derivative maps those forms into forms of the same type.  In that case $f^{(\w,-\w)}$ has $U(1)$ charge $q_u = 2\w$ and the modular transformation is a phase transformation $U(1)$.   The derivative $D_{\hat \tau}$ maps charge $q_u$ to $q_u+2$ (and the complex conjugate derivative maps $q_u$ to $q_u-2$).  
 
\sm

 This derivative $U(1)$-covariant $D_{\hat \tau}$ has proved useful in considering the S-duality of Type IIB superstring theory, where the duality group is $SL(2,\ZZ)$  and the  massless fields are charged under the $U(1)$ R-symmetry group.   The higher derivative interactions in the low energy expansion of scattering amplitudes violate conservation of $q_u$ by even integers.  The
 coefficients of these interactions are modular forms that are related to each other by covariant differentiation.   For example, the coefficient of the $U(1)$-conserving $\cR^4$ interaction, where $\cR$ stands for the Riemann tensor,  is the non-holomorphic Eisenstein series $E_{\frac{3}{2}}(\tau)$ (where $\tau$ is the complex scalar of the Type IIB superstring theory).   The interaction $\lambda^{16}$, where $\lambda$ is the spin-$1/2$ dilatino that has $q_u = -3/2$,  violates the $U(1)$ charge by  $\Delta q_U=-24$.  The coefficient of $\lambda^{16}$  is a $(12,-12)$ modular form given by 
 $(D_{\hat \tau})^{12} \, E_{\frac{3}{2}}$. More generally,  the coefficients of other processes that have $0 <-\Delta q \le 24$ are given by $ (D_{\hat \tau})^{-\Delta q_u/2} \, E_{\frac{3}{2}}$.

\section{Review of Eisenstein regularized sums}
\setcounter{equation}{0}
\label{app:B}

In this appendix, we shall evaluate the Eisenstein regularized sums for some of the ingredients needed in the evaluation of holomorphic subgraph reduction in section \ref{sec:5}. In particular, we need to evaluate $Q_2(0)$ and $Q_1(p_0)$, which we do in turn.

\subsection{Evaluating $Q_2(0)$} 

By definition of the Eisenstein regularized  sums, the value of $Q_2(0)$ is given by the following expression,
\bea
Q_2(0) = \lim _{P \to \infty} \left ( 
\sum _{ - P \leq m \leq P} ' { 1 \over m^2} 
+ \sum _{ - P \leq n \leq P} ' ~
\sum _{ - P \leq m \leq P}  { 1 \over (m  + n \tau )^2}  \right )
\, .
\eea
Note that there is no prime on top of the second $m$-sum, so that the term with $m=0$ must be included in the sum. The sum in the first term gives $2 \zeta (2)= \pi^2/3$, while the $m$-sum in the second term is performed using the summation formula,
\bea
\sum _{m \in \ZZ} { 1 \over (z+m)^2} = - 4 \pi ^2 { e^{2 \pi i z} \over (1 - e^{2 \pi i z} )^2}
\, .
\eea
The remaining summation over $n$ is absolutely convergent, so that the limit $P \to \infty$ may be taken directly, and we find,
\bea
Q_2(0) = { \pi^2 \over 3} - 8 \pi ^2  \sum _{n =1} ^ \infty { q^n \over (1-q^n)^2}
= { \pi^2 \over 3} - 8 \pi ^2  \sum _{n =1} ^ \infty { n q^n \over 1-q^n}
\, .
\eea
where $q= e^{2 \pi i \tau}$.
It is possible to identify this quantity with a derivative involving the Dedekind $\eta$-function,
\bea
Q_2(0) = - 4 \pi i \p_\tau \ln \eta (\tau)
\hskip 1in 
\eta (\tau) = q^{1 / 24} \prod _{n=1}^\infty (1-q^n)
\, .
\eea

\subsection{Evaluating $Q_1(p_0)$}

Given that the Eisenstein regulator is symmetric under $p_+ \to - p_+$, we readily have $ Q_1(-p_0) = - Q_1(p_0)$.  Next, we decompose $p_0$ and $p_+$ as follows,
\bea
p_0 & = & m_0  + n_0 \, \tau \hskip 1in (m_0,n_0) \not= (0,0) \qquad m_0 > 0
\no \\
p_+ & = & m_+  + n_+ \, \tau \hskip 1in (m_+ ,n_+) \not= (0,0)
\, .
\eea
The regularized sums are given by the limit $P \to \infty$ of the following expression,
\bea
2 Q_1(p_0) & = & 
\sum _{{n_+ \not= 0 \atop - P \leq n_+ \leq P}}
\sum _{m_+ = - P } ^P
\left ( { 1 \over m_+ + n_+ \tau } - {1 \over m_+ -m_0 +(n_+-n_0) \tau} \right )
\\ &&
+ \sum _{{m_+ \not= 0 \atop - P \leq m_+ \leq P}}
\left ( { 1 \over m_+} - { 1 \over m_+ - m_0 - n_0 \tau} \right )
+ \sum _{{m_+ \not= m_0 \atop - P \leq m_+ \leq P}}
\left ( { 1 \over m_+ + n_0 \tau } - { 1 \over m_+ - m_0 } \right )
\, .
\no
\eea
The sums over $1/m_+$ and $1/(m_+-m_0)$ cancel in the limit $P \to \infty$.
To perform the sums in $m_+$, we use the summation formula,
\bea
\lim_{P \to \infty} \sum _{m_+ = - P } ^P  { 1 \over z+m_+} = - i \pi \, { 1+ e^{2 \pi i z} \over 1 - e^{2 \pi i z} }
\, .
\eea
The resulting terms may be arranged as follows,
\bea
Q_1(p_0) = -{ 1\over p_0} - { i \pi \over 2} \lim _{P \to \infty}  
\sum _{n_+ =- P} ^P \left (  { 1 +q^{n_+} \over 1 - q^{n_+}} (1-\delta_{n_+,0})
- { 1 +q^{n_+-n_0} \over 1 - q^{n_+-n_0}} (1-\delta_{n_+,n_0}) \right )
\, .
\eea
Shifting the summation variable in the second term $ n_+$ to $n_+ + n_0$, we have
\bea
Q_1(p_0) = -{ 1\over p_0} - { i \pi \over 2} \lim _{P \to \infty}  
\left ( \sum _{n_+ =- P} ^P - \sum _{n_+ =- P -n_0} ^{P -n_0}  \right )
  { 1 +q^{n_+} \over 1 - q^{n_+}} (1-\delta_{n_+,0})
  \, .
\eea
Using the fact that the limit of the $q$-dependent factor is $+1$ for $n_+ \gg 1$,
and $-1$ for  $n_+ \ll -1$, we find, 
\bea
Q_1(p_0) = -{ 1 \over p_0 } - i \pi n_0
\, .
\eea
While this formula was derived for $n_0>0$, it is actually valid for all $n_0$ in view of the odd parity of $Q_1(p_0)$.  It may be recast as follows,
\bea
Q_1(p_0) = - { 1 \over p_0} - i \pi \, { p_0 - \bar p_0 \over \tau - \bar \tau}
\, .
\eea

\subsection{Modular transformations}

The functions $Q_k(p_0)$ are modular forms for $k \geq 3$, but neither $Q_2(p_0)$, nor $Q_1(p_0)$ are separately modular forms. Modular transformations act on $\tau$ as given in (\ref{A2}),
The basic rule for the transformation of the momenta $p_0$, and their components $m_0, n_0$, is as follows,
\bea
p_0 \to p_0' = { p_0 \over \gamma \tau + \delta}
& \hskip 1in & 
m_0  \to  m_0' = + \alpha \, m_0 - \beta \, n_0
\no \\ &&
n_0  \, \to  \, n_0' =  \, - \gamma \, m_0 + \delta \, n_0
\, .
\eea
The Dedekind $\eta$-function transforms as follows, 
\bea
\eta (\tau') = \ep \, (\gamma \tau + \delta)^{\half}  \eta (\tau) 
\hskip 1in \ep ^{24}=1
\, .
\eea
where $\ep$ depends on the transformation in the full modular group.
As a result, the modular transformation laws of $Q_1$ and $Q_2$ are given by,
\bea
Q_2'(p_0') & = & (\gamma \tau + \delta )^2 Q_2(p_0) - 2 \pi i \gamma ( \gamma \tau + \delta)
\no \\
Q_1'(p_0') & = & (\gamma \tau + \delta) Q_1(p_0) + \pi i \gamma p_0
\, .
\eea
The combination entering $\cG_{a_+, \, a_-}(p_0)$ is given by,
$2 Q_1(p_0)+ p_0 Q_2(p_0)$, and it manifestly transforms as a modular form of weight 1,
\bea
2 Q_1'(p'_0)+ p' _0 Q_2'(p'_0) = (\gamma \tau + \delta ) \Big (2 Q_1(p_0)+ p_0 Q_2(p_0) \Big )
\, .
\eea
In view of this, we may introduce a combination of $Q_2(0)$ which transforms like a modular form of weight 2, but which is not holomorphic, 
\bea
G_2  = Q_2(0) - { \pi \over \tau_2}
\, .
\eea
We then have,
\bea
2p_0Q_1(p_0) + p_0^2 Q_2(p_0) = - 3 + p_0^2 G_2 + {\pi \over \tau_2} p_0 \bar p_0
\, .
\eea

\section{Auxiliary identities}
\setcounter{equation}{0}
\label{app:D}

The following identities, which have been derived with the help of the holomorphic  subgraph reduction identities of subsection \ref{sec:54}, are used in the proof the three weight 5 conjectures,
\bea
\nabla   \cC^+ \! \left [ \matrix{ 6 \, 1 \, 1 \cr 0 \, 1 \, 1 \cr } \right ]
& = & 
+ 40 \,  \cC^+ \! \left [ \matrix{ 9 \, 0 \cr 1 \, 0 \cr } \right ] 
- 6 \,  \cC^+ \! \left [ \matrix{ 4 \, 0 \cr 0 \, 0 \cr } \right ]\,  \cC^+ \! \left [ \matrix{ 5 \, 0 \cr 1 \, 0 \cr } \right ]
- 10 \,  \cC^+ \! \left [ \matrix{ 6 \, 0 \cr 0 \, 0 \cr } \right ]\,  \cC^+ \! \left [ \matrix{ 3 \, 0 \cr 1 \, 0 \cr } \right ]
\no \\
\nabla   \cC^+ \! \left [ \matrix{ 5 \, 2 \, 1 \cr 1 \, 0 \, 1 \cr } \right ]
& = & 
-26  \,  \cC^+ \! \left [ \matrix{ 9 \, 0 \cr 1 \, 0 \cr } \right ] 
+ 9  \,  \cC^+ \! \left [ \matrix{ 4 \, 0 \cr 0 \, 0 \cr } \right ]\,  \cC^+ \! \left [ \matrix{ 5 \, 0 \cr 1 \, 0 \cr } \right ]
+ 5 \,  \cC^+ \! \left [ \matrix{ 6 \, 0 \cr 0 \, 0 \cr } \right ]\,  \cC^+ \! \left [ \matrix{ 3 \, 0 \cr 1 \, 0 \cr } \right ]
\no \\
\nabla   \cC^+ \! \left [ \matrix{ 5 \, 2 \, 1 \cr 0 \, 1 \, 1 \cr } \right ]
& = & 
+ 14  \,  \cC^+ \! \left [ \matrix{ 9 \, 0 \cr 1 \, 0 \cr } \right ] 
- 6  \,  \cC^+ \! \left [ \matrix{ 4 \, 0 \cr 0 \, 0 \cr } \right ]\,  \cC^+ \! \left [ \matrix{ 5 \, 0 \cr 1 \, 0 \cr } \right ]
\no \\
\nabla   \cC^+ \! \left [ \matrix{ 4 \, 3 \, 1 \cr 1 \, 0 \, 1 \cr } \right ]
& = & 
- 19  \,  \cC^+ \! \left [ \matrix{ 9 \, 0 \cr 1 \, 0 \cr } \right ] 
+ 9  \,  \cC^+ \! \left [ \matrix{ 4 \, 0 \cr 0 \, 0 \cr } \right ]\,  \cC^+ \! \left [ \matrix{ 5 \, 0 \cr 1 \, 0 \cr } \right ]
\no \\
\nabla   \cC^+ \! \left [ \matrix{ 4 \, 3 \, 1 \cr 0 \, 1 \, 1 \cr } \right ]
& = & 
+ 23  \,  \cC^+ \! \left [ \matrix{ 9 \, 0 \cr 1 \, 0 \cr } \right ] 
- 9  \,  \cC^+ \! \left [ \matrix{ 4 \, 0 \cr 0 \, 0 \cr } \right ]\,  \cC^+ \! \left [ \matrix{ 5 \, 0 \cr 1 \, 0 \cr } \right ]
\, .
\eea

\end{document}